\documentclass[12pt]{article}
\pdfoutput=1
\usepackage{jheppub}

\usepackage{subfigure}
\usepackage{amsmath,amssymb,amsthm,amsfonts,array,bbm}
\usepackage{graphicx}
\usepackage{color}
\usepackage[american]{babel}

\usepackage[mathscr]{euscript}
\usepackage{MnSymbol}

\usepackage{tikz}
\usetikzlibrary{automata}
\usetikzlibrary{arrows}
\usetikzlibrary{calc}
\usetikzlibrary{decorations.markings}
\usetikzlibrary{decorations.pathreplacing}
\usetikzlibrary{intersections}
\usetikzlibrary{positioning}
\usetikzlibrary{topaths}
\usetikzlibrary{shapes.geometric}
\usetikzlibrary{shapes.misc}
\tikzset{->-/.style = {
    decoration = {markings, mark = at position #1 with {\arrow{>}}},
    postaction = {decorate}}}
\tikzset{color-group/.style = {
    shape = circle,
    minimum size = 2.5ex,
    inner sep = .5ex,
    draw}}
\tikzset{flavor-group/.style = {
    shape = rectangle,
    minimum size = 2.5ex,
    inner sep = .5ex,
    draw}}
\tikzset{cf-group/.style = {
    shape = rounded rectangle,
    rounded rectangle right arc = none,
    draw}}
\tikzset{fc-group/.style = {
    shape = rounded rectangle,
    rounded rectangle left arc = none,
    draw}}
\tikzset{cross/.style={minimum width=1pt, path picture={
      \draw[black, very thick]
               (path picture bounding box.south east)
            -- (path picture bounding box.north west)
               (path picture bounding box.south west)
            -- (path picture bounding box.north east);
          }}}

\newcommand{\lp}{\left(}
\newcommand{\rp}{\right)}
\newcommand{\ti}{\widetilde}
\newcommand{\tr}{\textrm{Tr} \,}

\newcommand{\p}{\partial}
\newcommand{\ch}{\text{ch}}
\newcommand{\sh}{\text{sh}}

\def\CP1{\bC\bP^1}

\numberwithin{equation}{section}

\newcommand{\be}{\begin{equation}} \newcommand{\ee}{\end{equation}}
\newcommand{\bea}{\begin{equation} \begin{aligned}} \newcommand{\eea}{\end{aligned} \end{equation}}

\newcommand{\vev}[1]{{\langle {#1} \rangle}}
\newcommand{\dd}{\text{d}}

\newcommand{\ceil}[1]{\lceil #1 \rceil}

\newcommand{\cC}{\mathcal{C}}

\newcommand{\cH}{\mathcal{H}}
\newcommand{\cI}{\mathcal{I}}

\newcommand{\cN}{\mathcal{N}}

\newcommand{\cW}{\mathcal{W}}

\newcommand{\bC}{\mathbb{C}}

\newcommand{\bP}{\mathbb{P}}

\newcommand{\bR}{\mathbb{R}}
\newcommand{\bZ}{\mathbb{Z}}

\def\su{\mathfrak{su}}

\def\repa{\raise4pt\hbox{$\square$}\mkern-14mu\raise-4pt\hbox{$\square$}}
\def\repab{\overline{\raise4pt\hbox{$\square$}\mkern-14mu\raise-4pt\hbox{$\square$}\mkern-1mu}}

\numberwithin{equation}{section}       

\begin{document}

\begin{titlepage}

\vspace*{-2cm} 
\begin{flushright}
	{\tt KIAS-P19013 \\ CERN-TH-2019-022} 
\end{flushright}
	
	\begin{center}

		\vskip .5in 
		\noindent

		{\Large \bf{On Monopole Bubbling Contributions \\  
		 \vspace{0.4 cm} to  't Hooft Loops}}
		
		\bigskip\medskip
		
		Benjamin Assel$^1$ and Antonio Sciarappa$^2$\\
		
		\bigskip\medskip
		{\small 
			$^1$ Theory Department, CERN, CH-1211, Geneva 23, Switzerland \\
			$^2$ School of Physics, Korea Institute for Advanced Study \\
85 Hoegiro, Dongdaemun-gu, Seoul 130-722, Republic of Korea			
		}
		
		\vskip .5cm 
		{\small \tt benjamin.assel@gmail.com, asciara@kias.re.kr}
		\vskip .9cm 
		{\bf Abstract }
		\vskip .1in
\end{center}
	
\noindent
Monopole bubbling contributions to supersymmetric 't Hooft loops in 4d $\cN=2$ theories are computed by SQM indices. As recently argued, those indices are hard to compute due to the presence of Coulomb vacua that are not captured by standard localization techniques.  We propose an algorithmic method to compute the full bubbling contributions that circumvent this issue, by considering SQM with more matter fields and isolating the bubbling terms as residues in flavor fugacities. The enlarged SQMs are read from brane configurations realizing the bubbling sector of a given 't Hooft loop. We apply our technique to loop operators in $\cN=2$ conformal SQCD theories. In addition we embed this discussion in the larger setup of a 5d-4d system interacting along a line, associated to the brane systems previously discussed. The bubbling terms arise from residues of specific instanton sectors of 5d line operators in this context.


\vfill

\end{titlepage}

\setcounter{page}{1}

\noindent\hrulefill
\tableofcontents

\noindent\hrulefill

\bigskip



\section{Introduction}
\label{sec:Intro}

't Hooft loops are one of the most basic and fundamental line operators of gauge theories. They are defined in the path integral formulation of a theory by imposing specific boundary conditions on the fields along a line. In particular, there is a quantized magnetic flux emanating from every point along the line. They are the magnetic cousins of Wilson loops and one can think of them as the worldline of a heavy magnetically charged particle. The vacuum expectation value (vev) of 't Hooft loops and Wilson loops together are parameters which control the low energy non-perturbative behaviour of gauge theories \cite{tHooft:1977nqb}.
't Hooft loops play a prominent role in many deep aspects of supersymmetric gauge theories, such as S-duality \cite{Kapustin:2005py}, wall-crossing phenomenon \cite{Gaiotto:2008cd} or the AGT duality \cite{Alday:2009aq,Gomis:2010kv}. 

\medskip

In 4d $\cN=2$ Lagrangian theories, the vev of  half-BPS 't Hooft loops wrapped on $S^1$ in $S^1 \times \bR^3$ in the Coulomb phase was computed exactly in \cite{Ito:2011ea} using the technique of supersymmetric localization. This followed earlier localization computation in \cite{Gomis:2011pf} for 't Hooft loops placed at the equator of $S^4$. More precisely the background considered in \cite{Ito:2011ea} is $S^1 \times \bR^2_{\epsilon_+} \times \bR$, where $\epsilon_+$ is the parameter of an Omega background deformation of the $\bR^2$ plane \cite{Nekrasov:2010ka}. Supersymmetric loops are then wrapping $S^1$, sitting at the origin in $\bR^2_{\epsilon_+}$ and placed at any point on $\bR$. This setup preserves 1d $\cN=(0,2)$ supersymmetry at non-zero $\epsilon_+$ (and 1d $\cN=(0,4)$ at $\epsilon_+=0$).  By a standard argument the vev of a BPS loop is independent of its position along $\bR$. It takes the form of an index which counts {\it framed} BPS states \cite{Gaiotto:2010be}, which are the BPS states of the theory in the presence of the line defect. Additional results for the $\cN=2^\ast$ theory were presented in \cite{Brennan:2018yuj}.

The result of the localization computation has an interesting non-trivial structure related to the monopole bubbling phenomenon \cite{Kapustin:2006pk}. This is a subtle phenomenon of non-abelian gauge theories, where the magnetic charge $B$ (an element from the cocharacter lattice) emanating from the loop is screened by a smooth 't Hooft-Polyakov monopole of ``smaller" magnetic charge $B-v$, collapsing on the defect. The resulting configuration is that of a 't Hooft defect of smaller magnetic charge $v$.
The exact vev of a 't Hooft loop $L_B$  is organized as a sum of terms associated to the bubbling magnetic sectors $v$. Schematically,
\be
\vev{L_{B}} = \sum_{ ||v|| \le ||B|| }  e^{v.b} \, Z_{\rm 1-loop}(a;v) \, Z_{\rm mono}(a;B,v) \,,
\ee
where $a$ and $b$ refer to the Coulomb vevs of Cartan vector multiplet fields on $S^1 \times \bR^3$ (see section \ref{ssec:Generalities}), and play the role of chemical potentials for electric and magnetic charges respectively. For each monopole bubbling sector the term $Z_{\rm 1-loop}(a;v)$ arises from a one-loop determinant in the localization computation, whereas the term $Z_{\rm mono}(a;B,v)$ is a weight that is computed as the index of an ADHM supersymmetric quantum mechanics (SQM), similarly to the instanton weight $Z_{\rm inst}$ of the Nekrasov instanton partition function \cite{Nekrasov:2002qd}. 

All the pieces entering in the above formula are well-understood and easy to express, except for the bubbling factors $Z_{\rm mono}$. Each term $Z_{\rm mono}(B,v)$ is equal to the supersymmetric index of an SQM which localizes to a matrix integral. In many instances, the contour of integration for this matrix integral is given by the Jeffrey-Kirwan (JK) prescription \cite{Hori:2014tda, Hwang:2014uwa}, which sums over the {\it Higgs} vacua of the SQM.  It was pointed out in \cite{Brennan:2018rcn} that in some cases, and in particular in conformal SQCD theories, this recipe does not yield the correct result, because it misses contributions from {\it Coulomb} SQM vacua that belong to a continuum of states. The observation of \cite{Brennan:2018rcn} is that the extra contributions are necessary to match the AGT dual observables in Liouville/Toda 2d CFT that were computed in \cite{Gomis:2010kv}. To compute the SQM index correctly one then has to study the supersymmetric ground states of the SQM theory and this was carried out in \cite{Brennan:2018rcn} for the simplest cases, for instance for the bubbling factor of the minimal 't Hooft loop in the $SU(2)$, $N_f=4$ theory. Unfortunately such an analysis is discouragingly tedious and one would like to use a simpler method for practical purposes.
The main point of this paper is to provide such a method. 

\medskip

We make progress on this situation by proposing an algorithmic method which computes the correct bubbling factors using only the standard JK prescription. The main idea is to embed the ADHM SQM of a given bubbling term into a larger SQM theory, which has some extra matter fields and for which the supersymmetric index can be computed reliably using JK residues. The bubbling term is then obtained by taking specific residues in the flavor fugacities of the extra matter fields. One can think of the larger, or ``improved", SQM as a UV theory with massive matter fields, whose low-energy effective theory is the original SQM. The flavor residues of the UV SQM index then isolate the BPS vacua contributing to the low energy SQM index, capturing the Higgs and Coulomb contributions. The reason why the JK prescription can be used reliably in the improved SQM is related to the fact that the potential of this theory is unbounded, due to the presence of the extra matter fields, and there is no Coulomb vacua there.
We study conformal SQCD theories, namely $SU(N)$ (or $U(N)$) theories with $N_f=2N$ fundamental hypermultiplets and we focus on the minimal 't Hooft loop, dyonic loop and next-to-minimal 't Hooft loop. Our method can be applied to any higher charge loops as well.

The original SQM is defined by the type IIB brane realization of the 't Hooft loop bubbling sector. Indeed 't Hooft loops in 4d $\cN=2$ SQCD theories can be realized in type IIB branes systems by adding NS5 branes to the D3-D7 system and the bubbling sectors arise from D1 strings stretched between D3 branes (orientations are given in Table \ref{tab:orientations}).\footnote{To be more precise, the IIB brane setup realizes the loop insertion in the $\cN=2^\ast$ theory, which has an extra massive adjoint hypermultiplet (the mass arises from a geometric deformation in the space transverse to the D3 branes). We always think of the limit of infinite mass, when we integrate out the massive adjoint hypermultiplet.} This was studied in \cite{Brennan:2018moe} and in \cite{Brennan:2018yuj}.
The ADHM SQM computing the bubbling factor is read from such a brane setup as the low-energy theory on the D1 strings worldvolume. 
The simplest example is in the $SU(2)$, $N_f=4$ theory for the minimal 't Hooft loop, with the brane setup realizing the bubbling factor and the associated SQM given in Figure \ref{U2_0}. 

We argue that the brane setup considered so far are incomplete because they do not take into account the bending effect and charge-changing effect on the 5 branes due to the presence of the D7 branes. Taking into account these effects leads to setups with intersecting $(1,q)$ 5-branes which needs to be resolved by adding 5-brane junctions with D5 segments \cite{Aharony:1997bh}. In the completed brane setup the D3 and D7 branes sit inside a 5-brane web. The improved SQM is then read from the completed brane setup as the worldvolume theory on the D1-strings corresponding to a given bubbling sector. The additional matter fields come from D1-D5 and D3-D5 strings, and the residues to be taken are residues in the D5 flavor fugacities. Therefore our method arises naturally from the complete brane realization of the bubbling terms in IIB string theory. For the minimal 't Hooft loop bubbling in the $SU(2)$, $N_f=4$ theory, the complete brane setup is shown in Figure \ref{U2_2}, with the improved SQM. Computing the index of this improved SQM by JK residues and taking the residues over the two flavor fugacities, we reproduce the full bubbling factor found in \cite{Brennan:2018rcn} through tedious computations.
 We compute bubbling factors in $SU(N)$ (and $U(N)$) SQCD theories for the minimal and next-to-minimal 't Hooft loops and we apply it also to the computation of a minimal dyonic loop. We emphasize that the new method is easy and rapid to perform (with sufficient computer assistance). The only restriction arises from the complexity of the (improved) SQM, and the number of residues one has to compute by the JK prescription, which is rapidly growing with the magnetic charge of the 't Hooft loop. This is a standard limitation in such computations. 

\medskip

As a check, we compare our results with the OPE between line operators, which is computed by a non-commutative Moyal product between the vevs of the individual loops. In the presence of the Omega deformation with parameter $\epsilon_+$, the loops are inserted along a line $\bR$ with a certain ordering. The vev of this operator with multiple insertions depends on the positions of the loops only through their ordering. The OPE between two loops then defines a non-commutative product on the algebra of BPS loop operators. It turns out that this non-commutative product is realized by a Moyal product based on the Fenchel-Nielsen coordinates $a$ and $b$. We verify that the Moyal product of minimally charged loops expands as linear combinations of other loops, using our results. In particular we check that the product of the vevs  (or the OPE) of two minimally charged 't Hooft loops yields the vev of the next-to-minimal 't Hooft loop. 

Along the way we clarify some points about the OPE between line operators and the vev of loops computed by supersymmetric localization. The results of \cite{Ito:2011ea} and the results that we present in this paper for the bubbling factors are the vev of loop operators defined by singular boundary conditions in the path integral along a single line where the defect is inserted. This is by definition invariant under a $\bZ_2$ symmetry that sends $\epsilon_+ \to -\epsilon_+$. Indeed this operation can be regarded as a reflection along the $\bR$ line (where the operator sits at a point) and a reflection in the R-symmetry group, and it turns out that the BPS loop operators are invariant under this $\bZ_2$ symmetry. This implies that all 't Hooft loop vevs, and even all bubbling factors, must have this symmetry. As pointed out in \cite{Brennan:2018rcn}, this symmetry is respected for the full bubbling term (including SQM Coulomb vacua). The symmetry is far from obvious in the actual expressions one obtains and it constitutes a powerful check of the results.
On the other hand the OPE between two colliding loops depends on the ordering between the loops  along $\bR$ (which is why it defines a non-commutative product) and thus is, in general, not invariant under $\epsilon_+ \to -\epsilon_+$. It can be expanded in a linear combination of loop vevs, which are themselves $\bZ_2$ invariant, but with coefficients depending on $\epsilon_+$ (responsible for the global $\bZ_2$ non-invariance).\footnote{Trying to {\it define} higher charge loop operators through the OPE of smaller charge loops is unnatural in this context since this has ordering ambiguities.}

\medskip

Finally we relate our construction to the study of 5d loop operators that was carried out in \cite{Assel:2018rcw} (building on \cite{Kim:2016qqs, Nekrasov:2015wsu, Tong:2014yna, Tong:2014cha}), by regarding the complete brane setups as a coupled 5d-4d-1d systems. The 5-brane web that arises in the complete brane setup supports at low energies a 5d $\cN=1$ theory which is the Coulomb phase of a deformed 5d SCFT (the undeformed CFT is at infinite YM coupling). The 5d theory is read from the rules found in \cite{Aharony:1997bh}. In addition there is still the 4d theory living on the D3 branes. The 5d and 4d theories do not live on the same spacetime, rather they share only one space direction, where 1d fermions sourced by D3-D5 branes live. From the point of view of the 5d theory or of the 4d theory, this interaction corresponds to a half-BPS line operator. The presence of D1 strings, which are associated to bubbling sectors of the 4d theory, corresponds to instanton sectors in the 5d theory. 
 Such brane systems and the 5d loop operators $L_{\rm SQM}$ (or 5d-4d line defect) that they define were studied in \cite{Assel:2018rcw} (SQM here refers to the 1d theory of fermion matter fields living at the intersection of the 5d and 4d theories). In particular their vev was computed in specific cases as an expansion in the instanton sectors of the 5d theory. One of the main result was that BPS Wilson loops of the 5d theory could be obtained by taking residues of $\vev{L_{\rm SQM}}$ in the D3 flavor fugacities, circumventing the unsolved problem of computing the Wilson loop vevs directly. Now we find that 't Hooft loop bubbling terms are obtained from the same object $\vev{L_{\rm SQM}}$, by first selecting the instanton sector corresponding to the bubbling sector (specified by an array of D1 strings) and then taking residues in the flavor fugacities associated to the D5 branes. The ADHM quiver of the specific 5d instanton sector corresponds to the improved SQM. We find that every bubbling factor can be thought of as the D5 flavor residue in an instanton sector of a $L_{\rm SQM}$ operator, which is the underling deeper object associated to the 5-brane web/D3 configuration.

\bigskip

The paper is organized as follows. In section \ref{sec:tHLandBranes} we review some basics about 't Hooft loops, their brane realization and the computation of bubbling terms as presently known. In section \ref{sec:Main} we present our brane-based algorithm to compute simply and reliably bubbling terms in 't Hooft loop vevs and apply it in the cases mentioned above. In section \ref{sec:starprod} we show that our results are compatible with the OPE, or Moyal product, between loops, and in section \ref{sec:5d4d} we discuss the relation to 5d-4d-1d coupled systems and 5d line operators. We conclude in section \ref{sec:Discussion} with some comments and future directions to continue this work.  In appendix \ref{app:MM} we provide the details about the matrix models computing the index of $\cN=(0,4)$ (or $\cN=(0,4)^\ast$) SQM theories.

\section{'t Hooft loops and brane picture}
\label{sec:tHLandBranes}

\subsection{Generalities}
\label{ssec:Generalities}

We study supersymmetric 't Hooft loop operators in 4d $\cN=2$ $SU(N)$ theories with $N_f = 2N$ flavors of hypermultiplets. The vector multiplet contains  the gauge field $A_\mu$ with field strength $F$, a complex adjoint scalar field $\Phi$ and fermionic fields. Considering Euclidean space with cylindrical coordinates $(\tau, r, \theta, \varphi)$, the 't Hooft loop inserted at radial coordinate $r =0$ and spanning the $\tau$ direction is defined by prescribing a supersymmetric Dirac monopole singularity as an asymptotic behavior for the bosonic fields in the vector multiplet at every point of the straight line \cite{Gomis:2011pf},\footnote{Note that under a shift of $\vartheta \to \vartheta + 2\pi$ the 't Hooft loop singularity is also shifted by a singular piece. This extra singular piece corresponds to the singularity of a (supersymmetric) Wilson loop as described in \cite{Kapustin:2005py}. This encodes the fact that under this shift of the theta angle, the 't Hooft loop acquires electric charge and becomes a dyonic loop. This is the Witten effect \cite{Witten:1979ey} for loop operators in 4d.}
\bea
F &= \frac{B}{2} \sin\theta \, \dd\theta\wedge \dd\varphi -  \frac{i g^2 \vartheta B}{16\pi^2} \frac{\dd \tau \wedge \dd r}{r^2}  \cr
\Phi &= -\frac{B}{2r} \Big( 1 - i \frac{g^2\vartheta}{8\pi^2} \Big)  \,,
\eea
where $g^2$ is the Yang-Mills coupling and $\vartheta$ the theta angle of the $SU(N)$ gauge theory. $B$ is an element of a Cartan subalgebra $\mathfrak{t}$ of the gauge algebra $\su(N)$ and takes values in the lattice of magnetic weights $\Lambda_{mw} =\Lambda_{\rm cochar} = \{ H \in \mathfrak{t} \, | \, \exp(2\pi i H) = 1 \}$, which is the cocharacter lattice of the gauge group \cite{Kapustin:2005py}. Magnetic charges $B$ and $B'$ related by the action of the Weyl group $\cW$ define identical loops.
We represent $B$ as a traceless diagonal matrix with quantized diagonal coefficients $\vec B = (B_1,B_2,\cdots, B_N)$ giving the magnetic charges of the loop. For $SU(N)$, the allowed magnetic charges satisfy $B_i \in \bZ$, $B_1 \ge B_2 \ge \cdots \ge B_N$ and $\sum_i B_i =0$. 
This defines a 't Hooft loop $L_B$. Such a 't Hooft loop satisfies the BPS equation $F_{ij} = \epsilon_{ij}{}^k D_k\Phi$, with $i,j$ indices in $\bR^3$ transverse to the line. It preserves half of the supercharges.

In \cite{Ito:2011ea} (following \cite{Gomis:2011pf}), the VEV of  't Hooft loops $L_B$ in Coulomb vacua of 4d $\cN=2$ theories were computed using supersymmetric localization. In order to do so, the loops were placed in $S^1_{\beta} \times \bR^2_{\epsilon_+}\times \bR$, with $\beta$ the radius of $S^1$ and $\epsilon_+$ an Omega background deformation parameter. In this geometry the loop wraps $S^1$ and is placed at the center in $\bR^2_{\epsilon_+}\times \bR$, although the position along $\bR$ does not matter. The VEV of the loops can be thought of as the Witten index of the theory in the presence of the loop insertion, refined with fugacities,
\be
\vev{L} = \tr_{\cH(L)} (-1)^F e^{-\beta H} e^{\epsilon_+ (J_{12}+R)} e^{a.G} e^{b.\check G} e^{m.G_F} \,,
\ee
where $H$ is the Hamiltonian, $J_{12}$ the generator of rotations in a plane $\bR^2 \subset \bR^3$, $R$ a Cartan generator in the $SU(2)_R$ R-symmetry, $G$ denotes the Cartan generators of global $SU(N)$ gauge transformations, $\check G$ the Cartan generators of the magnetic dual group, and $F$ the Cartan generators of flavor symmetries of the theory. This index receives only contributions from ground states with $H=0$ and is therefore independent of $\beta$. 
The parameters $a = (a_i)$ correspond to the asymptotic holonomy of Cartan gauge field around $S^1$, complexified with the asymptotic value of a chosen scalar in the vector multiplet.  The parameters $b = (b_i)$ correspond to the  asymptotic values of compact scalar fields defined as the dual of Cartan gauge fields on $\bR^3$. They are complexified with the asymptotic value of another scalar in the vector multiplet. $a$ and $b$ are chemical potentials for electric and magnetic charges of the states, respectively. Finally the fugacities $m = (m_k)$ correspond to background (flavor) gauge field holonomies around $S^1$, complexified with real hypermultiplet mass parameters.

The exact result of the localization computation takes the form
\be
\vev{L_{B}} = \sum_{ ||v|| \le ||B|| \atop  v \in B + \Lambda_{\rm cr} }  e^{v.b} \, Z_{\rm 1-loop}(\epsilon_+,a_i,m;v) \, Z_{\rm mono}(\epsilon_+,a,m;B,v) \,.
\ee
This is a sum over monopole bubbling sectors labelled by magnetic charges $v \in \Lambda_{mw}$ such that $B-v$ is an element of the coroot lattice $\Lambda_{\rm cr}$ and the norm $||v||$ (defined by the Killing form on the gauge algebra) is smaller or equal to $||B||$. For $SU(N)$, this means $\sum_{i=1}^N (v_i)^2 \le \sum_{i=1}^N (B_i)^2$, and all the relevant lattices are the same: $\Lambda_{\rm cr} = \Lambda_{\rm cochar} (= \Lambda_{\rm root})$. This lattice is generated by simple roots of $\su(N)$ which we take as $N$-vectors of the form $\pm(0, \cdots, 0, 1 , -1, 0 ,\cdots ,0)$.

The physical interpretation of this sum is that the VEV of the 't Hooft loop with magnetic charge $B$ receives contributions from sectors where the Dirac magnetic singularity is screened by smooth 't Hooft-Polyakov monopoles, whose charges are elements of $\Lambda_{\rm cr}$, which decrease the total magnetic charge of the configuration. Each sector is then labelled with the asymptotic magnetic charge $v$. The contributions from sectors with $||v|| < ||B||$ are called ``monopole bubbling contributions".

The contribution of the $v$-sector is weighted with the coefficient $e^{v.b} = e^{\sum_i v_i b_i}$ and is the product of a 1-loop contribution $Z_{\rm 1-loop}(\epsilon_+,a,m;v)$, which is known,\footnote{We refer the reader to \cite{Ito:2011ea} for explicit expressions.} 
 and a monopole bubbling contribution $Z_{\rm mono}(\epsilon_+,a,m;B,v)$. 
The computation of $Z_{\rm mono}$ proved to be subtle and it was the central topic of \cite{Brennan:2018rcn}. 
 
 $Z_{\rm mono}$ is evaluated as the Witten index of a certain ADHM quiver quantum mechanical theory. One way to find the ADHM quiver is to realize the 't Hooft loop insertion and the monopole screening in a brane system and then recognize the ADHM theory as the theory living on D1 branes. 
 
Let us comment more about the $\epsilon_+$ deformation. This arises from an Omega background deformation in a plane $\bR^2 \subset \bR^3$. To preserve supersymmetry the line operators must sit at the origin in this plane, and at any point along the third direction $\bR$. This introduces an ordering between the loop insertions along $\bR$. The vev of a collection of 't Hooft loop insertions does not depend on the positions of the insertions along $\bR$, except for the ordering between these insertions. This promotes the OPE between loops to a non-commutative product, that we discuss further in section \ref{sec:starprod}. Going back to the insertion of a single 't Hooft loop, we emphasize that the vev of the line operator must be invariant under the $\bZ_2$ symmetry $\epsilon_+ \to -\epsilon_+$. This $\bZ_2$ symmetry can be expressed as a reflection, or orientation reversal, along $\bR$, plus an R-symmetry reflection. The line operator is invariant under these reflection\footnote{The 't Hooft loop is a Lorentz scalar from the point of view of the $\bR^3$ space and the $\bZ_2$ R-symmetry involved acts on Higgs branch operators and not on 't Hooft operators.} and thus its vev should be invariant under $\epsilon_+ \to -\epsilon_+$. This will turn out to be an important consistency requirement.

 \subsection{Brane realization and ADHM quivers}
\label{ssec:BranesADHM}

We review now the brane construction presented in \cite{Brennan:2018rcn,Brennan:2018moe}. We focus on 't Hooft loops and monopole bubbling in the 4d $\cN=2$ $SU(N)$ (or $U(N)$) theories with $N_f = 2N$ fundamental hypermultiplets.

\medskip

We consider a stack of $N$ D3 branes filling the directions $x^{0123}$. The low energy worldvolume theory is the $\cN=4$ $U(N)$ SYM theory.  To obtain the $\cN=2$ SYM theory we place the D3 branes at the center of an Omega background along $x^{6789}$. More precisely the background is $\bR^2_{m} \times \bR^2_{-m}$ with $\pm m$ the Omega background parameter in the two planes. This gives a mass $m$ to the adjoint hypermultiplet in the SYM theory. The limit of large mass $m$ corresponds to  the $\cN=2$ SYM theory (by integrating out the massive adjoint hypermultiplet).
The $N_f$ flavor hypermultiplets are realized by adding $N_f$ D7 branes filling $x^{01236789}$. Their positions in $x^{45}$ correspond to complex mass parameters for the hypermultiplets. The brane arrangement is described in Table \ref{tab:orientations}.
The positions of the D3 branes in $x^{45}$ correspond to the VEVs of the Cartan complex scalars in the $\cN=2$ vector multiplet and are associated with motion on the Coulomb branch of vacua.

\begin{table}[h]
\begin{center}
\begin{tabular}{|c||cccc|cc|cccc|}
  \hline
      & 0 & 1 & 2 & 3 & 4 & 5 & 6 & 7 & 8 & 9 \\ \hline
  D3  & X & X & X & X &   &   &   &   &   &   \\
  D7  & X & X & X & X  &  &  & X & X  &  X &  X \\ \hline
  NS5 & X &  &  &  &  & X & X  & X & X & X \\ 
  D1 & X &  &  &  & X  &   &   &  &  &  \\  \hline
  D5 & X &  &  &  & X &  & X  & X & X & X \\ 
  F1 & X &  &  &  &   & X  &   &  &  &  \\  \hline
\end{tabular}
\caption{\footnotesize Brane array for 4d $\cN=2$ theories (D3,D7), BPS 't Hooft loops (NS5,D1) and BPS Wilson loops (D5,F1).}
\label{tab:orientations}
\end{center}
\end{table}

Half-BPS 't Hooft loops along $x^0$ are realized by adding NS5 branes along $x^{056789}$ placed in-between the D3 branes in the $x^4$ direction.
In the $U(N)$ theory 't Hooft loops have magnetic charges $B = (b_i)$, with $b_i \ge b_{i+1}$ and $b_i-b_j \in \bZ$, for all $i,j$. Placing an NS5 brane between the $i$th and $(i+1)$th D3 brane realizes a 't Hooft loop with magnetic charge $h_i = (\frac 12,\cdots,\frac 12,-\frac 12,\cdots,-\frac 12)$, where $\frac 12$ appears $i$ times and $-\frac 12$ appears $N-i$ times. This is because the NS5 brane induces a magnetic flux on the D3 brane worldvolumes and the sign depends on whether the D3 is placed to the left or to the right of the NS5. The value $\pm \frac 12$ of the flux will be explained below.
In general a 't Hooft loop with magnetic charge $B = \sum_{i} n_i h_i$ is realized by placing $n_i$ NS5 branes between the $i$th and $(i+1)$th D3 brane. In this construction we should allow for $n_0$ NS5 branes placed to the left of all D3 branes and $n_{N}$ NS5 branes placed to the right of all D3 branes, inducing positive or negative magnetic charge in the $U(1)$ center of $U(N)$. For instance the loop with charge $(1,0,\cdots,0)$ is realized with two NS5 branes, one placed between the first and second D3s and one placed to the right of all D3s, according to the decomposition $(1,0,\cdots,0) = (\frac 12,-\frac 12,\cdots,-\frac 12) + (\frac 12,\frac 12\dots,\frac 12)$.

\medskip

\begin{figure}[t]
\centering
\includegraphics[scale=0.8]{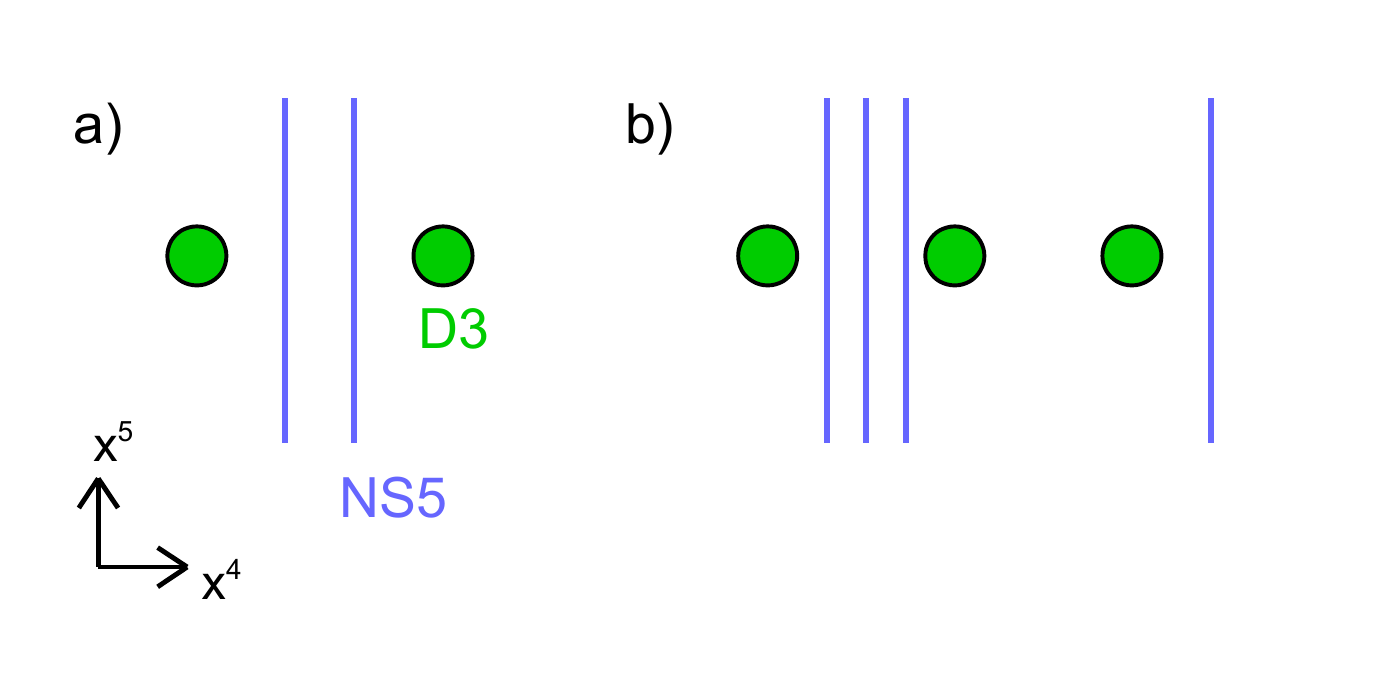}
\vspace{-.5cm}
\caption{\footnotesize{Brane setups realizing: a) a 't Hooft loop with magnetic charge $(1,-1)$ in the $[S]U(2)$ theory. b) a 't Hooft loop with charge $(2,-1,-1)$ in the $[S]U(3)$ theory. We can ignore the external NS5 in the $SU(3)$ theory.}}
\label{tHLBranes0}
\end{figure}

To select the loops of the $SU(N)$ theory, one just considers brane arrangements which realize magnetic charges of the $SU(N)$ magnetic lattice. We provide two examples in Figure \ref{tHLBranes0} for the 't Hooft loop of charge $(1,-1) = 2\times (\frac 12,-\frac 12)$ in the $SU(2)$ theory and the 't Hooft loop of charge $(2,-1,-1) = 3\times (\frac 12,-\frac 12,-\frac 12) + (\frac 12,\frac 12,\frac 12)$ in the $SU(3)$ theory. Actually the external NS5 branes -- those that are on the left or on the right of all D3 branes -- do not play a role for the $SU(N)$ theory, since they are inducing a magnetic flux in the diagonal $U(1) \subset U(N)$. We can simply remove them from the construction. For instance in the case of Figure \ref{tHLBranes0}-b, we can realize the 't Hooft loop of charge $(2,-1,-1)$ of the $SU(3)$ theory simply with three NS5 branes placed between the leftmost and middle D3 branes. In this convention, the magnetic charges of the $SU(N)$ 't Hooft loop is computed by projecting out the $U(1)$ component of the $U(N)$ magnetic charges.

For an $SU(2)$ theory the 't Hooft loops have charges $B=(b,-b)$, $b\in \bZ_{>0}$, and are realized with $2b$ NS5 branes placed in-between the two D3 branes.

\medskip

\begin{figure}[t]
\centering
\includegraphics[scale=0.7]{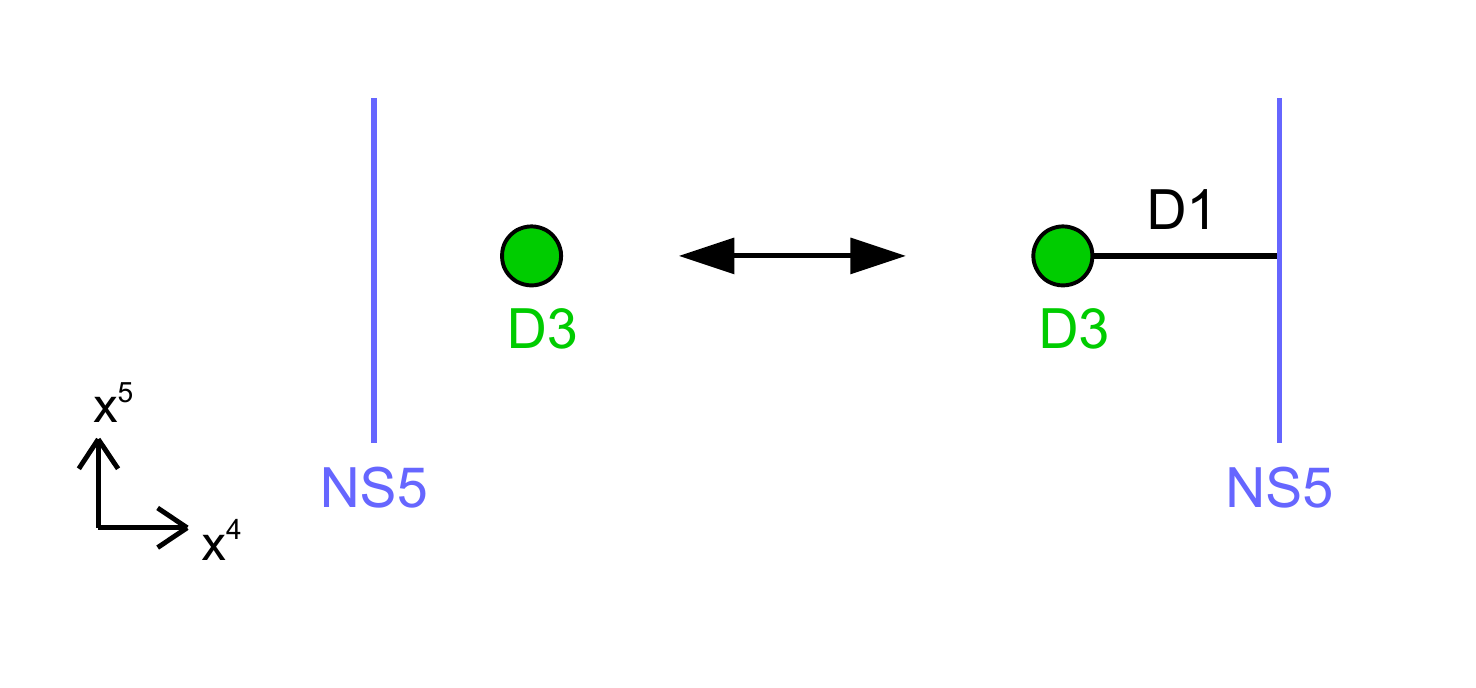}
\vspace{-.5cm}
\caption{\footnotesize{Hanany-Witten transition: as the NS5 passes through the D3, a D1 string is created.}}
\label{HWmove}
\end{figure}
The system of NS5 and D3 branes is of Hanany-Witten type \cite{Hanany:1996ie} with respect to D1 strings stretched along $x^{04}$, namely as an NS5 brane is pushed through a D3 brane, a D1 string is created stretched between them, as illustrated in Figure \ref{HWmove}. A D1 string is the source of a unit magnetic charge for the worldvolume gauge field of the D3 brane. In the process of an NS5 crossing a D3, the magnetic charge  on the D3 brane does not change (since it can be measured ``at infinity'' on the D3 worldvolume and does not depend on the local NS5 crossing. If we denote by $Q$, respectively $-Q$, the magnetic charge that the NS5 induces on the D3 brane when it is on its right, respectively on its left, we find that the magnetic charge conservation in the Hanany-Witten transition satisfies $Q = 1- Q$, namely $Q=\frac 12$. This justifies the claim above that the NS5 induces a 't Hooft loop with magnetic charge $\pm\frac 12$ for each D3 brane.

\medskip

In addition there are smooth monopoles, which are realized with D1 segments stretched between D3 branes. A D1 string stretched between the $i$th and $(i+1)$th D3 branes realizes a BPS smooth monopole with magnetic charge $H_i = (0,\cdots, 0,-1,1,0,\cdots,0)$, where the 1 is in $i$th position.

Monopole bubbling arises when one or more D1 segments are brought on top of (at least two) NS5 branes. The resulting configuration supports an $\cN =(0,4)$ SQM theory which is the ADHM quantum mechanics associated to $Z_{\rm mono}$. 
In order to read the (0,4) ADHM quiver theory it is useful to implement some Hanany-Witten moves, shuffling the NS5 branes around, so that in the resulting configuration the D1 strings are stretched between NS5 branes. 
In addition, D3 branes should be reordered so that their {\it linking numbers} are non-increasing from left to right. We will explain this point below.

\begin{figure}[t]
\centering
\includegraphics[scale=1]{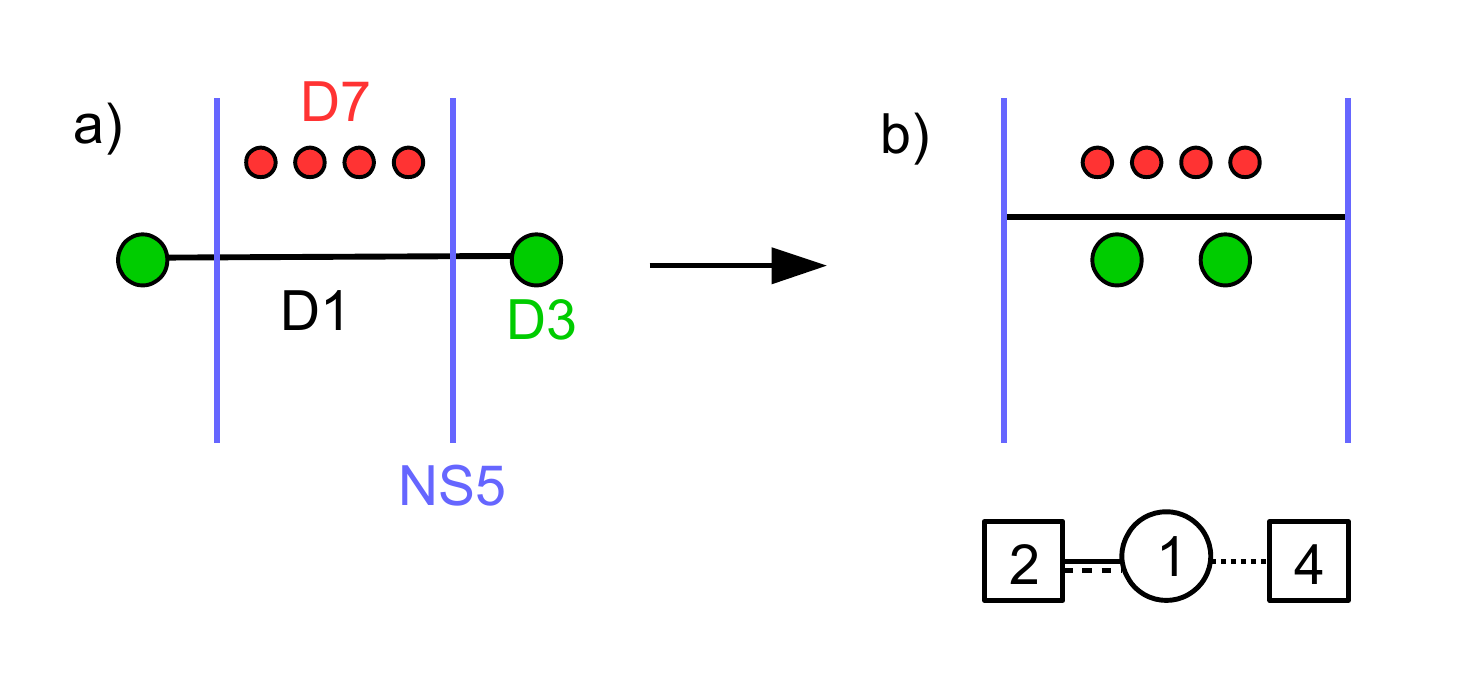}
\vspace{-.5cm}
\caption{\footnotesize{Bubbling configuration for the minimal 't Hooft loop in $SU(2)$ with $N_f=4$ flavors. The ADHM quiver is read after moving the NS5 branes to the sides.}}
\label{U2_0}
\end{figure}

The simplest example arises in the $SU(2)$ theory for the 't Hooft loop of minimal magnetic charge $(1,-1)$, which can be completely screened by a smooth monopole with magnetic charge $(-1,1)$. In the brane description this happens when there is a D1 segment stretched  between the D3s that comes on top of the two NS5 branes. This is illustrated in Figure \ref{U2_0}.
In this situation, one can read the ADHM quiver by pushing the NS5 branes on the sides. In the resulting configuration there is a single D1 stretched between the two NS5s, supporting an $\cN=(4,4)$ $U(1)$ vector multiplet. This is a vector multiplet, plus a twisted hypermultiplet in $\cN=(0,4)$ language. In addition the D1-D3 modes make a (4,4) hypermultiplet for each D3 brane and the D1-D7 strings make a (0,4) Fermi multiplet (which has only a single fermion) for each D7 brane. This leads to the ADHM quiver of Figure \ref{U2_0}.

The mass deformation of the setup that gives the mass $m$ to the 4d adjoint scalar, giving the $\cN=2^\star$ SYM theory on the D3 branes, also affects these (4,4) multiplets, which we will call $\cN =(0,4)^\star$ multiplets. The (4,4) vector multiplet becomes a $(0,4)^\star$ vector multiplet, in which the (0,4) adjoint twisted hypermultiplet has mass $m$. The (4,4) hypermultiplets become $(0,4)^\star$ hypermultiplets, in which the (0,4) Fermi multiplet has mass $m$. In the limit $m\to \infty$, we obtain only pure (0,4) multiplets.

\bigskip

\noindent{\bf Quiver notation}:  Our $\cN=(0,4)$ SQM quiver notation is as follows. A circle with number $n$ denotes a $U(n)$ gauge node for a $(0,4)^\star$ vector multiplet, which contains a (0,4) vector multiplet and a massive (0,4) adjoint twisted hypermultiplet. A box with number $m$ connected by a doubled solid/dashed line to a $U(n)$ node denotes $m$ fundamental $(0,4)^\star$ hypermultiplets of the $U(n)$ node, which contains a (0,4) hypermultiplet and a massive (0,4) Fermi multiplet with several fermions. A box with number $m$ connected by a dashed line to a $U(n)$ node denotes $m$ fundamental (0,4) Fermi multiplets with a single fermion of the $U(n)$ node. 

\bigskip

We should now explain the point about the {\it linking numbers}. For each D3 brane we define a linking number $\ell$ by
\be
\ell = \frac 12(n_R(NS5) - n_L(NS5)) + (n_L(D1) - n_R(D1)) \,,
\ee
where $n_L(NS5)/n_R(NS5)$ is the number of NS5 branes standing on the left/right of the D3, and $n_L(D1)/n_R(D1)$ is the number of D1 strings ending on the left/right of the D3. The linking number corresponds to the total quantized flux induced on the D3 worldvolume by the NS5s and the D1s. It is invariant under Hanany-Witten moves.

Before bubbling, the linking numbers of the D3 branes are ordered non-increasingly from left to right by construction. However once we add the D1 string responsible for the bubbling, this might not be the case and we need to reorder the D3 branes in non-increasing order. This is in a sense a way to avoid a redundancy in the brane description.\footnote{See \cite{Gaiotto:2008ak} for a related discussion on the ordering of branes in increasing/decreasing linking number order.}  We may equivalently say that we consider monopole bubbling configurations which do not alter the linking number ordering of the D3 branes.

\medskip

In this discussion we have been elusive as to what happens to D7 branes in the picture. In the configurations there are also $2N$ D7 branes sourcing the $2N$ fundamental hypermultiplets of the 4d theory. D7 branes appear as points in the $x^{45}$ plane of the brane picture. As we will see later, the positions of the D7 branes with respect to NS5 branes is important, since the 7 branes have a {\it bending effect} on NS5 branes. So in principle we should give a prescription as to where to put the D7 branes with respect to NS5 branes. In Figure \ref{U2_0} we have placed D7 branes in-between the NS5 branes, so that we do not need to move NS5s across D7s when we move the branes to read the ADHM quiver. 
In general we expect that placing the D7s/NS5s branes in different arrangements leads to different contributions for different loop operators. 

The 't Hooft loops of the $SU(N)$ theories are always realized with an even number of NS5 branes. A natural prescription is to always put the D7 branes in the middle of the NS5s, namely with as many NS5s on their left as on their right along $x^4$. We will adopt this prescription in our construction and comment further on this issue in the discussion section \ref{sec:Discussion}. The placement of the D7 branes along the vertical direction, with respect to the D1 strings, is also important and we will give a prescription for this as we proceed.

\subsection{$Z^0_1$ in conformal SQCD}
\label{ssec:Z10}

We illustrate the discussion with the simplest example: the bubbling contribution to the monopole of minimal magnetic charge in the  $\cN=2$ $SU(N)$ SQCD theory with $2N$ flavor hypermultiplets, still following \cite{Brennan:2018yuj,Brennan:2018rcn}. The minimally charged 't Hooft loop has $B=(1,0,\cdots,0,-1):= B_1$, we will denote it $L_1$. It has one bubbling sector with zero total magnetic charge $v = (0,\cdots,0)$. The VEV of $L_1$ takes the form
\be
\vev{L_1} =  \sum_{i\neq j} e^{b_i - b_j}   Z_{\rm 1-loop}(B^{(i,j)}_1) + Z^0_1 \,,
\ee
with $B^{(i,j)}_1$ the $N-$vector with entries $(B^{(i,j)}_1)_n = \delta_{ni} - \delta_{nj}$, and $Z^0_1 := Z_{\rm mono}(B_1,\vec 0)$. Here we have used that $ Z_{\rm 1-loop}(\vec 0) =1$.

\begin{figure}[t]
\centering
\includegraphics[scale=0.8]{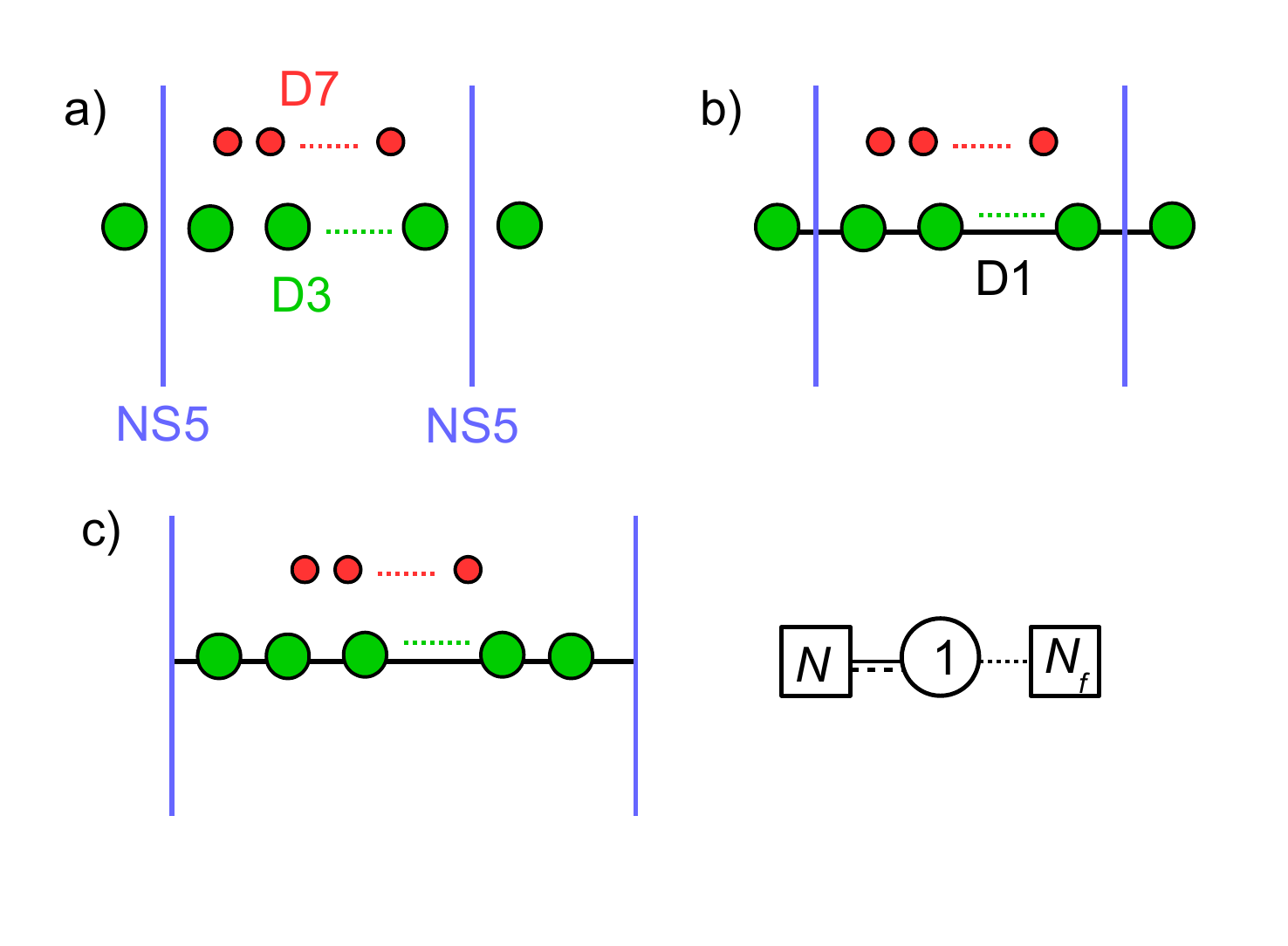}
\vspace{-1cm}
\caption{\footnotesize{a) Brane realization of $L_1$ loop in $SU(N)$ with $2N$ flavors. There are $N$ D3s and $2N$ D7s. b) Bubbling configuration. c) Same configuration after moving NS5s and ADHM SQM.}}
\label{UN_0}
\end{figure}

The bubbling contribution $Z^0_1$ is computed as the Witten index of an ADHM (0,4) SQM that is read from the brane construction, as explained in the previous section and in more details in Appendix \ref{app:MM}. The brane realization for the $L_1$ loop has two NS5 branes and is shown in Figure \ref{UN_0}-a. The bubbling sector corresponds to the (full) screening of $L_1$ by a D1 string as shown in Figure \ref{UN_0}-b. The ADHM SQM is the D1 theory as read from Figure \ref{UN_0}-c. It is a $U(1)$ theory with a $(0,4)^\star$ vector multiplet, $N$ $(0,4)^\ast$ fundamental hypermultiplets and $2N$ (0,4) fundamental Fermi multiplets. $Z^0_1$ is the Witten index of this SQM.

One usually expects that the Witten index, which is a sum over the BPS vacua, reduces to the contributions counted by the Jeffrey-Kirwan contour in the matrix model associated to the SQM. The main observation of \cite{Brennan:2018rcn} is that the Jeffrey-Kirwan contour of integration counts BPS {\it Higgs} vacua, but sometimes misses contributions from BPS {\it Coulomb} vacua that belong to a continuum of vacuum states, in some theories when this continuum exists.  This phenomenon arises when the effective potential for the SQM scalar field is bounded at infinity (instead of divergent), which happens in the computation of $Z^0_1$ in conformal SQCD theories. Here we can rephrase this as follows. The JK prescription captures the full result when the FI parameter of the SQM is non-zero and is taken as the JK parameter. In that case the Coulomb vacua are lifted.  In the brane picture the FI parameter is the separation of the two NS5 branes along the flat $\bR$ direction. The bubbling configuration arises when the NS5s are exactly aligned, therefore the bubbling term is captured by the SQM index at zero FI parameter. In that case the JK prescription (with any choice of non-zero JK parameter) does not capture the full answer. One must add contributions from ``poles at infinity".\footnote{The situation with non-zero FI parameter corresponds to a two-point function of 't Hooft loops. We discuss this further in section \ref{sec:starprod}.}

Therefore we have
\be
Z^0_1 = Z_{\rm JK} + Z_{\rm extra} \,,
\ee
with the JK contribution given by (see Appendix \ref{app:MM}) \footnote{We have $\epsilon_{\pm} := \frac 12 (\epsilon_1 \pm \epsilon_2)$, $\sh(x) := 2 \sinh(\frac x2)$ and $f(x\pm y) := f(x+y)f(x-y)$.}
\be
Z_{\rm JK} (\epsilon_-) = \int_{JK(\zeta)} \frac{\dd \phi}{2\pi i} \frac{(-1)\sh(2\epsilon_+)}{\sh(\epsilon_1)\sh(\epsilon_2)} \prod_{i=1}^N \frac{ \sh [\pm (\phi - a_i) + \epsilon_- ] }{\sh [\pm (\phi - a_i )+ \epsilon_+] }  \prod_{k=1}^{2N} \sh (\phi - m_k)  \,.
\ee
Here $a_i$ are the masses of the fundamental hypermultiplets and are identified with the Coulomb branch coordinates (in the Cartan subalgebra of $U(N)$) of the 4d theory. In the 4d $SU(N)$ theory they obey $\sum_{i=1}^N a_i =0$. The Fermi multiplet masses $m_k$ are identified with the masses of the $2N$ flavor hypermultiplets of the 4d theory.

The parameter $\epsilon_-$ is identified with the adjoint mass $m$ of the 4d $\cN=2^\ast$ SYM theory and appears here as the mass parameter of the $\cN=(0,4)^\star$ deformation as explained in the appendix. It is to be sent to infinity to reach the final result for the $\cN=2$ SQCD theory.

The integrals are evaluated with the JK selection of poles, with the JK parameter $\zeta$. Importantly the evaluation depends on the sign of $\zeta$.
Taking $\zeta >0$, we find\footnote{The JK prescription selects the residues at $\phi=a_i - \epsilon_+$.}
\be 
Z_{\rm JK} (\epsilon_-) =  \sum_{i=1}^N  \prod_{j\neq i} \frac{ \sh [\pm (a_i - a_j - \epsilon_+) + \epsilon_- ] }{\sh [\pm (a_i - a_j - \epsilon_+ )+ \epsilon_+] } \prod_{k=1}^{2N} \sh (a_i - m_k - \epsilon_+) \,. \qquad (\zeta>0)
\ee
We now send $\epsilon_- \to \infty$. To reach a finite result we should normalize by a contribution $(\sh(s\epsilon_-))^{\pm 1}$ for each (0,4) multiplets whose mass scales as $s\epsilon_-$, with $\pm$ being $-$ for a hypermultiplet and $+$ for a Fermi multiplet. This amounts to removing the massive (0,4) multiplets in the matrix model (in this case we could simply have done that from the beginning).  In the end we normalize by $Z_{\rm norm}(\epsilon_-) = (-1)^N \sh(\epsilon_-)^{2N-2}$.
We have no good explanation for the factor $(-1)^{N}$ in the normalization of the result. We do not know how to fix the overall sign and our prescription follows from consistency requirement, when relating the result to OPEs between loops, as we discuss in section \ref{sec:starprod}.  We obtain
\be
Z_{\rm JK} = \lim_{\epsilon_- \to \infty} \frac{Z_{\rm JK} (\epsilon_-)}{Z_{\rm norm}(\epsilon_-)} =  - \sum_{i=1}^N  \frac{  \prod_{k=1}^{2N} \sh (a_i - m_k - \epsilon_+)  }{\prod_{j\neq i} \sh(a_i - a_j)\sh(a_i - a_j - 2 \epsilon_+) }\,.  \qquad (\zeta>0)
\label{Z10JK}
\ee
Finally there is the extra contribution $Z_{\rm extra}$, missed by the JK prescription. To our knowledge, there is no computation of this term in general. For the simplest case $N=2$ it was computed in \cite{Brennan:2018rcn}. The minimal 't Hooft loop has a non-zero $Z_{\rm extra}$ in the $SU(2)$ theory with $N_f =4$ flavors, which is, for $\zeta >0$, \footnote{We define $\ch(x) := 2\cosh(\frac x2)$.}
\be
\underline{SU(2),N_f =4}: \quad  Z_{\rm extra}  =  \, \ch(\sum_{k=1}^{2N} m_k + 2\epsilon_+) \,.  \qquad (\zeta>0)
\ee
If we had chosen and negative $\zeta$ instead, we would have found the same results but with a flip of sign $\epsilon_+ \to - \epsilon_+$ in $Z_{JK}$ and in $Z_{\rm extra}$. As discussed at the end of section \ref{ssec:Generalities}, the result should be invariant under $\epsilon_+ \to - \epsilon_+$.  Although it is not obvious, it happens that the total bubbling factor $Z^0_1$ is indeed invariant under the $\bZ_2$ symmetry.

\medskip

Unfortunately, the method proposed in \cite{Brennan:2018rcn} for computing $Z_{\rm extra}$ is not straightforward and difficult to apply to more complicated theories.  What we propose in this paper is a method to compute directly the full answer for monopole bubbling contributions like $Z^0_1$, including the extra piece $Z_{\rm extra}$, from a modified ADHM quiver.

\section{Complete brane systems and improved SQM}
\label{sec:Main}

\subsection{The minimal 't Hooft loop}
\label{ssec:tHL}

To begin with we will focus on the computation of the bubbling contribution $Z^0_1$ for the 't Hooft loop $L_{1}$ of minimal magnetic charge $B=(1,0,\cdots,-1)$ in 4d $\cN=2$ $SU(N)$ theories with $N_f = 2N$ fundamental hypermultiplets.

\subsubsection{Brane system for $N=2$}

\begin{figure}[t]
\centering
\includegraphics[scale=0.8]{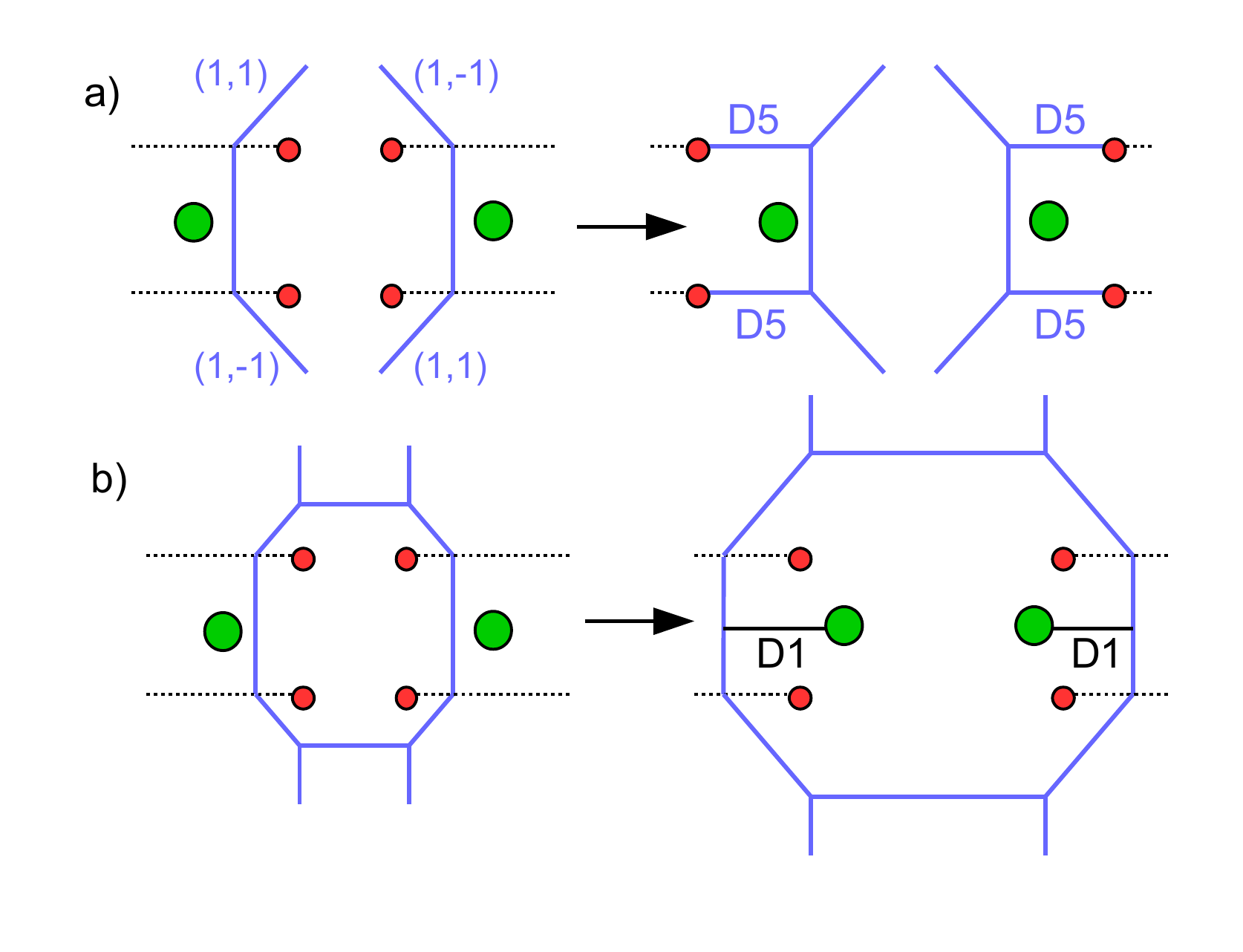}
\caption{\footnotesize{a) Brane setup for $L_1$ taking into account the effect of the D7-branes on the NS5 branes, here with four D7 branes, for the $N_f=4$ $SU(2)$ theory. b) Completion of the 5-brane web to resolve intersections.}}
\label{U2_1}
\end{figure}

The computation that we want to propose is based on a completion of the brane system realizing the 't Hooft loop. We have already presented the brane setup for $L_1$ in Figure \ref{UN_0}-a. 
We observe that this brane picture is not accurate enough because the D7 branes induce a bending and a change of type on the NS5-branes.  We should remember that 7-branes have a branch cut in their transverse plane, across which type IIB string theory enjoys a certain $SL(2,\bZ)$ duality transformation. The $SL(2,\bZ)$ element depends on the type of 7-brane. The branch cuts are not physical and can be moved wherever we like. A common way to represent such a brane configuration is to let the branch cuts run horizontally to the left or to the right. As a $(p,q)$ 5-brane crosses a D7 cut, its type changes to a $(p,q \pm 1)$ 5-brane, with the sign $\pm$ depending on which side of the cut the $(p,q)$ 5-brane stands. In addition the brane ``bends" in the sense that the $(p,q \pm 1)$ 5-brane segment has a different orientation in the $x^{45}$ plane. Precisely, a $(p,q)$ 5-brane spans a line in the $x^{45}$ plane with slope $\frac{q}{p}$. Here we take the convention that an NS5 brane is a $(1,0)$ 5-brane, and a D5 brane is a $(0,1)$ 5-brane.
We illustrate this for the case $N=2$ with 4 D7 branes in the left of Figure \ref{U2_1}-a. 
From this type of setup, it is common to push the 7 branes to the sides as shown in the right of Figure \ref{U2_1}-a. When the 7-brane crosses the 5-brane segments a third type of 5-branes is created, here a D5 brane, resulting in a 5-brane junction. In this paper however, we will not move the D7-branes to the sides because we find it more convenient.

The brane configuration that we have reached is not yet complete, since the $(1,1)$ and $(1,-1)$ 5-brane lines intersect, if we follow them far enough upstairs, or downstairs. 5-brane intersections can support degrees of freedom and to understand fully the brane configuration we must resolve these intersections. Concretely we need to find a completion of the brane setup upstairs and downstairs such that there is no intersection any more. For the case with four D7 branes, a minimal way to do so is shown on the left of Figure \ref{U2_1}-b. Here we were required to add extra D5 and NS5 segments to complete the 5-brane web. The final configuration is rather similar to the initial one, but there are two extra D5 segments. There exists other more complicated ways to complete the brane setup, but we do not need to look at them.

We are now satisfied with our brane realization of the minimal 't Hooft loop $L_1$. As it is it may be regarded as a fancy construction that has no effect on the loop insertion at low energies, since it is possible to send the 5-brane web away from the D3-D7 system and still realize the 't Hooft loop. In this process D1-strings are created and realize the 't Hooft loop on the D3 worlvolume. This is shown on the right of Figure \ref{U2_1}-b.

\subsubsection{Improved ADHM for $N=2$}
\label{sssec:tHLU2}

\begin{figure}[t]
\centering
\includegraphics[scale=0.8]{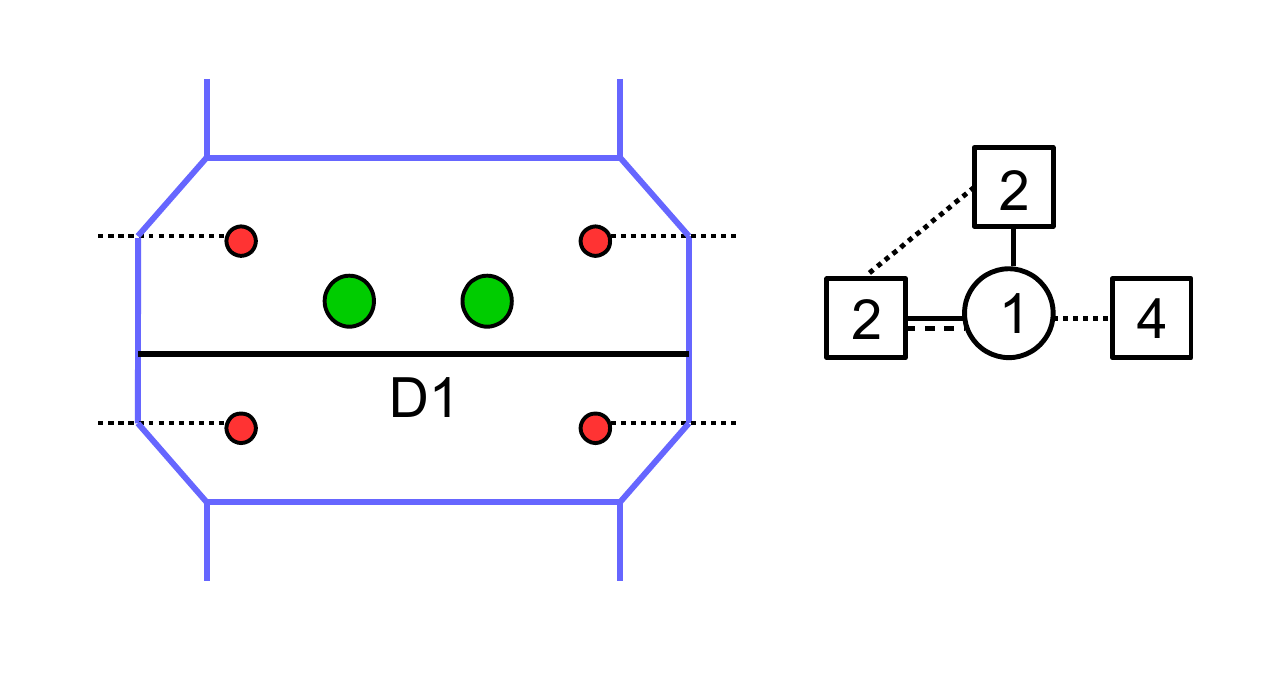}
\vspace{-1cm}
\caption{\footnotesize{Complete brane setup for the $L_1$ bubbling and improved SQM quiver.}}
\label{U2_2}
\end{figure}

The effect of the refined brane construction is better appreciated when we consider the bubbling sector and the computation of $Z^0_1$. The bubbling sector arises when an extra D1 segment is stretched between the two external D3s and recombine with the NS5-D3 segments into a single D1 string stretched between the two NS5s, as in Figure \ref{U2_2}. The ADHM associated with $Z^0_1$ is read as the (0,4) SQM living on the D1 string. In the complete 5-brane web of the $L_1$ insertion we have two extra D5 segments and the D1-D5 strings modes give two extra (0,4) hypermultiplets. In addition the D3-D5 modes gives four extra Fermi multiplets which are not charged under the $U(1)$ SQM gauge group. There are also superpotential terms ($J$-term and $E$-term) but we do not need to know them to compute the index. The resulting quiver SQM is given in Figure \ref{U2_2}. We will call it the {\it improved} (ADHM) SQM. One virtue of the improved SQM is that the potential of the matrix model on the integrand variable $z$ is divergent at $z \to \pm\infty$, which means that the Witten index does not have extra contributions from continuum of states and can be computed reliably using the Jeffrey-Kirwan prescription alone, with any choice of JK parameter (it should be independent of the choice).

On the other hand, the Witten index $\cI^0_1$ of this improved SQM is not directly equal to the bubbling contribution $Z^0_1$. For instance it depends on two more fugacities $w_1 =e^{-v_1}, w_2 = e^{-v_2}$, where $v_1,v_2$ are the masses of the two extra (0,4) hypermultiplets. There is no such fugacity in $Z^0_1$. Explicitly we have
\bea
 \cI^0_1(\epsilon_-) &= \int_{JK_{\zeta>0}} \frac{\dd \phi}{2\pi i} \frac{(-1)\sh(2\epsilon_+)}{\sh(\epsilon_1)\sh(\epsilon_2)} \prod_{i=1,2} \frac{ \sh [\pm (\phi - a_i) + \epsilon_- ] }{\sh [\pm (\phi - a_i )+ \epsilon_+] }  \frac{\prod_{k=1}^{4} \sh (\phi - m_k) \prod_{i,n} \sh(a_i - v_n)}{\prod_{n=1,2} \sh[\pm ( \phi - v_n) - \epsilon_+]} \,.
\eea
Now we have extra JK residues at $\phi = v_n + \epsilon_+$, for $n=1,2$. Evaluating the residues and normalizing as in \eqref{Z10JK}, we obtain
\bea
\cI^0_1 &= \lim_{\epsilon_- \to \infty} \frac{\cI^0_1(\epsilon_-)}{Z_{\rm norm}(\epsilon_-)} \cr
&=  - \frac{ \prod_{k=1}^{4} \sh (a_1 - m_k - \epsilon_+) \prod_{n=1,2} \sh( a_2 - v_n) }{\sh(a_{12})\sh(a_{12} - 2\epsilon_+) \prod_{n=1,2} \sh(v_n - a_1 + 2\epsilon_+)}  \ + [a_1 \leftrightarrow a_2]  \cr
& \quad + \frac{\prod_{i=1,2}\sh(a_i -v_2) \prod_{k=1}^{4} \sh (v_1 - m_k +\epsilon_+)}{\prod_{i=1,2}\sh(v_1 -a_i +2\epsilon_+ )\sh(v_{12}) \sh(v_{12} + 2\epsilon_+)}  \ + [v_1 \leftrightarrow v_2] 
\eea
Taking the residues in $w_1$ and then in $w_2$ around zero, we find 
\bea
\int \frac{dw_2}{2\pi i w_2} \int \frac{dw_1}{2\pi i w_1} \cI^0_1 &= -  \frac{ \prod_{k=1}^{4} \sh (a_1 - m_k - \epsilon_+) }{\sh(a_{12})\sh(a_{12} - 2\epsilon_+)}  \ + [a_1 \leftrightarrow a_2] \cr
& \ \  + 2 \ch(\sum_{k=1}^4 m_k - 2(a_1+a_2) + 2\epsilon_+)  \,.
\eea
Taking into account the $SU(2)$ constraint $a_1+a_2=0$, this is precisely $Z^0_1$ for the $SU(2)$ theory with $N_f=4$, including the extra piece $Z_{\rm extra}$! 
We thus find the relation
\be
Z^0_1 = \oint_{\cC} \frac{dw_1}{2\pi i w_1}\frac{dw_2}{2\pi i w_2} \, \cI^0_1(w_1,w_2) \,,
\label{rel1}
\ee
with $\cC = \cC_1 \times \cC_2$ the integration contours for $w_1$ and $w_2$ around the origin, defined with $|w_1| < |w_2 e^{-\epsilon_+}| \, \lp< |w_2 e^{\epsilon_+}| \rp$ on the contours. 
This choice of contour effectively imposes that we take the residues in $w_1$ first and $w_2$ after.\footnote{This implies that we do not take residues from the poles at $w_1 = w_2 e^{\pm\epsilon_+}$.}

The logic being this result is the following. Among the various terms contributing to the index $\cI^0_1$, one should isolate the one contributing to $Z^0_1$. The contributions to $\cI^0_1$ can be organised into sectors of fixed $U(1)^2$ flavor charges, where the $U(1)^2$ refers to flavor symmetries associated to the D5 branes, with fugacities $w_1$ and $w_2$. The terms with weight $w_1^n w_2^m$ belong to the charge $(n,m)$ sector. The states contributing to $Z^0_1$ are in principle not charged under the $U(1)^2$ D5 symmetries, therefore they should belong to the $(0,0)$ sector. The residue computation that we perform extracts this $(0,0)$ charge sector. What we find experimentally is that $Z^0_1$ is the full $(0,0)$ charge sector of $\cI^0_1$.

\medskip

We should comment that there is an alternative, and arguably simpler, computation that gives the same result. Since the original setup does not have the extra D5 segments, we can think of recovering $Z^0_1$ by sending the upper D5 segment to $+\infty$ and the lower D5 segment to $-\infty$. This means taking the limit $w_1 \to 0$ and $w_2 \to \infty$ in $\cI^0_1$. Indeed we find, without even the need for extra normalization,
\be
\lim_{w_1 \to 0 \atop w_2 \to +\infty} \cI^0_1 = Z^0_1 \,,
\label{rel2}
\ee
including the extra piece $Z_{\rm extra}$. This simple limit works in this case, but may not work in general due to the possible presence of diverging contributions.

\subsubsection{Bubbling in $SU(N)$ $N_f=2N$ theory}
\label{ssec:L1Gen}

\begin{figure}[t]
\centering
\includegraphics[scale=0.75]{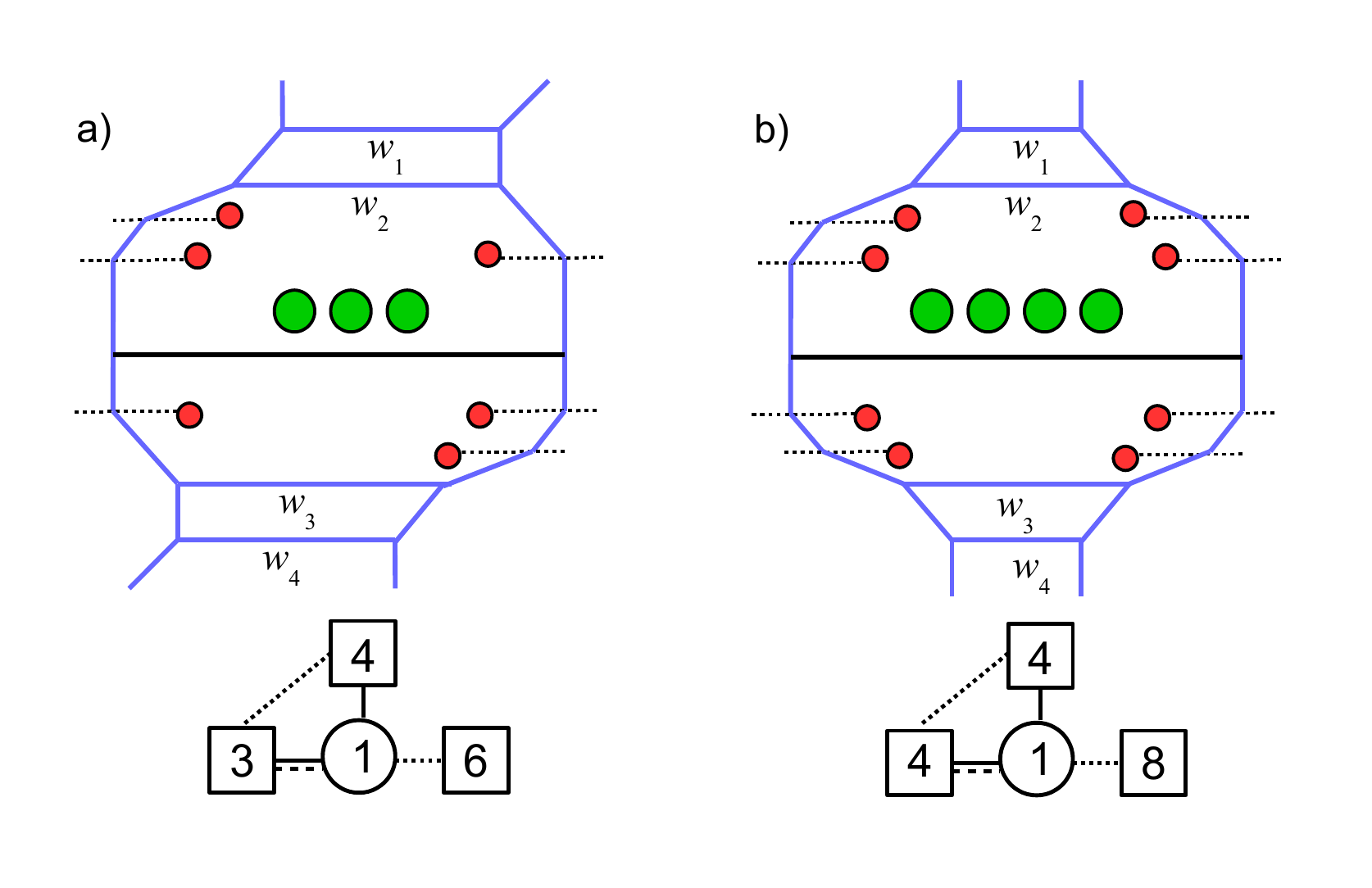}
\vspace{-0.5cm}
\caption{\footnotesize{Complete brane systems for the $L_1$ bubbling in: a) the $SU(3)$ $N_f=6$ theory; b) the $SU(4)$, $N_f=8$ theory.}}
\label{U3_U4}
\end{figure}

For $N >2$ we have $N_f = 2N > 4$ D7 branes in the brane system and the completion of the brane setup for the minimal 't Hooft loop requires more than two extra D5 segments. The cases for the $SU(3)$ $N_f =6$ and $SU(4)$ $N_f=8$ theories are shown in Figure \ref{U3_U4}. Here the brane completions require  four extra D5 segments. Consequently the improved ADHM has four extra (0,4) hypermultiplets and $4N$ extra Fermi multiplets.

In general, one can complete the brane setup for arbitrary $N$ with a brane web involving $h$ extra D5 segments, where $h =N$ for $N$ even and $h=N +1$ for $N$ odd, so we have $h = 2\ceil{\frac{N}{2}}$. $h$ is always an even integer with this choice of completion.
As a rule we always take the distribution of D7 branes, and thus D5 segments, as even as possible between the upper and lower part of the brane configuration. This means that we take $N$ D7 branes (and $\frac{h}{2}$ D5 segments) above the D3s, and $N$ D7 branes (and $\frac{h}{2}$ D5 segments) below the D3s. 
 This leads to an improved ADHM with $h$ extra fundamental (0,4) hypermultipets and $hN$ extra Fermi multiplets.  Let us denote again $\cI^0_1$ the supersymmetric index of this improved ADHM SQM. Explicitly the mass deformed $\cN=(0,4)^\star$ SQM index $\cI^0_1(\epsilon)$ is given by
 \bea
 \cI^0_1(\epsilon_-) &= \int\limits_{JK_{\zeta>0}} \frac{\dd \phi}{2\pi i} \frac{(-1)\sh(2\epsilon_+)}{\sh(\epsilon_1)\sh(\epsilon_2)} \prod_{i=1}^N \frac{ \sh [\pm (\phi - a_i) + \epsilon_- ] }{\sh [\pm (\phi - a_i )+ \epsilon_+] }  \frac{\prod_{k=1}^{2N} \sh (\phi - m_k) \prod_{i =1}^N\prod_{n=1}^{h} \sh(a_i - v_n)}{\prod_{n=1}^h \sh[\pm ( \phi - v_n) - \epsilon_+]} \,.
\eea
Evaluating the residues at $\phi = a_i -\epsilon_+$ and $\phi = v_n + \epsilon_+$, and taking the normalized limit $\epsilon_- \to \infty$, we find
\bea
\cI^0_1 &= \lim_{\epsilon_- \to \infty} \frac{\cI^0_1(\epsilon_-)}{(-1)^N\sh(\epsilon_-)^{2N-2}} \cr
&=  - \sum_{i=1}^N \frac{ \prod_{k=1}^{2N} \sh (a_i - m_k - \epsilon_+) \prod_{j\neq i,n} \sh( a_j - v_n) }{\prod_{j\neq i}\sh(a_{ij})\sh(a_{ij} - 2\epsilon_+) \prod_{n=1}^h \sh(v_n - a_i + 2\epsilon_+)}    \cr
& \quad +  \sum_{n=1}^h \frac{\prod_{i, m\neq n} \sh(a_i -v_m) \prod_{k=1}^{2N} \sh (v_n - m_k +\epsilon_+)}{\prod_{i=1}^N\sh(a_i -v_n - 2\epsilon_+ ) \prod_{m\neq n}\sh(v_{nm}) \sh(v_{nm} + 2\epsilon_+)}  \,.
\label{I10gen}
\eea
Generalizing the result of the previous sections we propose that the following relation holds:
\bea
Z^0_1 &=  \oint_{\cC} \prod_{n=1}^h \frac{dw_n}{2\pi i w_n} \, F(w) \, \cI^0_1(w) \,, \cr
\text{with} & \quad F(w) = \lp \frac{\prod_{n=1}^{h/2} w_n}{\prod_{n'=h/2+1}^{h} w_{n'}} \rp^{\frac{N-2}{2}} \,,
\label{rel4}
\eea
with $\cC = \prod_n \cC_n$ the integration contours for $w_1,w_2,\cdots, w_h$ around the origin, defined with $|w_n| < |w_{n+1} e^{-\epsilon_+}| \, \lp< |w_{n+1} e^{\epsilon_+}| \rp$ on the contours. Effectively it means that we take the residues at zero in $w_1$ first, then in $w_2$, ... etc.

The relation \eqref{rel4} is a direct generalization of the results of the previous sections except for the factor $F$ in the integrand, that we need to explain.
 The presence of this factors means that, instead of selecting the sector of zero $U(1)^h$ charge in $\cI^0_1$, where $U(1)^h$ is the (Cartan) flavor symmetry associated with the $h$ extra D5 segments, we are selecting the sector of $U(1)^h$ charge $(-\frac{N-2}{2},\cdots, -\frac{N-2}{2}, \frac{N-2}{2},\cdots, \frac{N-2}{2})$, with $h/2$ negative charges and $h/2$ positive charges. So we are saying that $Z^0_1$ corresponds to this charge sector in $\cI^0_1$.

Integrating out a Fermi multiplet of mass $m$ induces a shift of the SQM Chern-Simons level $\kappa = \frac{\text{sign}(m)}{2}$ for the $U(1)$ flavor symmetry, and results in a factor $e^{\kappa m} = e^{\frac{|m|}{2}}$ in the matrix model. Similarly integrating out a (0,4) hypermultiplet of mass $m$ will induce a 1d background Chern-Simons level $\kappa' = - \text{sign}(m)$ in the matrix model, resulting in a factor $e^{-|m|}$.\footnote{These properties can be deduced from looking at the large mass limit $m \to \pm\infty$ of the matrix model factors for these multiplets.} The presence of these 1d Chern-Simons term can be understood as turning on a background charge for the $U(1)$ flavor symmetry.

When we completed the ADHM quiver we added $hN$ Fermi multiplets, which give a background charge for the $U(1)^h$ flavor symmetry and $SU(N)$ gauge symmetry, corresponding to a factor $\prod_{n=1}^h \prod_{i=1}^N e^{\frac 12|v_n -a_i|}$. The sign of $v_n -a_i$ is set by the relative position of the $n$th D5 segment and $i$th D3 brane. The distribution of the D5s in the upper and lower regions (above or below the D3s) is such that the resulting factor measuring the background $U(1)^h$ charge is $\prod_{n=1}^{h/2} (w_n)^{-N/2} \prod_{n'=h/2+1}^h (w_{n'})^{N/2}$ (while the $SU(N)$ charge is zero). Saying it differently, we take the first $h/2$ $v_n$ to be large and positive and the last $h/2$ $v_n$ to be large and negative.
Similarly the presence of the $h$ (0,4) extra hypermultiplets induces a background charge for the $U(1)^h$ flavor symmetry and the gauge $U(1)$ symmetry which is measured by a factor $\prod_{n=1}^h e^{-|v_n -\phi|}$. In particular the contribution to the $U(1)^h$ charge is $\prod_{n=1}^{h/2} w_n \prod_{n'=h/2+1}^h (w_{n'})^{-1}$. 

Combining the two factors we obtain that the presence of the additional matter fields in the improved SQM induces a $U(1)^h$ charge $(-\frac{N-2}{2},\cdots, -\frac{N-2}{2}, \frac{N-2}{2},\cdots, \frac{N-2}{2})$, measured by a factor $\prod_{n=1}^{h/2} (w_n)^{1-N/2} \prod_{n'=h/2+1}^h (w_{n'})^{N/2-1}$ in the matrix model. Therefore, in order to recover $Z^0_1$, we need to pick the sector with this charge. This explains the factor $F$ in the integrand of \eqref{rel4}.\footnote{There is also a related explanation of such factors from the brane setup, with the D5 branes inducing a flux on the D3 brane worldvolume. We refer the reader to \cite{Assel:2018rcw}.} 

\medskip

By explicit computations for low values of $N$ ($N=2,3,4,5$) we find that the relation \eqref{rel4} yields
\bea
Z^0_1 &= - \sum_{i=1}^N \frac{ \prod_{k=1}^{2N} \sh (a_i - m_k - \epsilon_+) }{\prod_{j\neq i}\sh(a_{ij})\sh(a_{ij} - 2\epsilon_+)} \cr
& \quad + \ch(\sum_{k=1}^{2N} m_k + 2\epsilon_+ - 2\sum_{i=1}^{N} a_i) \,.
\label{Z10final}
\eea
The terms on the first line match the JK contribution $Z_{\rm JK}$ of the original/standard index computation \eqref{Z10JK}  with $\zeta>0$. We propose that the term on the second line is the missing $Z_{\rm extra}$ contribution, for arbitrary $N$. One can check that this expression is (non-trivially) invariant under the $\bZ_2$ symmetry $\epsilon_+ \to -\epsilon_+$.

Here we have given the result for the $U(N)$ theory.\footnote{Note that in the $U(N)$ theory $L_1$ is not the minimally charged 't Hooft loop. The minimal loops have magnetic charges $B=(\pm 1,0,\cdots,0)$ and have no bubbling sector.} For the $SU(N)$ result, one simply sets $\sum_{i=1}^{N} a_i =0$.

\subsubsection{A simplification}
\label{sssec:simpler}

\begin{figure}[t]
\centering
\includegraphics[scale=0.75]{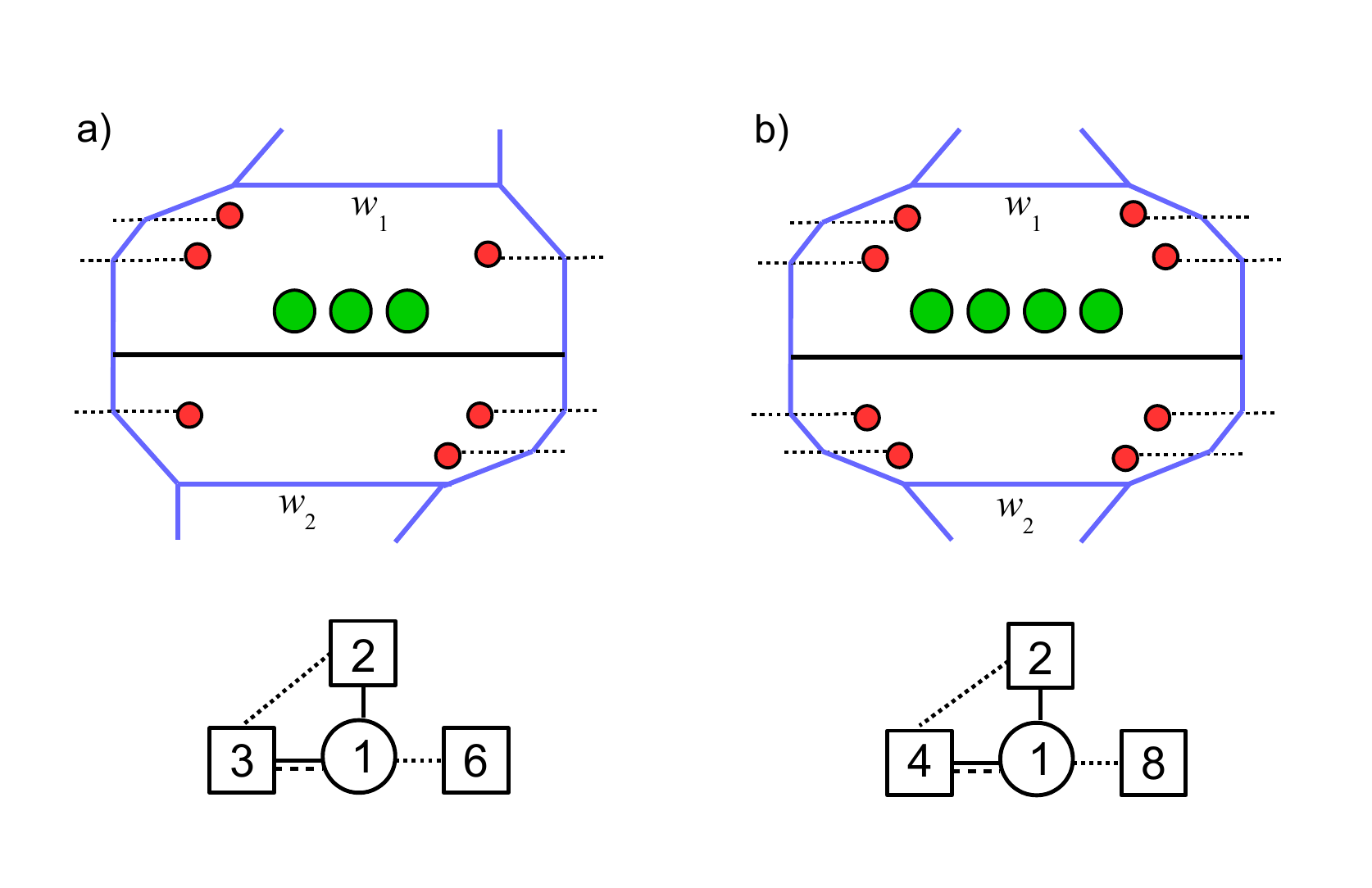}
\vspace{-0.5cm}
\caption{\footnotesize{Partial completion of the brane systems for the $L_1$ bubbling in the $SU(3)$ and $SU(4)$ theories and semi-improved SQM.}}
\label{U3_U4_2}
\end{figure}

It turns out that the residue computation can be simplified. Indeed we observe that the same result is reached if, instead of completing the brane system to a full 5-brane web, we only partially complete it with the addition of two D5 segments as in Figure \ref{U3_U4_2}.
The semi-improved SQM now has only $2$ new (0,4) hypermultiplets and $2N$ new Fermi multiplets. The index is computed by 
\bea
 \ti\cI^0_1(\epsilon_-) &= \int\limits_{JK_{\zeta>0}} \frac{\dd \phi}{2\pi i} \frac{(-1)\sh(2\epsilon_+)}{\sh(\epsilon_1)\sh(\epsilon_2)} \prod_{i=1}^N \frac{ \sh [\pm (\phi - a_i) + \epsilon_- ] }{\sh [\pm (\phi - a_i )+ \epsilon_+] }  \frac{\prod_{k=1}^{2N} \sh (\phi - m_k) \prod_{i =1}^N\prod_{n=1,2} \sh(a_i - v_n)}{\prod_{n=1,2} \sh[\pm ( \phi - v_n) - \epsilon_+]} \,,
 \label{I10simple}
\eea
and the bubbling contribution is
\be
Z^0_1 =  \oint_{\cC} \frac{dw_1 dw_2}{(2\pi i)^2w_1w_2} \lp \frac{w_1}{w_2} \rp^{\frac{N-2}{2}} \ti\cI^0_1(w_1,w_2) \,,
\ee
with the residue in $w_1$ taken first, and then the residue in $w_2$, and with $\ti\cI^0_1(w_1,w_2) = \lim_{\epsilon_- \to \infty} \frac{(-1)^N}{\sh(\epsilon_-)^{2N-2}}   \ti\cI^0_1(\epsilon_-, w_1,w_2)$.

This computation reproduces the result \eqref{Z10final}.
This means that additional completion of the brane setup giving the full improved ADHM does not change the evaluation of $Z^0_1$ beyond the effect brought by the semi-improved ADHM. In general we expect such a simplification to occur, in the sense that it may not be necessary to fully complete the brane setup, however we do not have a criteria for determining when to stop the brane completion.

\subsection{Minimal dyonic loop}
\label{ssec:DL}

We expect that the technique of brane completion can be used to compute the vev of other loops of the $\cN=2$ theories (possibly all loops). In this section we study a dyonic loop $L_{1,1}$ which has minimal non-zero magnetic and electric charges.

\subsubsection{Brane setups}

The first thing to ask is: what is the brane setup realizing the minimal dyonic loop (without bubbling)? To our knowledge this has not been studied in the literature. We know that the presence of a magnetic loop ('t Hooft loop) is related to adding NS5 branes and that the presence of an electric loop (Wilson loop) is related to adding D5 branes. Here we propose that dyonic loops are related to the presence of $(p,q)$ 5-branes.

\begin{figure}[t]
\centering
\includegraphics[scale=0.8]{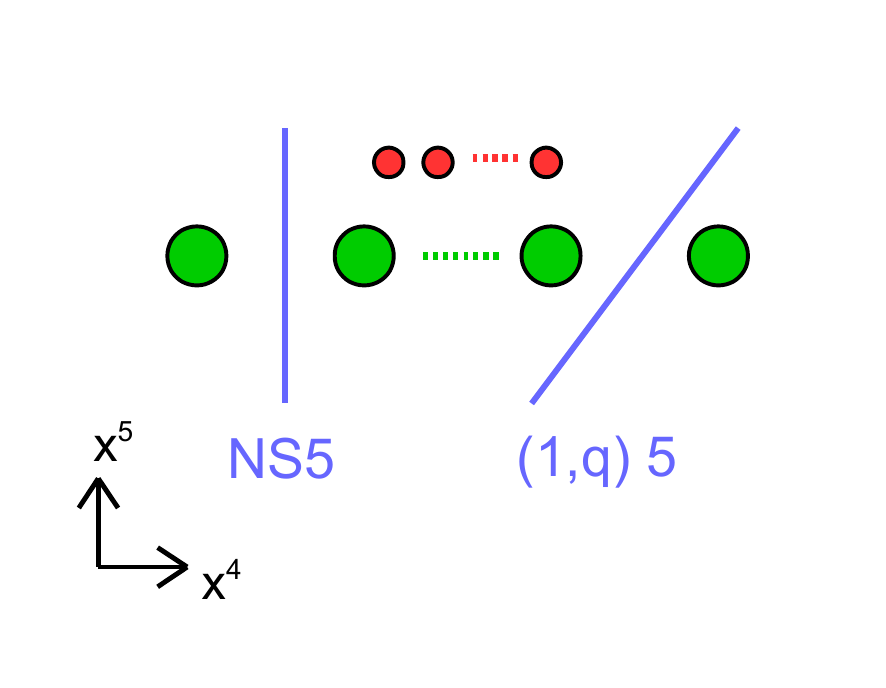}
\vspace{-1cm}
\caption{\footnotesize{Initial brane setup for the dyonic loop $L_{1,q}$, with an NS5 brane and a $(1,q)$ 5-brane.}}
\label{DLoop0}
\end{figure}

The idea is that a $(p,q)$ 5-brane induces a magnetic charge $\pm p$ and electric charge $\pm q$ on the D3 worldvolume gauge theory, with $\pm$ depending on whether the D3 brane is placed to the left/top or to the right/bottom of the 5-brane. Based on this principle we can realize a dyonic loop of arbitrary electric and magnetic charge in the $U(N)$ theory. For the $SU(N)$ theory one restricts to the allowed subset of magnetic and electric charges. For instance, adding an NS5 brane ($=(1,0)$ 5-brane) and a $(1,q)$ 5-brane as described in Figure \ref{DLoop0} realizes a dyonic loop operators $L_{1,q}$ of magnetic charge $B=(1,0^{N-2},-1)$ and electric charge $E=\frac{q}{2}\times (1^{N-1}, -1)$.  Here the electric charge $E$ should be thought of as defining a representation of the stabilizer of $B$ in $U(N)$. In this case the stabilizer of $B$ is  $U(1)^2\times U(N-2) \sim U(1)^3\times SU(N-2) $. The electric charge $E$ decomposes as $E =  \frac q2\times (1,0^{N-1}) + \frac q2\times (0^{N-1},-1) + \frac q2 \times (0,1^{N-2},0)$. It corresponds to a representation of charge $(\frac q2, -\frac q2, \frac{N-2}{2} q)$ under the $U(1)^3$ factors and the trivial representation under $SU(N-2)$. For the theory with $SU(N)$ gauge group the overall $U(1)$ electric charge is unphysical and we have $E \simeq q\times (0^{N-1}, -1)$.

\medskip

\begin{figure}[t]
\centering
\includegraphics[scale=1]{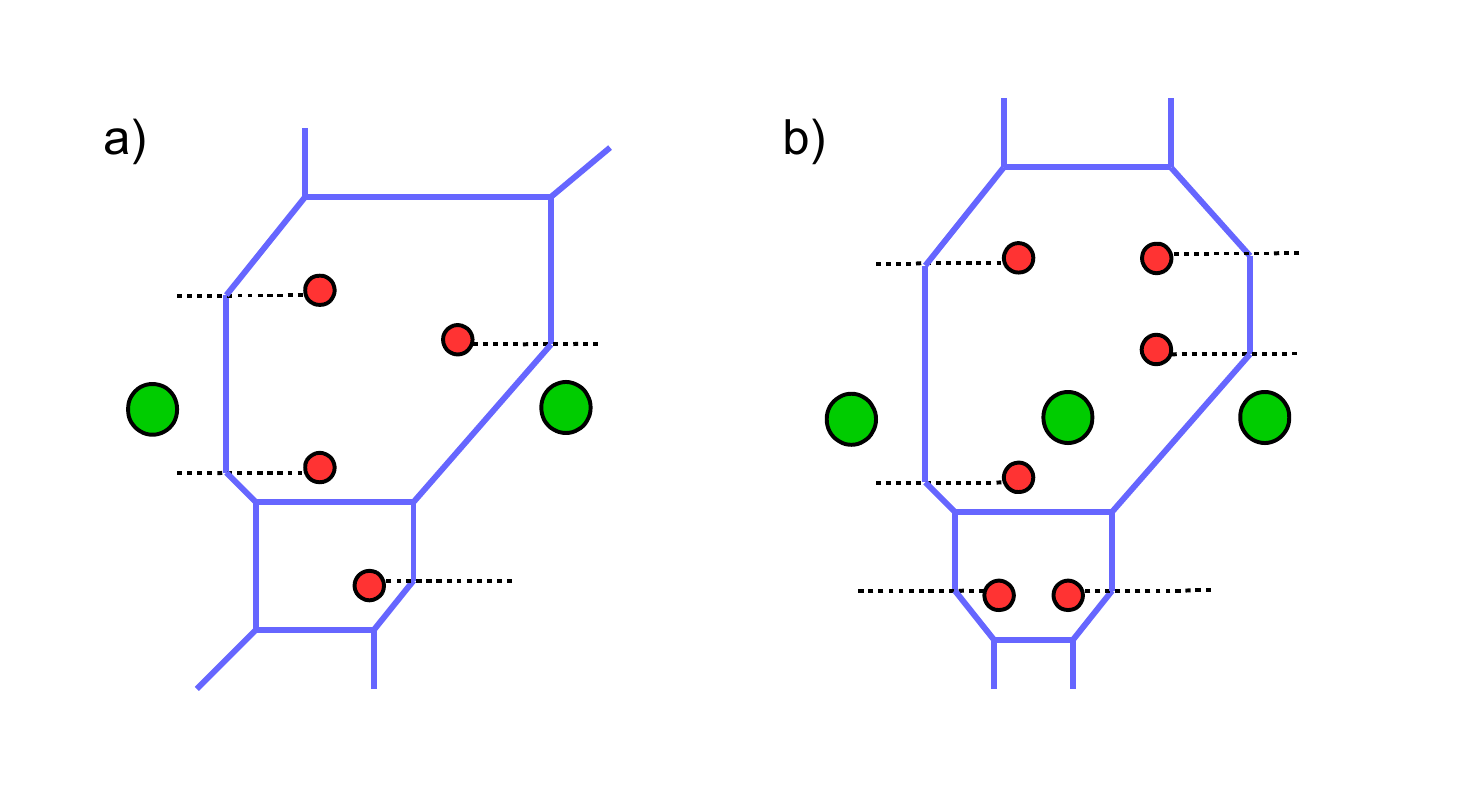}
\vspace{-1cm}
\caption{\footnotesize{Complete brane system for the dyonic loop $L_{1,1}$ in the (a) $SU(2)$ and (b) $SU(3)$ theories.}}
\label{U2_U3}
\end{figure}

We now focus on the minimal loop $L_{1,1}$, whose charges are $B=(1,0^{N-2},-1)$ and $E = \frac 12 \times (1,1^{N-2}, -1)$.\footnote{ For $N>2$, this is not the only dyonic loop  that one would like to call ``minimal". For instance the loop with $B=(1,0^{N-2},-1)$ and $E = \frac 12\times(1, 1^{N-3}, - 1, 1) = \frac 12 \times (1^N) + (0,0^{N-3},-1,0)$ is also a minimal dyonic loop. In this case the magnetic loop is dressed with a Wilson loop in the anti-fundamental representation of $SU(N-2)$.} In Figure \ref{U2_U3} we show how to complete the brane setups for $L_{1,1}$ in the conformal SQCD theories, for $N=2$ and $N=3$. We have arranged the D7 branes so that half of them is above the D3 branes and half of them is below the D3 branes (as they stand in the initial setup of Figure \ref{DLoop0}). Under this constraint, they are otherwise placed at convenience. The setups for higher values of $N$ can be worked out in a similar fashion.

\begin{figure}[t]
\centering
\includegraphics[scale=0.8]{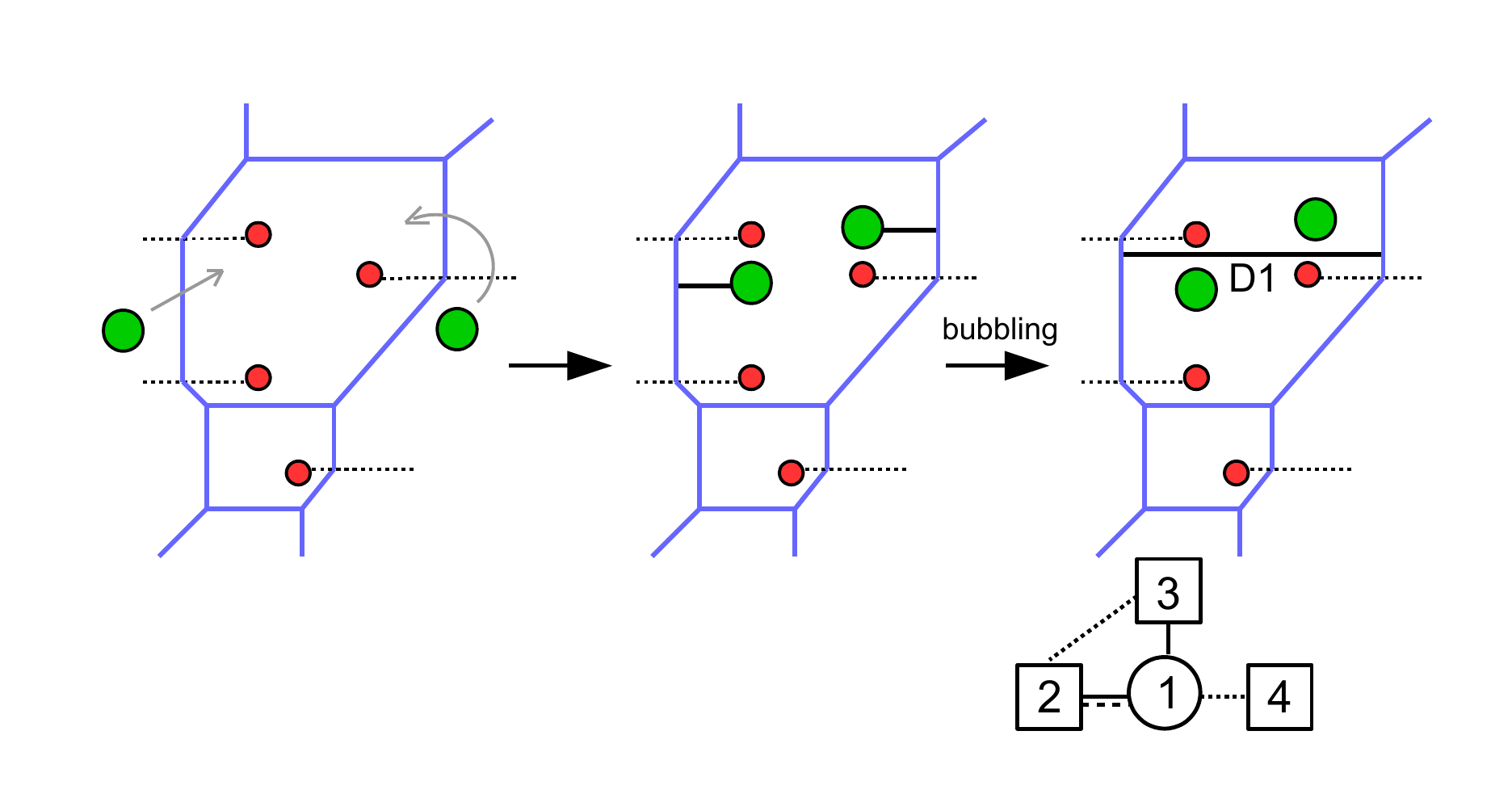}
\vspace{-0.5cm}
\caption{\footnotesize{Moving the D3s inside the web, through NS5 branes, we reach a configuration that can bubble. On the right: bubbling configuration for $L_{1,1}$ in the $SU(2)$ $N_f=4$ theory.}}
\label{U2DL}
\end{figure}

The bubbling contribution arises from D1 segments stretched between two D3s and combining with other D1 segments to form a D1 string stretched between two NS5 branes. It is less direct to see how this is realized in the dyonic context. We show in Figure \ref{U2DL} how this happens for the $SU(2)$ theory. First we move the D3s around and push them inside the 5-brane web by crossing NS5 branes. This creates D1 segments between the D3s and the NS5s in such a way that they can combine with the bubbling D1 segment to reach the desired configuration. In the process one of the D3 branes had to cross a D7 cut. This creates an excitation of an F1 string stretched between the D3 and the D7. However this excitation has no effect on the bubbling term since in the resulting ADHM quiver one has to sum over D3-D7 excitations. Therefore, for the computation of the bubbling term, we can simply ignore this effect.

\begin{figure}[t]
\centering
\includegraphics[scale=0.8]{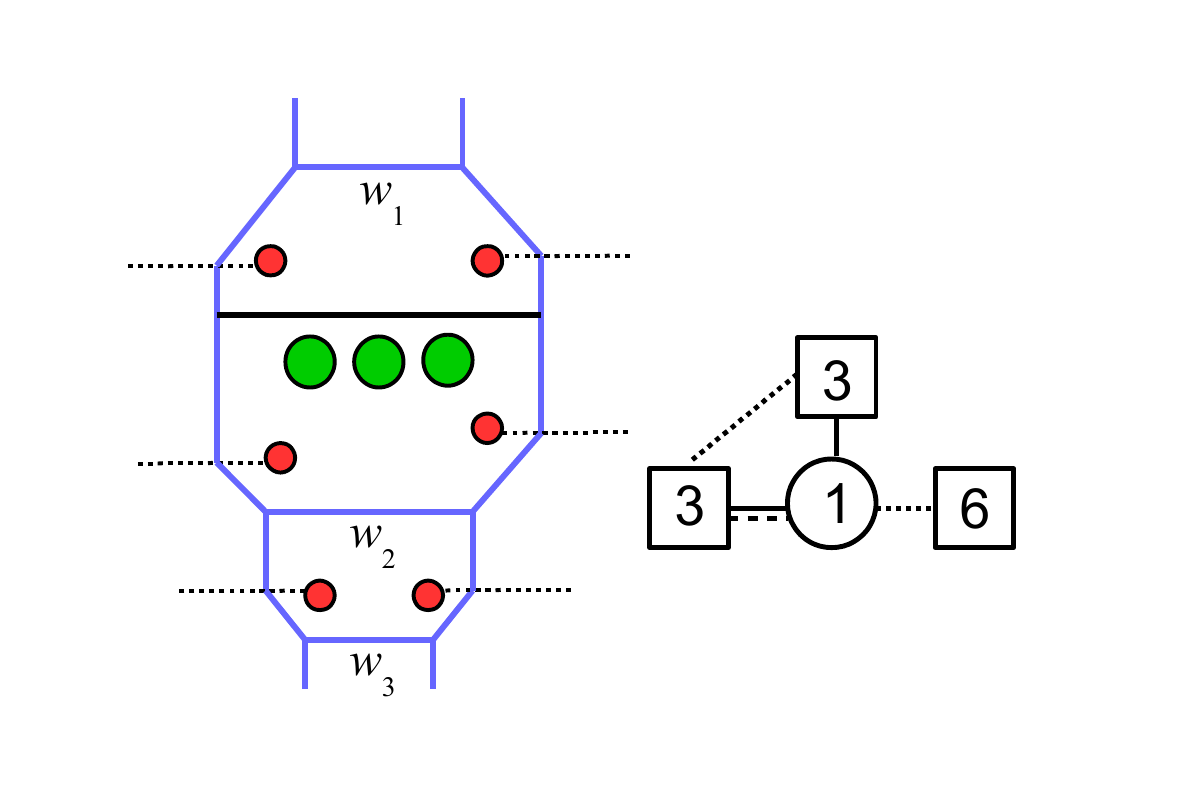}
\vspace{-1cm}
\caption{\footnotesize{Bubbling configuration for the dyonic loop $L_{1,1}$ in the $SU(3)$ $N_f=6$ theory.}}
\label{U3DL}
\end{figure}

The same manipulation leads to the bubbling configuration of Figure \ref{U3DL} for the $L_{1,1}$ loop in the $SU(3)$ theory. Here we have arranged the D7 branes so as to make the web relatively good-looking. In this form we realize that the same bubbling configuration could have been reached starting from the minimal 't Hooft loop incomplete setup and completing it in a non-symmetrical way, by putting four D7 below the D3s (and D1) and two D7 above the D3s (and D1). Similarly the configuration on the right in Figure \ref{U2DL} could be reached by completing the $SU(2)$ minimal 't Hooft loop setup with three D7s below and one D7 above. Therefore we see that the repartition of D7 branes in the process of completing the brane web is important. Our claim here is that the configurations of Figure \ref{U2DL} and \ref{U3DL} correspond to the dyonic loop bubbling (and not the minimal 't Hooft loop bubbling). 

\subsubsection{$Z_{1,1}^{0,1}$ bubbling}

\begin{figure}[t]
\centering
\includegraphics[scale=1]{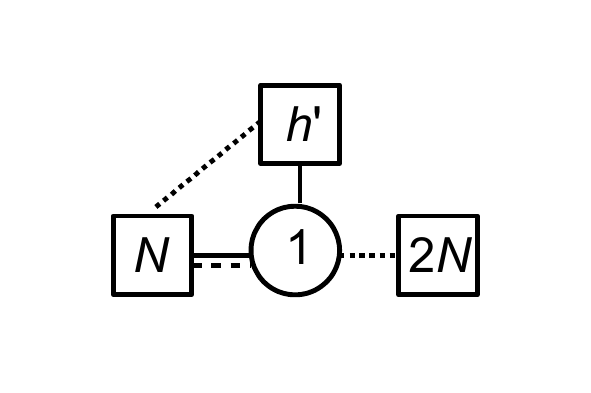}
\vspace{-1cm}
\caption{\footnotesize{ADHM SQM associated with the $L_{1,1}$ bubbling. $h'=N+1$ for $N$ even and $h'=N$ for $N$ odd.}}
\label{DL_SQM}
\end{figure}

The bubbling sector screens the magnetic charge, but leaves the electric charge unscreened. Let us denote $Z_{1,1}^{0,1}$ the bubbling factor in the $L_{1,1}$ loop vev. From the complete brane setups of the bubbling contribution we read the improved ADHM SQM for $Z_{1,1}^{0,1}$. For the $SU(2)$ and $SU(3)$ theories the improved SQM quivers are indicated in Figures \ref{U2DL} and \ref{U3DL}. For the general $SU(N)$ $N_f=2N$ theory the improved ADHM is given in Figure \ref{DL_SQM}. It has a $U(1)$ gauge node (with $(0,4)^\star$ vector multiplet), $2N$ fundamental Fermi multiplets, $N$ $(0,4)^\star$ fundamental hypermultiplets, $h'$ (0,4) fundamental hypermultiplets and $h'N$ uncharged Fermi multiplets, with $h' = N+1$ for $N$ even and $h' =N$ for $N$ odd. Here $h'$ is always odd. This is the same as the improved SQM for the minimal 't Hooft loop $L_1$, except for the replacement of $h$ by $h'$.

Note that we are calling the SQM an ``improved" SQM, but we did not have an initial ADHM theory to start with. Such an initial ADHM SQM does exist though. It can be read from an incomplete brane setup where the effect of the extra D5 segments on the SQM matter is dismissed. In such a case the SQM has a non-zero Chern-Simons level (we will discuss this in the next subsection).

\medskip

From the improved SQM we can compute the Witten index $\cI_{1,1}^{0,1}$ with say $\zeta >0$ (the result does not depend on the choice of sign). The mass deformed index is given by
\bea
\cI_{1,1}^{0,1}(\epsilon_-) &=  \int\limits_{JK_{\zeta>0}} \frac{\dd \phi}{2\pi i} \frac{(-1)\sh(2\epsilon_+)}{\sh(\epsilon_1)\sh(\epsilon_2)} \prod_{i=1}^N \frac{ \sh [\pm (\phi - a_i) + \epsilon_- ] }{\sh [\pm (\phi - a_i )+ \epsilon_+] }  \frac{\prod_{k=1}^{2N} \sh (\phi - m_k) \prod_{i =1}^N\prod_{n=1}^{h'} \sh(a_i - v_n)}{\prod_{n=1}^{h'} \sh[\pm ( \phi - v_n) - \epsilon_+]} \,.
\eea
We now extract the bubbling contribution by the same manipulation as in the 't Hooft loop case,
\bea
Z_{1,1}^{0,1} &=   \oint_{\cC} \prod_{n=1}^{h'} \frac{dw_n}{2\pi i w_n} \, F'(w) \, \lim_{\epsilon_- \to \infty} \frac{(-1)^{N+1}}{\sh(\epsilon_-)^{2N-2}}  \cI_{1,1}^{0,1}(\epsilon_-,w) \,, \cr
\text{with} & \quad F'(w) =    \lp \frac{\prod_{n=1}^{(h'-1)/2} w_n}{\prod_{n'=(h'+1)/2}^{h'} w_{n'}} \rp^{\frac{N-2}{2}} \,,
\label{rel5}
\eea
with the residue at the origin in $w_1$ taken first, and then the residue at the origin in $w_2$, ... etc. 

Once again the factor $F'$ select a sector of a definite charge under the $U(1)^{h'}$ flavor symmetry associated with the D5 segments in the brane web. This charge sector can be worked out by computing the $U(1)^{h'}$ background charge due to the extra multiplets associated with the D5s in the improved SQM, taking into account that the repartition of the D5 segments is such that $(h'-1)/2$ of them are above the D3 branes and D1 string and the rest is below (see Section \ref{ssec:L1Gen}).

We have no good justification for the presence of the factor $(-1)^{N+1}$ in the normalization of the result. We do not know how to fix the overall sign and our prescription follows from consistency requirement, when relating the full dyonic loop vev to OPEs between loops, as we discuss in section \ref{sec:starprod}. Still there are several consistent choices of signs and we do not know how to fix it completely. 

\medskip

The evaluation of $\cI_{1,1}^{0,1}(\epsilon_-)$ and $Z_{1,1}^{0,1}$ is not very different from that of the minimal 't Hooft loop discussed in previous sections. From computations at low values of $N$ we infer the result
\bea
Z_{1,1}^{0,1} &=  - e^{\frac 12 \sum_i a_i + \epsilon_+}\sum_{i=1}^N e^{-a_i} \frac{ \prod_{k=1}^{2N} \sh (a_i - m_k - \epsilon_+) }{\prod_{j\neq i}\sh(a_{ij})\sh(a_{ij} - 2\epsilon_+)} \cr
& \quad - e^{\frac 12\sum_{k=1}^{2N} m_k + \epsilon_+ - \frac 12 \sum_{i=1}^N a_i } \lp \sum_{k=1}^{2N} e^{-m_k} - \sum_{i=1}^N e^{-a_i + \epsilon_+} \rp \,.
\label{Z1101final}
\eea
The terms in the first line correspond to what could be obtained from the SQM of an incomplete brane setup (without D5 segments) by taking JK residues, while the term on the second line would be the missing $Z_{\rm extra}$ piece. 
On the first line we recognized a sum of terms weighted by factors $e^{\frac 12( -a_i + \sum_{j\neq i} a_j)}$, which should be interpreted as the classical contributions due to the unscreened electric charge $E = (\frac 12, \cdots, \frac 12, - \frac 12)$. Each electric charge sector in the Weyl average is weighted with its own monopole bubbling factor. The reason why this electric charge is not visible in the terms on the second line is unclear.

One can check that this expression is invariant under the $\bZ_2$ symmetry $\epsilon_+ \to -\epsilon_+$. 
Equation \eqref{Z1101final} gives the bubbling contribution in the $U(N)$ theory. For the $SU(N)$ theory one imposes $\sum_i a_i =0$.

\subsubsection{Simplification}
\label{sssec:simplerDL}

\begin{figure}[t]
\centering
\includegraphics[scale=0.8]{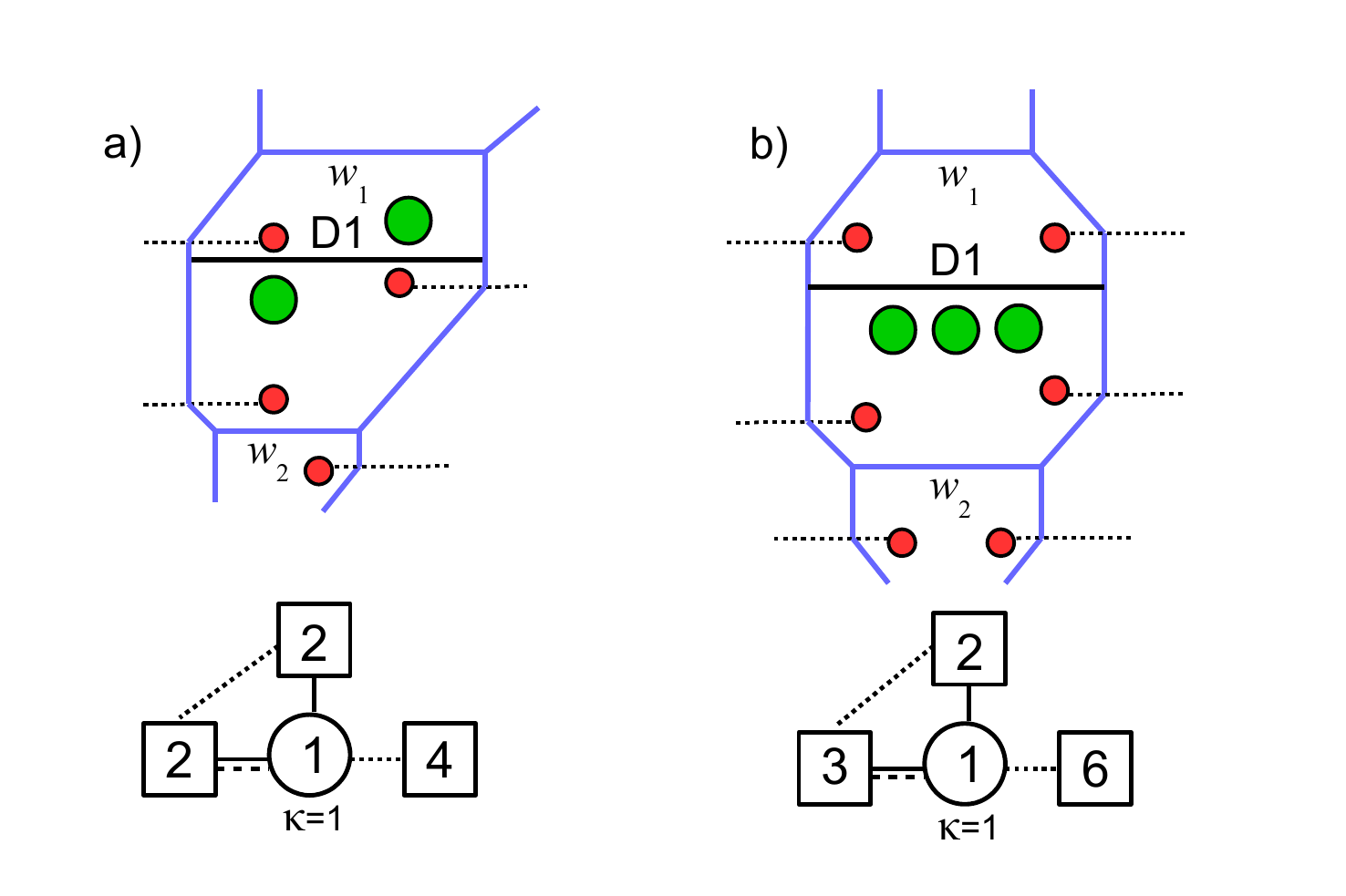}
\vspace{-0.5cm}
\caption{\footnotesize{Partial completion of the brane setup for the $L_{1,1}$ bubbling in the (a) $SU(2)$ and (b) $SU(3)$ theory. The resulting SQMs have Chern-Simons level $\kappa=1$.}}
\label{DL2}
\end{figure}

As in the $L_{1}$ 't Hooft loop case, we observe that the relation \eqref{rel5} can be simplified, namely $Z_{1,1}^{0,1}$ can be obtained by residues from a simpler SQM. This simpler SQM is obtained from a brane setup which is a partial completion of the initial brane setup. This partial completion is such that we add only two D5 segments, one above and one below the D1 string, as shown in Figure \ref{DL2} for $N=2$ and $N=3$. The resulting SQM is simpler in the sense that it has less matter fields. 

One subtlety here is that the SQM that one reads from these brane setup has a Chern-Simons level $\kappa=1$. The Chern-Simons level is associated with branes sourcing matter fields and is computed by the formula
\be
\kappa =  N^{\rm up}_{D5} - N^{\rm down}_{D5}  - \frac 12 \lp N^{\rm up}_{D7} - N^{\rm down}_{D7} \rp \,,
\label{CSformula}
\ee
with $N^{\rm up}_{D5}$, respectively $N^{\rm down}_{D5} $, the number of D5 segments placed above, respectively below, the D1 string, and similarly for $N^{\rm up}_{D7}$ and $N^{\rm down}_{D7}$. 
In all the brane systems that we have studied in previous sections, the Chern-Simons level was always vanishing. This is not the case any more for the SQM of Figure \ref{DL2}. It is the same SQM as for the $L_1$ 't Hooft loop bubbling in \eqref{I10simple}, except for the extra CS term. 

\medskip

The index of this SQM is computed by
\bea
\ti\cI^{0,1}_{1,1}(\epsilon_-) &= \int\limits_{JK_{\zeta>0}} \frac{\dd \phi}{2\pi i} \, e^{-\phi} \, \frac{(-1)\sh(2\epsilon_+)}{\sh(\epsilon_1)\sh(\epsilon_2)} \prod_{i=1}^N \frac{ \sh [\pm (\phi - a_i) + \epsilon_- ] }{\sh [\pm (\phi - a_i )+ \epsilon_+] }  \frac{\prod_{k=1}^{2N} \sh (\phi - m_k) \prod_{i =1}^N\prod_{n=1,2} \sh(a_i - v_n)}{\prod_{n=1,2} \sh[\pm ( \phi - v_n) - \epsilon_+]} \,.
\eea
We now observe that the result \eqref{Z1101final} for $Z_{1,1}^{0,1}$ can be reproduced by the simpler residue computation
\be
Z_{1,1}^{0,1} =   \oint_{\cC} \frac{dw_1 dw_2}{(2\pi i)^2w_1w_2}  \lp \frac{w_1}{w_2} \rp^{\frac{N-2}{2}} \lim_{\epsilon_- \to \infty} \frac{(-1)^N e^{\frac 12 \sum_i a_i}}{\sh(\epsilon_-)^{2N-2}}  \ti\cI^{0,1}_{1,1}(\epsilon_-, w_1,w_2) \,,
\ee
with the residue in $w_1$ taken first, and then the residue in $w_2$. Here the factor $e^{\frac 12 \sum_i a_i}$ takes into account the electric charge induced on the D3s by the missing D5 segments.\footnote{Since for general $N$ we remove $(h'-1)/2$ D5 segments downstairs and $(h'-1)/2 -1$ D5 segments upstairs, we compensate by multiplying by a single factor $e^{\frac 12 \sum_i a_i}$.}

\subsection{Non-minimal 't Hooft loops}
\label{ssec:L2}

Our method applies to 't Hooft loops of higher magnetic charge. In this section we provide the example of the bubbling contribution for a non-minimal 't Hooft loop in conformal SQCD, which we denote $L_2$, with magnetic charge $B = (2,0,\cdots, 0,-2) := B_2$. 

The vev of $L_2$ decomposes into three sectors: the unscreened sector of charge $B_2$, the partially screened bubbling sector of charge $v_1 = (1,0,\cdots,0,-1)$ and the fully screened sector of charge $v_2= \vec 0$:
\bea
\vev{L_2} &= \lp \sum_{i\neq j} e^{2(b_i-b_j)} \rp Z_{\rm 1-loop}(B_2) + \lp \sum_{i\neq j} e^{b_i-b_j} \rp Z_{\rm 1-loop}(v_1)Z_{\rm mono}(B_2,v_1) + Z_{\rm mono}(B_2,v_2)   \,,
\eea
where we used $ Z_{\rm 1-loop}(v_2=\vec 0)=1$. Our goal is to compute $Z^1_2 := Z_{\rm mono}(B_2,v_1)$ and $Z^0_2 := Z_{\rm mono}(B_2,v_2)$.

\medskip

\begin{figure}[t]
\centering
\includegraphics[scale=0.75]{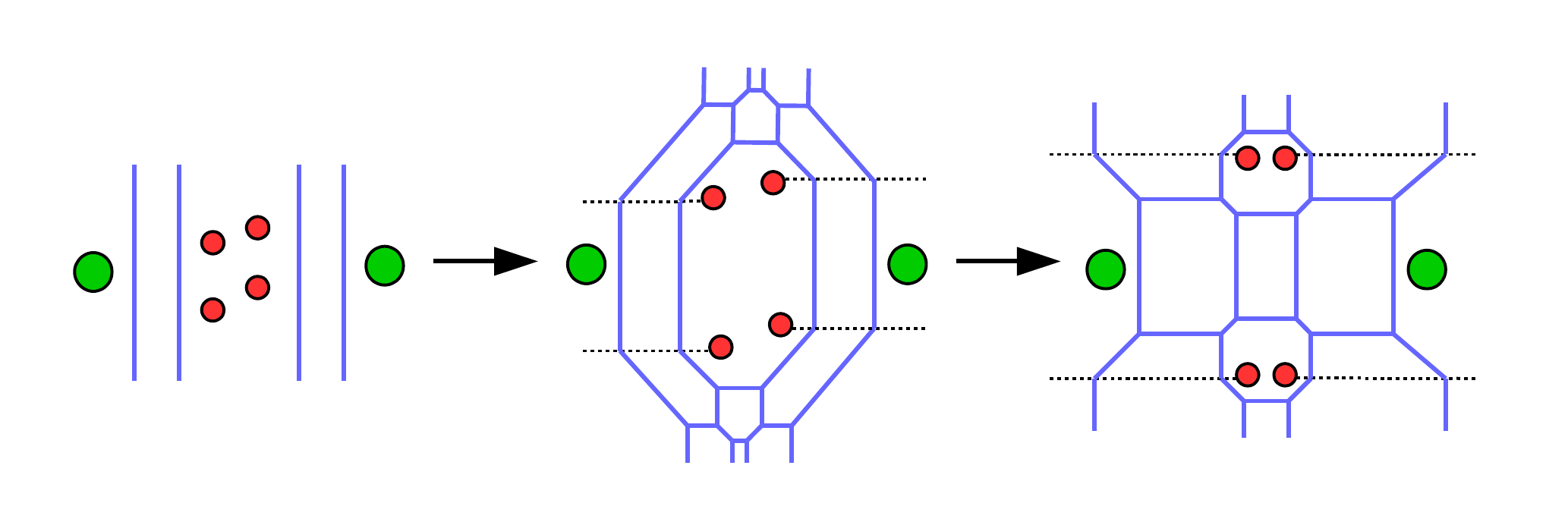}
\vspace{-1cm}
\caption{\footnotesize{Initial brane setup and completion for the $L_2$ loop in $SU(2)$ $N_f=4$ theory. Moving the branes we obtain a nicer configuration (on the right).}}
\label{U2L2}
\end{figure}
The brane setup realizing $L_2$ has now four NS5 branes. In Figure \ref{U2L2} we show the completed brane configuration for $L_2$ in the $SU(2)$ $N_f=4$ theory. Because there are more NS5 branes to begin with, the completion leads to a more involved 5-brane web.

\medskip

\begin{figure}[t]
\centering
\includegraphics[scale=0.75]{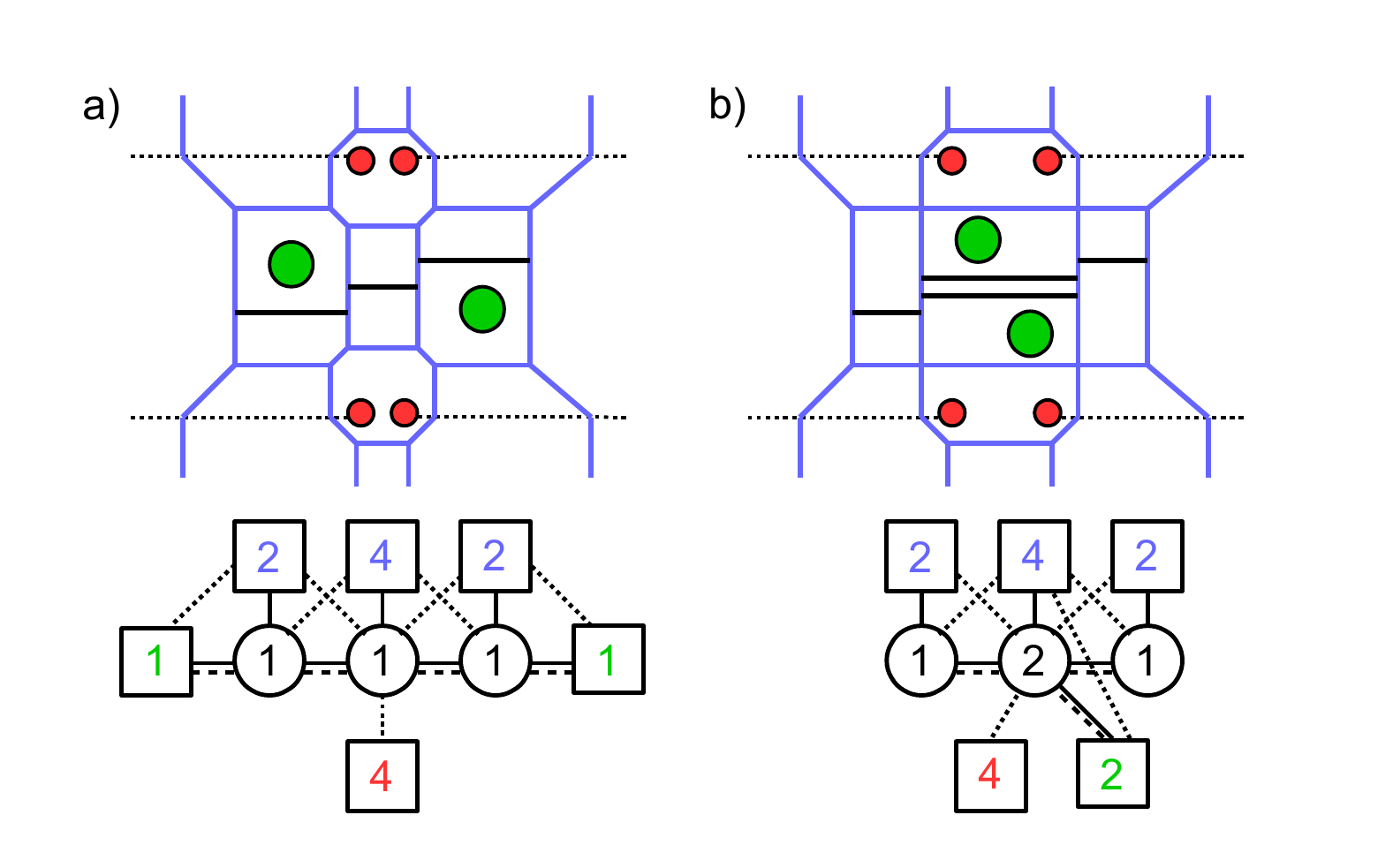}
\vspace{-0.5cm}
\caption{\footnotesize{Brane setup and improved SQM for the bubbling sectors (a) $v_1$ for $Z^1_2$, and (b) $v_2$ for $Z^0_2$.}}
\label{U2L2Bub}
\end{figure}

We will focus on the $SU(2)$ $N_f=4$ theory for simplicity. The bubbling sector $v_1$ arises when a D1 string is stretched between the two NS5s. The D3s can be moved inside the brane web and the D1 can break into three segments as shown in Figure \ref{U2L2Bub}-a.
The resulting improved SQM has three $U(1)$ nodes. It is given by the quiver of the Figure \ref{U2L2Bub}-a. We denote $\cI^1_2$ its Witten index.

The improved SQM now depends on 8 extra flavor fugacities for the $U(2)\times U(4) \times U(2)$ flavor symmetry associated with the D5 segments. We denote these fugacities $w^{(1)}_{n=1,2}$, $w^{(2)}_{m=1,2,3,4}$, $w^{(3)}_{n=1,2}$. The bubbling contribution $Z^1_2$ is obtained from $\cI^1_2$ by taking a residue in these fugacities. Precisely we have the formula
\be
Z^1_2 =    \oint_{\cC} \frac{d^2w^{(1)} d^4w^{(2)} d^2w^{(3)}}{(2\pi i)^8 \prod_{n} w^{(1)}_n w^{(3)}_{n}\prod_m w^{(2)}_{m} }  \,  \lim_{\epsilon_- \to \infty} \frac{1}{\sh(\epsilon_-)^{2}} \cI^1_2(\epsilon_-, w^{(1)}, w^{(2)}, w^{(3)} ) \,,
\label{rel6}
\ee
where the contour $\cC$ picks up the poles at the origin for each fugacity. 
Without further details let us give the final result of the computation:
\bea
Z^1_2 &=  - \sum_{i=1,2} \frac{\prod_{k=1}^{4} \sh(a_i - m_k -2\epsilon_+)}{\prod_{j\neq i}\sh(a_i-a_j)\sh(a_i-a_j - 2\epsilon_+)}  - \sum_{i=1,2} \frac{\prod_{k=1}^{4} \sh(a_i - m_k -2\epsilon_+)}{\prod_{j\neq i}\sh(a_i-a_j-2\epsilon_+)\sh(a_i-a_j - 4\epsilon_+)}  \cr
& \quad + \ch(\sum_{k=1}^{4} m_k - 2(a_1 + a_2)+ 2\epsilon_+ ) + \ch(\sum_{k=1}^{4} m_k - 2(a_1 + a_2) + 6\epsilon_+ ) \,.
\label{Z12fin}
\eea
This result is in agreement with the computation of \cite{Brennan:2018rcn} ($Z_{\rm mono}(2,1)$ in section 3.6.2), after imposing $a_1+a_2=0$. Once again the terms on the second line correspond to the contribution $Z_{\rm extra}$ in the computation that uses the non-improved SQM.

\medskip

The second bubbling sector arises when we stretch one more D1 segment between the two D3 branes, screening completely the magnetic charge. In this case the D3 branes can be moved to the central region of the web and the resulting configuration has one more D1 segment in the middle. The gauge group of the improved SQM is then $U(1)\times U(2) \times U(1)$. The explicit brane configuration and improved SQM are given in Figure \ref{U2L2Bub}-b. 

We denote $\cI^0_2$ the index of this SQM .
For the sake of clarity let us write down the matrix model explicitly in this more complicated case:
\bea
\cI^0_2(\epsilon_-) &= \int_{{\rm JK}>0} \frac{d\hat\phi d^2\phi d\ti\phi}{2(2\pi i)^4} \Big[\frac{\sh(2\epsilon_+)}{\sh(\epsilon_1)\sh(\epsilon_2)}\Big]^4 \frac{\sh(\phi_{12})^2\sh(\pm\phi_{12} - 2\epsilon_+)}{\sh(\pm\phi_{12} -\epsilon_1)\sh(\pm\phi_{12} -\epsilon_2)}  \prod_{m=1}^4\prod_{p=1,2}\sh(v^{(2)}_m - a_p) \cr
& \times \prod_{i=1,2}\frac{\sh(\pm(\phi_i -\hat\phi)+\epsilon_-)\sh(\pm(\phi_i -\ti\phi)+\epsilon_-)\prod_{p=1,2} \sh(\pm(\phi_i -a_k) +\epsilon_-) }{\sh(\pm(\phi_i -\hat\phi)+\epsilon_+)\sh(\pm(\phi_i -\ti\phi)+\epsilon_+)\prod_{p=1,2} \sh(\pm(\phi_i -a_k)+\epsilon_+) }  \prod_{k=1}^4 \sh(\phi_i - m_k) \cr
& \times \frac{\prod_{i=1,2}[\prod_{n=1,2} \sh(\phi_i - v^{(1)}_n)\sh(\phi_i - v^{(3)}_n)] \prod_{m=1}^4\sh(\hat\phi - v^{(2)}_m)\sh(\ti\phi - v^{(2)}_m)}{\prod_{i=1,2}\prod_{m=1}^4 \sh(\pm(\phi_i -v^{(2)}_m)-\epsilon_+) \prod_{n=1,2}\sh(\pm(\hat\phi -v^{(1)}_n)-\epsilon_+)\sh(\pm(\ti\phi -v^{(3)}_n)-\epsilon_+)} \,.
\eea
We have now the relation 
\be
Z^0_2 =  \oint_{\cC} \frac{d^2w^{(1)} d^4w^{(2)} d^2w^{(3)}}{(2\pi i)^8 \prod_{n} w^{(1)}_n w^{(3)}_{n}\prod_m w^{(2)}_{m}}  \,  \lim_{\epsilon_- \to \infty} \frac{1}{\sh(\epsilon_-)^{4}}   \cI^0_2(\epsilon_-, w^{(1)}, w^{(2)}, w^{(3)} ) \,,
\label{rel8}
\ee
which evaluates to
\bea
Z^0_2 &= \frac{\ch(2\epsilon_+)^2\prod_{i=1,2}\prod_{k=1}^4 \sh(a_i - m_k - \epsilon_+)}{\sh(a_{12} \pm 2\epsilon_+)^2} 
+ \sum_{i=1,2} \frac{\prod_{k=1}^4\sh(a_i - m_k - \epsilon_+)\sh(a_i - m_k - 3\epsilon_+)}{\prod_{j\neq i}\sh(a_{ij})\sh(a_{ij}-2\epsilon_+)^2 \sh(a_{ij}-4\epsilon_+)} \cr
& \quad 
- \ch(\sum_k m_k - 2a_1 -2 a_2 + 4\epsilon_+) \sum_{i=1,2} \frac{\ch(2\epsilon_+)\prod_{k=1}^4 \sh(a_i - m_k -\epsilon_+)}{\prod_{j\neq i}\sh(a_{ij}) \sh(a_{ij} - 2\epsilon_+)}  \cr
& \quad + \ch(\sum_k m_k - 2a_1 -2 a_2 + 2\epsilon_+)^2  \,.
\label{rel9}
\eea
The terms on the first line corresponds to the JK-evaluation of the ADHM quiver constructed from the incomplete brane configuration without D5s. The terms on the second and third line arise from our complete brane system.
\medskip

It can be checked that the expressions for $Z^1_2$ and $Z^0_2$ are invariant under $\epsilon_+ \to -\epsilon_+$ as they should. This happens only after one sums over the terms on the three lines.


\section{Non-commutative product and tests of the results}
\label{sec:starprod}

The line operators studied in this paper are placed at a point on $\bR^3$ and are wrapping an $S^1$ circle. In the localization computation one turns on an Omega deformation with parameter $\epsilon_+$ on $\bR^2 \subset \bR^3$ which forces the line to sit at the origin of $\bR^2$ (in order to preserve some supersymmetries). In this context there is a notion of ordering of the operators along the transverse $\bR$ line. We can insert loops $L_i$ at different positions $z_i$ along this line. By a standard argument the vev of a product of BPS loops depends on the positions $z_i$ only through their ordering.   So we can have 
\be 
\vev{L_1(z_1)L_2(z_2)}_{z_1 > z_2} := \vev{L_1.L_2} \neq \vev{L_2.L_1} \,, 
\ee
for $\epsilon_+ \neq 0$ and $L_1 \neq L_2$ two line operators.
The OPE of two line operators defines a non-commutative product acting on the line operator algebra.
This product turns out to have the form of a Moyal product
\be
\vev{L_1.L_2} = \vev{L_1}\star\vev{L_2} \,,
\ee
given by
\be
(f\star g)(a,b) := e^{\epsilon_+ \sum_i (\p_{b_i}\p_{a'_i} - \p_{a_i}\p_{b'_i} )} f(a,b) g(a',b') \Big|_{a'=a \atop b'=b} \,.
\ee
The Moyal product endows the line operator algebra with a Poisson structure, which can be used to quantize this algebra.

The star product of two line operators can be computed via a simple formula.
With the vev of a loop $L$ given by the expansion
\be
\vev{L} = \sum_{w,v} e^{w.a+v.b} Z_L (a;w,v) := \sum_{w,v} Z_{L,\text{tot}} (a,b;w,v) \,,
\ee
we then have
\be
\vev{L_1}\star\vev{L_2} = \sum_{w_1,v_1} \sum_{w_2,v_2}  Z_{L_1,\text{tot}} \big( a - \epsilon_+ v_2 ,b ;w_1,v_1 \big)Z_{L_2,\text{tot}} \big( a + \epsilon_+ v_1 ,b ;w_2,v_2 \big)  \,.
\label{starprod}
\ee
This structure naturally emerges from the localization computation of \cite{Ito:2011ea}. We refer to that paper for more details.

\medskip

We would like to check that our findings are compatible with this structure. For this, we will compute the two products $\vev{L_{0,1}}\star\vev{L_{1,0}}$ and $\vev{L_{1,0}}\star\vev{L_{1,0}}$, and see if they are expressed as linear combinations of others loops, in particular $\vev{L_{1,\pm 1}}$ and $\vev{L_{2,0}}$, that we have computed. For simplicity we will focus first on the $U(2)$ $N_f=4$ theory. In this theory, $\vev{L_{1,q}}$ refers to the dyonic loop with minimal magnetic charge $B=(1,-1)$ and electric charge $E=q(\frac 12, -\frac 12)$. The minimal 't Hooft loop is $L_{1} := L_{1,0}$ and the minimal Wilson loop is $L_{0,1}$.

The expressions for the vev of $L_{0,1}$ and $L_{1,0}$ are
\bea
\vev{L_{0,1}} &= e^{\frac 12 a_{12}} + e^{-\frac 12 a_{12}} \,, \cr
\vev{L_{1,0}} &= \lp e^{b_{12}} + e^{-b_{12}} \rp \lp \frac{\prod_{k=1}^4 \sh(a_1-m_k)\sh(a_2-m_k)}{\sh(\pm a_{12})\sh(\pm a_{12} + 2\epsilon_+)} \rp^{1/2}  \cr
& \quad - \frac{\prod_{k=1}^4 \sh(a_1-m_k-\epsilon_+)}{\sh(a_{12})\sh(a_{12} - 2\epsilon_+)} - \frac{\prod_{k=1}^4 \sh(a_2-m_k-\epsilon_+)}{\sh(a_{12})\sh(a_{12} + 2\epsilon_+)} + \ch(\sum_{k=1}^4 m_k - 2a_1 -2a_2 + 2\epsilon_+) \,,
\label{irredLoops}
\eea
with $a_{12} := a_1 -a_2$, $b_{12} := b_1 - b_2$ and $f(x\pm y) := f(x+y)f(x-y)$. 

Using formula \eqref{starprod},  we compute the star product between these two line operators
\bea
& \vev{L_{0,1}}\star\vev{L_{1,0}} = \cr 
& \lp e^{b_{12} + \frac{a_{12}}{2}  - \epsilon_+} + e^{-b_{12} + \frac{a_{12}}{2}  + \epsilon_+}  + e^{b_{12} - \frac{a_{12}}{2}  + \epsilon_+}  + e^{-b_{12} - \frac{a_{12}}{2}  - \epsilon_+}  \rp  \Big( \frac{\prod\limits_{i=1,2}\prod\limits_{k=1}^4 \sh(a_i-m_k)}{\sh(\pm a_{12})\sh(\pm a_{12} + 2\epsilon_+)} \Big)^{1/2} \cr
& + \Big( e^{\frac{a_{12}}{2}} + e^{-\frac{a_{12}}{2}}  \Big) \Big[ -\sum_{i=1,2}  \frac{ \prod_{k=1}^{4} \sh (a_i - m_k - \epsilon_+) }{\prod_{j\neq i}\sh(a_{ij})\sh(a_{ij} - 2\epsilon_+)} + \ch(\sum_{k=1}^4 m_k - 2a_1 -2a_2 + 2\epsilon_+) \Big] \,.
\label{prod1}
\eea

We expect this to be a combination of the loops $\vev{L_{1,\pm 1}}$.
We have computed the bubbling contribution to $L_{1,1}$ in section \ref{ssec:DL}. The computation for the bubbling of $L_{1,-1}$ is easily done following the same reasoning. The (partially complete) brane configuration for $L_{1,-1}$ is simply the upside-down reverse of Figure \ref{DL2} and the SQM Chern-Simons level is $\kappa=-1$.
The final result for the vevs of these dyonic loops, including non-bubbling and bubbling, is\footnote{In this relation $\pm$ is either plus or minus (not a product over two factors).}
\bea
\vev{L_{1,\pm 1}} &= \lp e^{b_{12} \pm \frac{a_{12}}{2} } +  e^{-b_{12} \mp \frac{a_{12}}{2} }\rp \Big( \frac{\prod\limits_{i=1,2}\prod\limits_{k=1}^4 \sh(a_i-m_k)}{\sh(a_{12})^2\sh(a_{12} + 2\epsilon_+)\sh(a_{12} - 2\epsilon_+)} \Big)^{1/2} \cr
& \quad - e^{\pm\frac{a_1+a_2}{2} \pm \epsilon_+} \sum_{i=1,2} e^{\mp a_i}  \frac{ \prod_{k=1}^{4} \sh (a_i - m_k - \epsilon_+) }{\prod_{j\neq i}\sh(a_{ij})\sh(a_{ij} - 2\epsilon_+)} \cr
& \quad - e^{\pm\frac 12\sum_{k=1}^{4} m_k \pm \epsilon_+ \mp \frac{a_1+a_2}{2}} \Big( \sum_{k=1}^{4} e^{\mp m_k} - e^{\pm \epsilon_+}( e^{\mp a_1} +  e^{\mp a_2}) \Big)  \,.
\eea
From here we find the non-trivial relation 
\be
\vev{L_{0,1}}\star\vev{L_{1,0}}  = e^{-\epsilon_+} \vev{L_{1,1}} + e^{\epsilon_+} \vev{L_{1,-1}} + \vev{W} \,,
\ee
with $\vev{W} = e^{\frac 12(\sum_k m_k - a_1 -a_2)}\sum_k e^{-m_k} +  e^{-\frac 12(\sum_k m_k - a_1 -a_2)}\sum_k e^{m_k}$, a sum of Wilson loop vevs for the flavor group and the central part of gauge group.  The existence of this relation is a non-trivial check of our computation of bubbling terms for the loops involved. 
Similarly we have 
\be
\vev{L_{1,0}}\star\vev{L_{0,1}}  = e^{\epsilon_+} \vev{L_{1,1}} + e^{-\epsilon_+} \vev{L_{1,-1}} + \vev{W} \,,
\ee
which corresponds to sending $\epsilon_+ \to -\epsilon_+$.

\medskip

We now look at the product (or OPE) between two $L_{1,0}$ loops. Using \eqref{starprod} we find
\bea
& \vev{L_{1,0}}\star\vev{L_{1,0}} = \big( e^{2b_{12}} + e^{-2b_{12}}  \big) \Big[ \frac{\prod_{i=1,2}\prod_{k=1}^4 \sh(a_i - m_k \pm\epsilon_+)}{\sh(\pm a_{12} + 2\epsilon_+)^2 \sh(\pm a_{12})\sh(\pm a_{12} + 4\epsilon_+)} \Big]^{1/2} \cr
& + \frac{\prod_{k=1}^4\sh(a_1 - m_k + \epsilon_+)\sh(a_2 - m_k - \epsilon_+)}{\sh(a_{12}+2\epsilon_+)^2\sh(a_{12})\sh(a_{12}+4\epsilon_+)} + \ ``a_1 \leftrightarrow a_2"  \  + (Z^0_1)^2 \cr
& - \big( e^{b_{12}} + e^{-b_{12}} \big) \Big[\frac{\prod_{k=1}^4 \sh(a_1-m_k)\sh(a_2-m_k)}{\sh(\pm a_{12})\sh(\pm a_{12} + 2\epsilon_+)} \Big]^{1/2}  \Big[ \sum_{i=1,2} \frac{\prod_{k=1}^{4} \sh(a_i - m_k)}{\prod_{j\neq i}\sh(a_{ij})\sh(a_{ij} + 2\epsilon_+)} \cr
&  \hspace{2cm} + \frac{\prod_{k=1}^{4} \sh(a_i - m_k -2\epsilon_+)}{\prod_{j\neq i}\sh(a_{ij}-2\epsilon_+)\sh(a_{ij} - 4\epsilon_+)} - 2\ch(\sum_{k=1}^{4} m_k - 2a_1- 2a_2 + 2\epsilon_+ )  \Big] \,,
\eea
with $Z^0_1$ the bubbling contribution to $\vev{L_{1,0}}$ in \eqref{irredLoops} (also given in \eqref{Z10final}, with $N=2$).

We want check that this is equal to the vev $\vev{L_{2,0}}$ of the next-to-minimal 't Hooft loop. This vev has an expansion
\be
\vev{L_2} = \big( e^{2b_{12}} + e^{-2b_{12}}  \big) Z_{\rm 1-loop}[B=(2,-2)] + \big( e^{b_{12}} + e^{-b_{12}}  \big) Z_{\rm 1-loop}[B=(1,-1)] Z^1_2 + Z^0_2 \,,
\ee
and we have computed the monopole bubbling contributions $Z^1_2$ and $Z^0_2$ in section \ref{ssec:L2}. 

We find that the relation 
\be
\vev{L_{1,0}}\star\vev{L_{1,0}}  = \vev{L_{2,0}} 
\label{rel20}
\ee
is correct if and only if
\bea
Z^1_2 &= \sum_{i=1,2} \frac{\prod_{k=1}^{4} \sh(a_i - m_k)}{\prod_{j\neq i}\sh(a_{ij})\sh(a_{ij} + 2\epsilon_+)}  - \frac{\prod_{k=1}^{4} \sh(a_i - m_k -2\epsilon_+)}{\prod_{j\neq i}\sh(a_{ij}-2\epsilon_+)\sh(a_{ij} - 4\epsilon_+)} \cr
& \quad + 2\ch(\sum_{k=1}^{4} m_k - 2a_1- 2a_2 + 2\epsilon_+ )  \,, \cr
Z^0_2 &= \frac{\prod_{k=1}^4\sh(a_1 - m_k + \epsilon_+)\sh(a_2 - m_k - \epsilon_+)}{\sh(a_{12}+2\epsilon_+)^2\sh(a_{12})\sh(a_{12}+4\epsilon_+)} + \frac{\prod_{k=1}^4\sh(a_1 - m_k - \epsilon_+)\sh(a_2 - m_k + \epsilon_+)}{\sh(a_{12}-2\epsilon_+)^2\sh(a_{12})\sh(a_{12}-4\epsilon_+)}  \cr
& \quad + (Z^0_1)^2 \,.
\label{rel21}
\eea
Although this is not obvious, one can check (with Mathematica for instance) that indeed our expressions \eqref{Z12fin} for $Z^1_2$ and \eqref{rel9} for $Z^0_2$ agree with \eqref{rel21}. 
This provides a strong check of the validity of our construction.

\medskip

\noindent{\bf Product of minimal loops in $U(N)$ theories}

Finally, if we think about 't Hooft loops in the $U(N)$ theory with $2N$ flavors, the minimally charged loop is not $L_{1,0}$, but rather there are two minimally charged loops $\ell^{\pm}$ of magnetic charges $B_\pm=(\pm 1,0^{N-1})$. These loops do not have bubbling sectors and their vev is simply given by
\be
\vev{\ell^+} = \sum_{i=1}^N e^{b_i}  \lp \frac{\prod_{k=1}^4 \sh(a_i-m_k)}{\prod_{j\neq i}\sh(\pm \, a_{ij} + \epsilon_+)} \rp^{1/2}  \,, 
\quad \vev{\ell^-} = \sum_{i=1}^N e^{-b_i}  \lp \frac{\prod_{k=1}^4 \sh(a_i-m_k)}{\prod_{j\neq i}\sh(\pm \, a_{ij} + \epsilon_+)} \rp^{1/2}  \,.
\ee
We expect that the OPE of $\ell^+$ and $\ell^-$ is related to $L_{1,0}$, which has magnetic charge $B=(1,0^{N-2},-1)$. We compute
\bea
\vev{\ell^-}\star \vev{\ell^+} &= \sum_{i\neq j} e^{b_{ij}} \lp \frac{\prod_{k=1}^{2N} \sh(a_i-m_k)\sh(a_j-m_k)}{\sh(\pm a_{ij})\sh( \pm a_{ij} + 2\epsilon_+) \prod_{k\neq i,j} \sh(\pm a_{ik} +\epsilon_+) \sh(\pm a_{jk} +\epsilon_+)} \rp^{\frac 12}  \cr
& \quad - \sum_{i=1}^N \frac{\prod_{k=1}^{2N} \sh(a_i-m_k-\epsilon_+)}{\prod_{j\neq i}\sh(a_{ij})\sh(a_{ij} - 2\epsilon_+)}  \,, \cr
\vev{\ell^+}\star \vev{\ell^-} &= \sum_{i\neq j} e^{b_{ij}} \lp \frac{\prod_{k=1}^{2N} \sh(a_i-m_k)\sh(a_j-m_k)}{\sh(\pm a_{ij})\sh( \pm a_{ij} + 2\epsilon_+) \prod_{k\neq i,j} \sh(\pm a_{ik} +\epsilon_+) \sh(\pm a_{jk} +\epsilon_+)} \rp^{\frac 12}  \cr
& \quad - \sum_{i=1}^N \frac{\prod_{k=1}^{2N} \sh(a_i-m_k+\epsilon_+)}{\prod_{j\neq i}\sh(a_{ij})\sh(a_{ij} + 2\epsilon_+)}  \,.
\label{OPEellpm}
\eea
This is to be compared with
\bea
\vev{L_{1,0}} &= \sum_{i\neq j} e^{b_{ij}} \lp \frac{\prod_{k=1}^{2N} \sh(a_i-m_k)\sh(a_j-m_k)}{\sh(\pm a_{ij})\sh( \pm a_{ij} + 2\epsilon_+) \prod_{k\neq i,j} \sh(\pm a_{ik} +\epsilon_+) \sh(\pm a_{jk} +\epsilon_+)} \rp^{\frac 12}  \cr
& \quad - \sum_{i=1}^N \frac{\prod_{k=1}^{2N} \sh(a_i-m_k-\epsilon_+)}{\prod_{j\neq i}\sh(a_{ij})\sh(a_{ij} - 2\epsilon_+)}  +  \ch(\sum_{k=1}^{2N} m_k - 2\sum_{i=1}^N a_i + 2\epsilon_+) \,,
\eea
which is invariant under $\epsilon_+ \to -\epsilon_+$.
Then we observe the relations
\bea
\vev{\ell^-}\star \vev{\ell^+} &= \vev{L_{1,0}} - \vev{W'}(\epsilon_+)  \,, \cr
\vev{\ell^+}\star \vev{\ell^-} &= \vev{L_{1,0}} - \vev{W'}(-\epsilon_+)  \,,
\eea
with $\vev{W'}(\epsilon_+) = e^{\epsilon_+} e^{\frac 12\sum_{k=1}^{2N} m_k - \sum_{i=1}^N a_i} +e^{-\epsilon_+} e^{-\frac 12\sum_{k=1}^{2N} m_k + \sum_{i=1}^N a_i}$ a linear combination of Wilson loop vevs. The OPE of the two loop operators closes on the loop operator algebra as it should.

Here the OPE between $\ell^+$ and $\ell^-$ can be realized with same brane configuration as $L_{1,0}$ with two NS5 branes, except that the two NS5 branes sit at different positions along the $z$ direction, which is the direction where operator insertions are ordered. Each NS5 sources one of the two $\ell^\pm$ line operators. This has a bubbling configuration which is as in Figure \ref{U2_0}, but with the NS5s separated along $z$. This separation of the NS5s is interpreted in the corresponding (non-improved) SQM as an FI parameter. Thus the bubbling term in the OPE $\vev{\ell^-}\star \vev{\ell^+}$ in \eqref{OPEellpm} should correspond to the JK residues of the non-improved SQM at positive FI parameter and the bubbling term in the OPE $\vev{\ell^+}\star \vev{\ell^-}$ in \eqref{OPEellpm} should correspond to the JK residues  at negative FI parameter. This is indeed the case. At non-zero FI parameter the extra Coulomb vacua are lifted, so there is no extra contribution. Moreover there is no $\epsilon_+ \to -\epsilon_+$ invariance here, rather this operation exchanges the $\ell^-$ and $\ell^+$ insertions. On the other hand, when the two NS5 branes are at the same position in $z$, realizing the insertion of $L_{1,0}$, the FI parameter of the SQM vanishes and Coulomb vacua contributes, requiring the more sophisticated method with the improved SQM. We see that the OPE structure is perfectly compatible with the brane picture and the SQM computations.


\section{Embedding in 5d-4d systems}
\label{sec:5d4d}

We have presented a method to compute monopole bubbling contributions to 't Hooft loops in which the loop is realized by a brane configuration in IIB string theory. In particular the 4d $\cN=2$ theory is realized by D3 and D7 branes and the loop operator is realized by embedding them in a 5-brane web. A 5-brane web by itself supports a 5d SCFT \cite{Aharony:1997bh}, which we have ignored so far.\footnote{At sufficiently low energies, the 5d theory can be considered as frozen, compared to the 4d dynamics.} 

From the point of view of the 5d theory, the presence of D3 branes introduces half-BPS line operators. This setup was studied in \cite{Assel:2018rcw}, where the 4d theory living on the D3 brane was considered non-dynamical. The line operators of the 5d theory realized by D3 branes are SQM coupled to the 5d theory in a supersymmetric way. It was found in \cite{Assel:2018rcw} that the vev of half-BPS Wilson loops of the 5d theory can be obtained by taking residues of $\vev{L_{\rm SQM}}$ in the fugacities associated to the D3-brane flavor symmetries, with $L_{\rm SQM}$ denoting an SQM loop.  This is in a sense the reversed operation to what we have been doing in this paper, where we took residues in fugacites associated to the D5-brane flavor symmetries to obtain the bubbling contribution to 't Hooft loops of the 4d theory.

In fact such D3-D7-5-brane setups really describe at low energies the coupled system of a 5d and a 4d SCFT. The two theories live on different spaces which intersect along a line. With the orientations described in Table \ref{tab:orientations}, the 4d theory lives on $x^{0123}$ and the 5d theory lives on $x^{06789}$. The two theories are coupled together by the presence of an $\cN=(0,4)$ SQM living on the $x^0$ line, which consists of bifundamental fermions charged under the 4d and 5d gauge fields, sourced by D3-D5 string. The presence of this SQM breaks half of the supersymmetries of the configuration. The partition function $\vev{L_{\rm SQM}}$ of such a system contains information about Wilson loops of the 5d theory and  't Hooft loop bubblings of the 4d theory, which can be obtained by residues in the D5 or in the D3 fugacities, therefore $\vev{L_{\rm SQM}}$ is the more fundamental object.

\medskip

The recipe to obtain Wilson loops of the 5d theory from $\vev{L_{\rm SQM}}$ is rather straightforward: one simply takes some residues in the D3 fugacities (we refer to \cite{Assel:2018rcw} for precise relations)
 The recipe to obtain monopole bubbling terms in the 4d theory is slightly more elaborate. The precise configurations for monopole bubbling in the 4d theory arise when there are stretched D1 strings between NS5 pairs. The (improved) ADHM SQM that we have discussed in previous sections is the theory living on a specific array of D1 strings. From the point of view of the 5d theory the D1 strings correspond to instanton configurations and the index $\cI$ of the SQM on the D1s corresponds to a certain instanton sector of the total partition function $\vev{L_{\rm SQM}}$. Therefore the recipe to obtain the monopole bubbling contribution is to first select the relevant instanton sector in $\vev{L_{\rm SQM}}$ and then to take residues in the D5 fugacities.
 
 \medskip

\begin{figure}[t]
\centering
\includegraphics[scale=0.75]{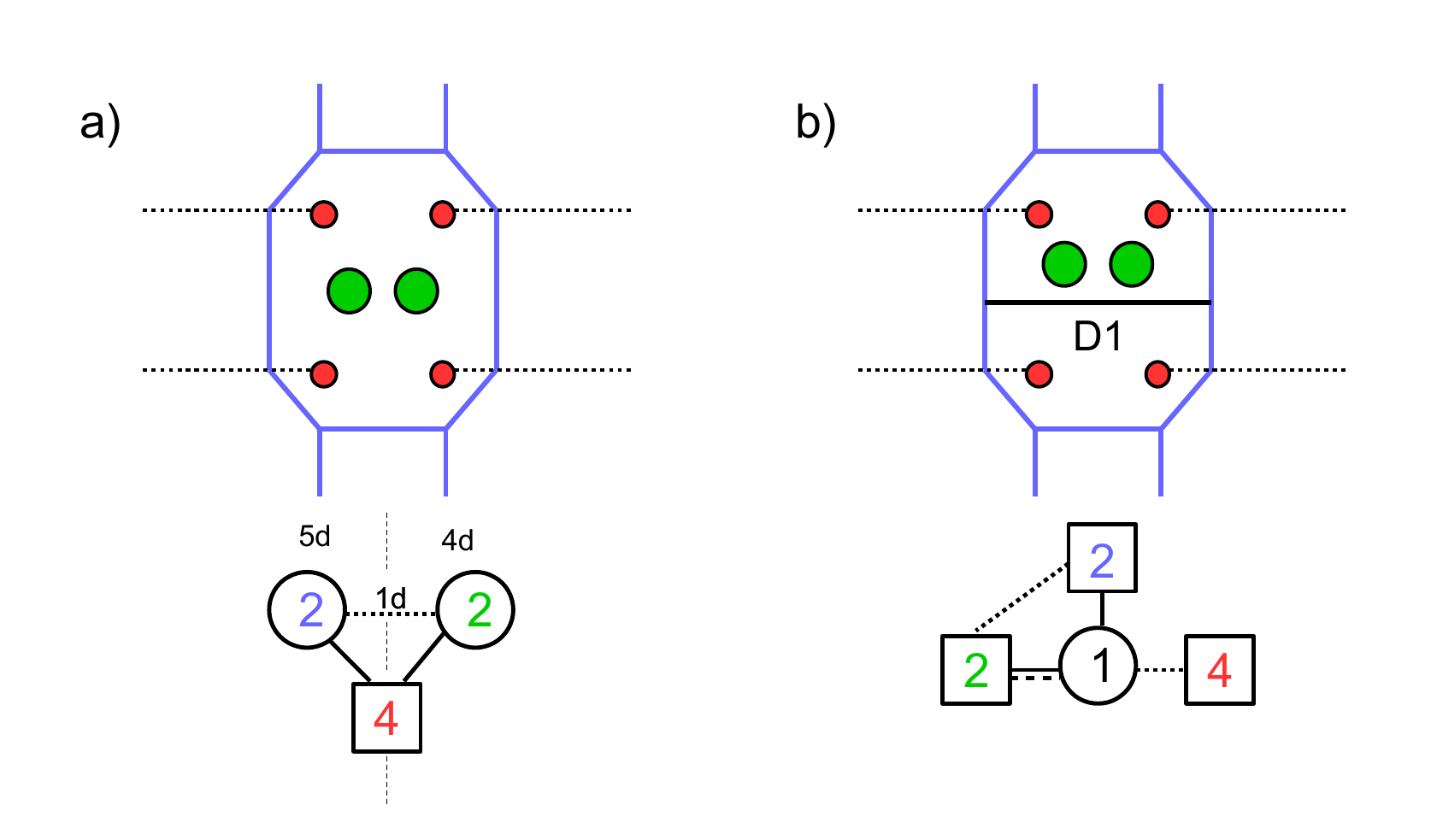}
\vspace{-0.5cm}
\caption{\footnotesize{a) Brane setup of 5d $U(2)$ $N_f=4$ coupled to 4d $U(2)$ $N_f=4$ through 1d Fermi multiplets. b) 1-instanton sector of the theory and corresponding ADHM $\cN=(0,4)^\ast$ SQM. This is the same as the monopole bubbling configuration of $L_{1}$.}}
\label{5d4dex1}
\end{figure}

Let us give an illuminating example. We consider the brane setup of Figure \ref{5d4dex1}-a. It represents a 5d-4d-1d coupled theory, where the 5d theory is an $\cN=1$ $U(2)$ SYM theory with four fundamental hypermultiplets -- this is the theory living on the D5 segments and the hypermultiplets are D5-D7 string modes -- and the 4d theory is $\cN=2$ $U(2)$ SYM with four fundamental hypermultiplets.\footnote{Truely this is an $\cN=2^\ast$ theory: there is massive adjoint hypermultiplet and we take of the decoupling limit of large mass, as explained in previous sections.}
The worldvolumes of the two theories intersect along a line which supports four 1d $\cN=(0,4)$ (single-fermion) Fermi multiplets transforming in the  bifundamental representation  of $U(2)^2$ (this means that 1d vector multiplets embedded in the 5d and 4d vector multiplets gauge the $U(2)^2$ flavor symmetry of the Fermi multiplets). These Fermi fields are D3-D5 string modes. In addition the 4d and 5d hypermultiplets are coupled by potential terms which break the flavor symmetry to the diagonal $SU(4)$ flavor symmetry. The corresponding 5d-4d-1d mixed quiver theory is given in Figure \ref{5d4dex1}-a.

We denote $\vev{L_{\rm SQM}}$ the partition function of this coupled theory, normalized by the 5d and 4d partition functions. The configuration of Figure \ref{5d4dex1}-a is not one that we have seen so far. From the point of view of the 4d theory living on the D3 branes, there is a line operator defined by the 1d Fermi multiplets living on a line, but otherwise there is no non-trivial magnetic or electric line operator. 

In the combined 5d-4d-1d theory the partition function receives non-perturbative contributions from D1 strings stretched between NS5 segments. These are seen as instanton configurations in the 5d theory. For instance the setup of Figure \ref{5d4dex1}-b corresponds to the one-instanton sector of the 5d theory. Accordingly, the observable $\vev{L_{\rm SQM}}$ admits an expansion in terms of instanton sectors weighted with a fugacity $Q$,
\be
\vev{L_{\rm SQM}} = \sum_{k\ge 0} Q^k L^{(k)}_{\rm SQM} \,.
\ee
From the point of view of the 4d theory these instanton sectors may correspond to bubbling sectors of 't Hooft (or dyonic) loops. For instance the brane setup for the one-instanton sector of Figure \ref{5d4dex1} is nothing but the complete brane setup for the bubbling sector of the minimal 't Hooft loop in the $U(2)$ $N_f=4$ theory, that we found in section \ref{sssec:tHLU2}. This is the same as Figure \ref{U2_2}. Therefore the ADHM for the one-instanton sector of $L_{\rm SQM}$ is the same as what we called the improved SQM for the bubbling contribution to $L_1$ and we have
\be
L^{(1)}_{\rm SQM} = \cI^0_1(\epsilon_-) \,.
\ee 
This index depends (as before) on the D3 fugacites $\alpha_i = e^{-a_i}$, $i=1,2$, and the D5 fugacites $w_n = e^{-v_n}$, $n=1,2$.

We can thus reformulate the residue formula \eqref{rel1} as
\be
Z^0_1 = \oint_{\cC} \frac{dw_1}{2\pi i w_1}\frac{dw_2}{2\pi i w_2} \, \lim_{\epsilon_- \to \infty} \frac{1}{\sh(\epsilon_-)^2}  L^{(1)}_{\rm SQM}(w,\alpha,\epsilon_-) \,.
\label{rel10}
\ee
We emphasize that this is not an new method to compute $Z^0_1$. It is the same computation, expressed in a larger context.

On the other hand, in \cite{Assel:2018rcw} it was found that taking residues in $\alpha_i$ computes the vev of BPS Wilson loop operators in the tensor product representation ${\bf 2 \otimes 2}$ of the $U(2)$ (or $SU(2)$) 5d gauge group\footnote{The notation in that paper are such that the $\alpha_i$ are called $x_i$, while the $w_n$ are called $\alpha_n$.}
\be
\vev{W_{\bf 2 \otimes 2}} = \oint_{\cC} \frac{d\alpha_1}{2\pi i \alpha_1}\frac{d\alpha_2}{2\pi i \alpha_2} \, L_{\rm SQM}(w,\alpha,\epsilon_-) \,,
\ee
where now the integrand is the full $L_{\rm SQM}$ vev and the residues are taken around the origin.
From the point of view of the 5d theory it makes sense to keep $\epsilon_-$ finite (it plays the role of another $\Omega$ background parameter). 

Therefore we see that the partition function of the 5d-4d system contains information both about 5d and 4d BPS line operators and constitutes a interesting object to study. It is all the more interesting that there is no simple way to isolate the 5d or 4d observables from a simpler object directly. Indeed, this object was constructed in \cite{Assel:2018rcw} in order to compute the vev of Wilson loops in 5d, for which no direct evaluation method is known. Similarly, in this paper, we have constructed the improved ADHM SQM, which is a sector of $L_{\rm SQM}$, in order to capture the full contribution of the monopole bubbling for 't Hooft loops in 5d. 

\medskip

Using the same reasoning we can associate to each monopole bubbling factor $Z_{\rm mono}(B,v)$ a given 5d-4d-1d theory from which it is extracted as a residue in a certain instanton sector. We will not work out this dictionary in general, but we will give the map for the bubbling sectors that we have studied. The 5d-4d-1d system is read directly from the complete brane setup that we have introduced and discussed at length.

The monopole bubbling contribution $Z^0_1$ of the minimal 't Hooft loop $L_1$ in the 4d $U(N)$ theory with $2N$ flavors is related to the 5d-4d-1d system, where the 5d theory is $\cN=1$ $U(h)$ theory with $2N$ flavors, with $h$ defined as in section \ref{ssec:L1Gen}. The 1d theory is composed of $Nh$ Fermi multiplets transforming in the bifundamental representation $(N,\overline h)$ of $U(N) \times U(h)$. $Z^0_1$ is obtained by residues in the D5 fugacities $w$ of the one-instanton sector of the $L_{\rm SQM}$ for this system,
\be
Z^0_1 = \oint_{\cC} \prod_{n=1}^h \frac{dw_n}{2\pi i w_n} \, F(w) \,  \lim_{\epsilon_- \to \infty} \frac{(-1)^N}{\sh(\epsilon_-)^{4N-2}} L^{(1)}_{\rm SQM}(w,\alpha,\epsilon_-) \,,
\label{rel11}
\ee
with $F(w)$ as in \eqref{rel4}.

In addition, we have seen in section \ref{sssec:simpler} that $Z^0_1$ can be obtained from a simpler 5d-4d-1d system that is obtained from the partially complete brane setup with only two D5 segments, like in Figure \ref{U3_U4_2}. The 5d theory that is realized by this setup is $\cN=1$ $U(2)$ SYM with $2N$ flavors (and Chern-Simons level $\kappa=0$). The 1d theory (in the absence of D1 strings) is a set of $2N$ Fermi fields transforming in $(N,\overline 2)$ of $U(N)\times U(2)$. The 5d theory is actually not always a genuine 5d SCFT, in the sense that the $U(2)$ theory with $N_f > 8$ flavors does not have an SCFT at `infinite coupling'. Rather it is not UV complete by itself. It requires a UV completion. For instance we can think of it as a piece of the 5d theory of the fully completed brane setup.  For the computation of the observable $L_{\rm SQM}$ there does not seem to be a difficulty in the computation and this pseudo 5d theory can be used for practical purposes.

These relations are summarized in the first row of Table \ref{tab:5d4dMap}.

\begin{table}[h]
\begin{center}
\setlength{\extrarowheight}{2pt}
\begin{tabular}{|c|c|c|c|c|c|c|}
\hline
Bubbling  & 4d  & 4d  & 5d  & 1d & Instanton  & reduced \cr
 term &  theory & loop & theory &  theory & sector &  5d theory \cr
 \hline 
$Z^0_1$ & $U(N)_{2N}$ & $L_{1}$ & $U(h)_{2N}$ & $(N,\overline h)$ Fermi & 1-inst & $U(2)_{2N}$, $\kappa=0$ \cr
\hline
$Z^{0,1}_{1,1}$ & $U(N)_{2N}$ & $L_{1,1}$ & $U(h')_{2N}$ & $(N,\overline h')$ Fermi & 1-inst & $U(2)_{2N}$, $\kappa=1$ \cr
\hline
$Z^1_2$ & $U(2)_{4}$ & $L_{2}$ & $U(2)^2\times U(4)_{4}$ & 2 bif. Fermi & (1,1,1) & $\times$ \cr
\hline
$Z^0_2$ & $U(2)_{4}$ & $L_{2}$ & $U(2)^2\times U(4)_{4}$ & $(2,4)$ Fermi & (1,2,1) & $\times$ \cr
\hline
\end{tabular}
\caption{\footnotesize Map between bubbling terms and 5d-4d-1d systems. $U(n)_{n_f}$ indicates a $U(n)$ gauge group with $n_f$ flavor hypermultiplets.}
\label{tab:5d4dMap}
\end{center}
\end{table}

\medskip

Similarly we can associate a 5d-4d-1d system to the monopole bubbling contribution $Z^{0,1}_{1,1}$ of the dyonic loop $L_{1,1}$ of 4d $U(N)$ $N_f=2N$ SYM.  The brane configuration (e.g. Figure \ref{U3DL}) indicates that the 5d theory is $U(h')$ SYM with $2N$ flavors with $h'$ defined as in section \ref{ssec:DL}. The 1d theory is composed of $Nh'$ Fermi multiplets transforming in the bifundamental representation $(N,\overline h')$ of $U(N) \times U(h')$. $Z^{0,1}_{1,1}$ is obtained by residues in the D5 fugacities $w$ of the one-instanton sector of $L'_{\rm SQM}$ for this system,
\be
Z^{0,1}_{1,1} = \oint_{\cC} \prod_{n=1}^{h'} \frac{dw_n}{2\pi i w_n} \, F'(w) \,  \lim_{\epsilon_- \to \infty} \frac{(-1)^N}{\sh(\epsilon_-)^{4N-2}} L'{}^{(1)}_{\rm SQM}(w,\alpha,\epsilon_-) \,,
\label{rel12}
\ee
with $F'(w)$ as in \eqref{rel5}. So $L'{}^{(1)}_{\rm SQM}$ is $\cI^{0,1}_{1,1}(\epsilon_-)$.
The simplified brane setup leads to a simplified pseudo 5d theory with $U(2)$ gauge group, $2N$ flavors and Chern-Simons level $\kappa=1$, with the 1d theory having a $U(N)\times U(2)$ bifundamental Fermi multiplet. This is summarized in the second row of Table \ref{tab:5d4dMap}.

\medskip

\begin{figure}[t]
\centering
\includegraphics[scale=0.8]{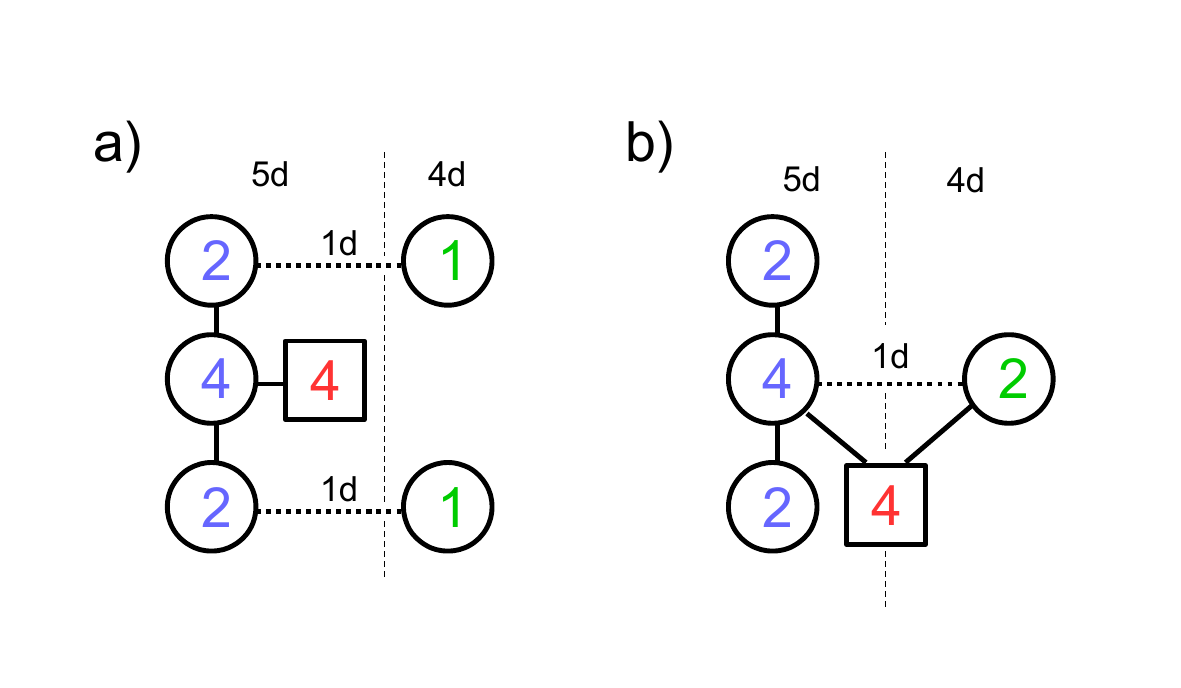}
\vspace{-0.5cm}
\caption{\footnotesize{a) 5d-4d-1d system for $Z^1_2$. b)  5d-4d-1d system for $Z^0_2$.}}
\label{5d4d1dQuiv}
\end{figure}

Finally for the bubbling terms of the non-minimal 't Hooft loop $L_{2}$ in the 4d $U(2)$ theory with four flavors, the brane setups of Figure \ref{U2L2Bub}, with the D1 strings taken out, indicate the associated 5d-4d-1d systems. For $Z^0_2$ the 5d-4d-1d system is given in Figure \ref{5d4d1dQuiv}-b. The 5d theory is a quiver with gauge group $U(2)\times U(4) \times U(2)$ and 4 flavors in the central node. 5d and 4d $U(4)$ flavor symmetries are broken to the diagonal $U(4)$ by potential terms. The 1d theory is a bifundamental Fermi multiplet between the 5d $U(4)$ and 4d $U(2)$ gauge nodes, understood in the sense described before. 
The bubbling contribution $Z^0_2$ is the $w$-residue of the instanton sector of charge $(1,2,1)$ (there are three instanton charges for three 5d gauge nodes), corresponding to the addition of the D1 strings,
\be
Z^0_2 = \oint_{\cC}  \frac{d^2w^{(1)} d^4w^{(2)} d^2w^{(3)}}{(2\pi i)^8 \prod_{n,n',n''} w^{(1)}_n w^{(2)}_{n'} w^{(3)}_{n''}}  \,  \lim_{\epsilon_- \to \infty} \frac{1}{\sh(\epsilon_-)^{2}}  L^{(1,2,1)}_{\rm SQM}(w,\alpha,\epsilon_-) \,.
\label{rel13}
\ee
So $L^{(1,2,1)}_{\rm SQM}$ is $\cI^0_2(\epsilon_-)$.

For the $Z^1_2$ bubbling the situation is slightly different because the configuration of Figure \ref{U2L2Bub}-a, with the D1 strings taken out, realize the $L_1$ loop in the 4d theory that lives on the two D3s. This is not surprising since in this bubbling sector, there is a remaining magnetic charge $v_1 = (1,-1)$ (in the previous cases the magnetic charge was completely screened). The presence of this remaining 't Hooft loop breaks the 4d gauge symmetry to $U(1)^2$ and thus the 4d theory of the corresponding 5d-4d-1d setup is a $U(1)^2$ gauge theory. The 5d-4d-1d system that is read from the brane setup is given in Figure \ref{5d4d1dQuiv}-a. The 5d theory is the same as for $Z^0_2$. The 1d theory is composed of two sets of Fermi bifundamentals, between the 5d $U(2)$ and 4d $U(1)$ nodes.

The bubbling contribution $Z^1_2$ is the $w$ residue of the 5d instanton sector of charge $(1,1,1)$, corresponding to the addition of the D1 strings. 
\be
Z^1_2 =   \oint_{\cC} \frac{d^2w^{(1)} d^4w^{(2)} d^2w^{(3)}}{(2\pi i)^8 \prod_{n,n',n''} w^{(1)}_n w^{(2)}_{n'} w^{(3)}_{n''}}  \, \lim_{\epsilon_- \to \infty} \frac{1}{\sh(\epsilon_-)^{2}}   \ti L^{(1,1,1)}(w,\alpha,\epsilon_-) \,,
\label{rel14}
\ee
So $\ti L{}^{(1,1,1)}_{\rm SQM}$ is $\cI^1_2(\epsilon_-)$. The relations for $Z^0_2$ and $Z^1_2$ are summarized in the last two rows of Table \ref{tab:5d4dMap}.

This completes the identification of the 5d-4d-1d systems and instanton sectors associated with the bubbling terms that we have studied. It is clear that this relation can be derived for any bubbling term of 't Hooft loops, once the corresponding complete brane setup is constructed. Of course, as the ranks of the 4d gauge group and the monopole magnetic charge increase, the 5d-4d-1d systems get more and more complicated.

\section{Comments and directions}
\label{sec:Discussion}

Let us make some comments about the results that we found.
We have proposed a method to compute monopole bubbling contributions to half-BPS 't Hooft loops (or dyonic loops) which relies on a brane realization of the line operator and the construction of an auxiliary ADHM SQM theory. In particular this method captures contributions to the monopole bubbling terms that are difficult to extract from other methods. 

The evidence for the validity of the results comes from several consistency checks. One check is the compatibility with the non-commutative star product discussed in section \ref{sec:starprod}. Another check is the invariance of the bubbling term under the $\bZ_2$ symmetry $\epsilon_+ \to - \epsilon_+$. Both of these tests are highly non-trivial. 

\medskip

\noindent{\bf Comparison with AGT}

Another important piece of evidence comes from the comparison with  
Verlinde loop operators of the 2d Toda CFT, which for theories of class $\mathcal{S}$ (such as $SU(N)$ conformal SQCD) relate to 't Hooft and dyonic loop vevs through the AGT correspondence \cite{Alday:2009aq, Alday:2009fs, Drukker:2009id}. As we already mentioned in the Introduction, the 2d Toda CFT computation of such loop operators vevs performed in \cite{Gomis:2010kv} does not agree with the 4d gauge theory results of \cite{Ito:2011ea}, due to the latter missing important contributions coming from the Coulomb vacua of the underlying SQM. Because of AGT, agreement between the two computations is however expected once these extra contributions are properly taken into account; this was indeed verified for the minimal 't Hooft loop of $SU(2)$ conformal SQCD in \cite{Brennan:2018rcn}. Given that our results contain also such SQM Coulomb branch contributions, we therefore expect agreement with the 2d CFT computation.

Such comparison can at the moment be performed only for minimal 't Hooft and dyonic loops in $SU(N)$ conformal SQCD, since all 2d CFT results in the literature we are aware of only consider minimal loops. Quite non-trivially, we indeed find that the monopole bubbling contribution to the minimal dyonic loop that we obtained in \eqref{Z1101final} perfectly agrees with its CFT counterpart computed in \cite{Gomis:2010kv}; nevertheless, our result for the monopole bubbling contribution to the minimal 't Hooft loop \eqref{Z10final} still differs from the CFT one presented in \cite{Gomis:2010kv} by an extra term. We however believe the correct bubbling contribution should be \eqref{Z10final}, for two simple reasons.\footnote{We thank Bruno Le Floch for a discussion on this point.} First of all, while our final result for the vevs of minimal 't Hooft and dyonic loops (and, trivially, Wilson loops) are invariant under the whole $U(2N)$ flavor symmetry group of $SU(N)$ conformal SQCD, in the computation of \cite{Gomis:2010kv} only dyonic and Wilson loops preserve the full $SU(2N)$ flavor symmetry, while the extra term in the 't Hooft loop bubbling contribution breaks this flavor symmetry to an $SU(N) \times SU(N)$ subgroup: given that electro-magnetic duality should relate Wilson, 't Hooft and dyonic loops without breaking the flavor symmetry, this is a strong hint on the fact that the result of \cite{Gomis:2010kv} must be incorrect. The second reason instead comes from realizing that, modulo an irrelevant overall prefactor, the SQM associated to the monopole bubbling contribution to the 4d $\mathcal{N} = 2$ $SU(N)$ conformal SQCD minimal 't Hooft loop is nothing else but the ADHM quantum mechanics for the 1-instanton contribution to the instanton partition function of the 5d $\mathcal{N} = 1$ $SU(N)$ $N_f = 2N$ theory, whose Coulomb vacua contribution was already derived in \cite{Hwang:2014uwa} and precisely coincides with the hyperbolic cosine in the second line of \eqref{Z10final}. A careful check of the various steps taken in \cite{Gomis:2010kv} should clarify where possible mistakes or misprint appear, but this would go beyond the scope of our work.

\medskip

\noindent{\bf Some remaining issues}

Although the method presented is quite successful, there are still some points that should be improved. 
One point has to do with the brane construction and the placement of D7 branes relative to NS5 branes in the completion of the 't Hooft loop brane setup. We have given prescriptions that work for minimal and next-to-minimal 't Hooft loops in $SU(N)$ conformal SQCD theories. The prescription is to place the D7 branes in the middle of the NS5 branes and to distribute them evenly in the upper and lower part of the picture. There are other possible choices though, in particular for 't Hooft loops realized with many NS5 branes. It would be desirable to understand precisely the dictionary between the loop realized and the distribution of D7 branes, relatively to the NS5 branes. This is a point that deserves further study. The same comment applies to the realization of dyonic loops, which we have not studied in detail.

Another issue is that we do not have a precise way to fix the overall sign of a given contribution $Z_{\rm mono}(B,v)$. In particular the ADHM SQM involve fermion determinants, which have sign ambiguities. The same is true of the one-loop determinants, which involve square-roots and thus have sign ambiguities. The way we have fixed some of the signs is by imposing compatibility with the non-commutative product (like in \eqref{rel20}).

\medskip

\noindent{\bf Lower $N_f$}

In this paper we have focused on the conformal SQCD theories, namely $SU(N)$ (or $U(N)$) theories with $2N$ flavor hypermultiplets. A natural extension is to study line operators in non-conformal theories, namely $SU(N)$ theories with $N_f < 2N$ flavors. Our construction naturally generalizes to these theories by looking at brane setups with $N$ D3 branes and $N_f < 2N$ D7 branes, placed in 5-brane webs. The computations are just the same and yield similar exact results for the bubbling terms. The issue of the placement of D7 branes is also present in these situations, in particular for odd $N_f$, namely odd number of D7s, one is forced to break the symmetry between the upper and lower parts of the brane configuration, by putting more D7s `upstairs', or `downstairs'. This yields different  choices brane setups, giving different answers for a given bubbling contribution. There is a priori no choice of brane setup that is preferable to the others and we obtain (at least) two answers for the bubbling term. Similarly it was observed in \cite{Brennan:2018rcn} that bubbling contributions at $N_f < 2N$ might be obtained by taking hypermultiplet masses to infinity in the $N_f=2N$ result. However the outcome of this procedure depends on whether one sends the mass(es) to $+\infty$ or $-\infty$. Actually the results obtained this way correspond to the different choices of D7 brane placements in our construction. It is not clear to us how this puzzle will be resolved, however one possibility is simply that the bubbling contribution to the 't Hooft loop has ambiguities. This would mean that the difference between the two bubbling answers is the vev of a line operator with no magnetic charge, which can be included or not in a regularization of the bubbling term. It would be interesting to study this issue further.\footnote{It was also suggested to us by D. Brennan, A. Dey and G. Moore that the limit of large mass from the conformal case may involve redefinitions/limits of the Fenchel-Nielsen coordinates $a_i,b_i$, in a way that would resolve the tension between the results.}

\medskip

\noindent{\bf Observables of 5d-4d systems}

An important lesson from this discussion is that one should study the 5d-4d(-1d) coupled system and the observables $L_{\rm SQM}$, which are the `parent observables' of BPS loops in both 5d SCFTs and 4d SCFTs (see section \ref{sec:5d4d}). These are more difficult to manipulate since they are given by infinite sums over instanton sectors, but they are the more fundamental objects. It is unclear to us at the moment whether one can define 5d-4d obervables $L_{\rm SQM}$ which contain the full 4d 't Hooft loops, instead of only the monopole bubbling contributions. We believe that such objects can be defined and are naturally associated to the D3-5web brane systems. It would be interesting to explore the mathematical properties of such objects, as for instance the invariance under S-duality transformation of the IIB brane setup \cite{Assel:2018rcw}.

\medskip

\noindent{\bf Quiver theories and type IIA setups}

Finally, a natural extension of our work is to carry out the computation of 't Hooft loop bubbling factors in quiver theories. The brane configurations for those are in type IIA string theory, with D4 branes suspended between NS5 branes and crossing D6 branes. It would be interesting to see how the construction works there.

\section*{Acknowledgements}

It is a pleasure to thank Stefano Cremonesi, Takuya Okuda for discussions, D. Brennan, A. Dey, G. Moore and Bruno Le Floch for email exchanges.


\appendix

\section{SQM matrix models}
\label{app:MM}

In this appendix we provide the explicit form of the matrix model computing the partition function of an $\cN=(0,4)$ SQM in terms of Jeffrey-Kirwan residues. 

The global symmetry of the SQM is $SU(2)_+\times SU(2)_1 \times SU(2)_2$, with chemical potentials $-2 \epsilon_+,\epsilon_1$ and $\epsilon_2$ respectively, and the R-symmetry is the $SU(2)_+\times {\rm diag}(SU(2)_1 \times SU(2)_2)$ subgroup. To preserve supersymmetry the deformation parameters obey $\epsilon_+ = \frac 12 (\epsilon_1 + \epsilon_2)$, so we have only two independent parameters. The parameter $\epsilon_- = \frac 12 (\epsilon_1 - \epsilon_2)$ corresponds to the adjoint mass $m$ of the 4d adjoint hypermultiplet of the $\cN=2^\star$ theory on the D3 branes. It should be send to infinity to obtain the final results. We do not send it to infinity immediately because it acts as a regulator in (some) computations, so it is convenient to send it to infinity only after performing residue computations.

In the brane picture the global symmetries are identified as $SO(3)_{123} \sim SU(2)_+$, $SO(4)_{6789} \sim SU(2)_1 \times SU(2)_2$. The chemical potentials $-2\epsilon_+,\epsilon_1$ and $\epsilon_2$ correspond to Omega background parameters in the planes $x^{23}$, $x^{67}$ and $x^{89}$, respectively.

The D1-D1 modes provide a $(0,4)^\star$ vector multiplet, which comprise a (0,4) vector multiplet and a massive (0,4) adjoint twisted hypermultiplet.
The D1-D3 modes provide a $(0,4)^\star$ fundamental hypermultiplet, which comprise a (0,4) hypermultiplet and a massive (0,4) Fermi multiplet (with multiple fermions). The D1-D7 modes provide a (0,4) Fermi multiplet (with a single fermion). The D1-D1 modes across an NS5 brane provide a $(0,4)^\star$ bifundamental hypermultiplet.
The massive multiplets in the $(0,4)^\star$ SQM have mass parameter $m = \epsilon_-$. Some of these masses are shifted in the background further deformed with $\epsilon_+$.

The partition function, or index, depends on the FI parameters of the SQM. When these FI parameters are non-zero, the index can be evaluated by a matrix model integral with the JK pole prescription by taking the JK parameters $\zeta_i$ to be the FI parameters.\footnote{We thank Stefano Cremonesi for discussions on this point.} When an FI parameter is zero, one can pick any sign for the JK parameter, but the evaluation may miss some contributions from ``poles at infinity" (Coulomb vacua). This happens when the potential of the SQM (or of the matrix model) is not sufficiently divergent.

The matrix model computing the index takes the form
\be
Z = \int \frac{d^r \phi}{(2\pi i)^r |\cW |} Z_{\rm vec} \, Z_{\rm f-hyp} \, Z_{\rm F} \, Z_{\rm bf-hyp} \, Z_{\rm CS} \,,
\ee
with $r$ the rank of the gauge group and $|\cW|$ the order for the Weyl group $\cW$. 
The term $Z_{vec}$ contains the contribution of a $(0,4)^\star$ vector multiplet. For a $U(k)$ gauge group, using the notation $\sh(x) := 2 \sinh( \frac x2)$, it is given by
\be
Z_{\rm vec} = \prod_{i\neq j} \sh(\phi_i - \phi_j)  \prod_{i,j} \frac{\sh(\phi_i-\phi_j - 2\epsilon_+)}{\sh(\phi_i - \phi_j -\epsilon_1)\sh(\phi_i-\phi_j - \epsilon_2)} \,.
\ee
The contribution of the $(0,4)^\star$ hypermultiplet with mass $m_{\rm h}$ in the fundamental representation of $U(k)$ is
\be
Z_{\rm f-hyp} = \prod_{i=1}^k \frac{ \sh [\pm (\phi_i + m_{\rm h} ) + \epsilon_- ] }{\sh [\pm (\phi_i + m_{\rm h} )+ \epsilon_+] } \,.
\ee
The contribution of the (0,4) Fermi multiplet with mass $m_{\rm F}$ in the fundamental representation of $U(k)$ is
\be
Z_{\rm F} = \prod_{i=1}^k \sh (\phi_i + m_{\rm F}) \,.
\ee
The contribution of the $(0,4)^\star$ bifundamental hypermultiplet is 
\be
Z_{\rm bf-hyp} = \prod_{i=1}^{k_1}\prod_{j=1}^{k_2} \frac{\sh [\pm(\phi_i- \hat \phi_j) + \epsilon_- ] }{\sh [\pm(\phi_i- \hat \phi_j) + \epsilon_+ ] } \,.
\ee
The Chern-Simons term contribution at level $\kappa$, takes the form
\be
Z_{\rm CS} = e^{-\kappa \sum_{i=1}^k \phi_i} \,.
\ee
We used the notation $f(a \pm b) = f(a+b)f(a-b)$.

In addition, when there are D5 branes in the construction, we have D1-D5 modes giving rise to (0,4) fundamental hypermultiplets, which contribute a factor
\be
Z^{(0,4)}_{\rm f-hyp} = \prod_{i=1}^k \frac{1}{\sh [\pm (\phi_i + m_{\rm h} ) - \epsilon_+] } \,.
\ee
There are also D3-D5 modes -- and possibly D1-D5 modes from string going across an NS5 brane --  giving a Fermi multiplet with contribution $Z_{\rm F}$.


\bibliography{5dbib}

\providecommand{\href}[2]{#2}\begingroup\raggedright\begin{thebibliography}{10}

\bibitem{tHooft:1977nqb}
G.~'t~Hooft, \emph{On the phase transition towards permanent quark
  confinement},
  \href{http://dx.doi.org/10.1016/0550-3213(78)90153-0}{\emph{Nucl.Phys.B} {\bf
  138} (1978) 1--25}.

\bibitem{Kapustin:2005py}
A.~Kapustin, \emph{{Wilson-'t Hooft operators in four-dimensional gauge
  theories and S-duality}},
  \href{http://dx.doi.org/10.1103/PhysRevD.74.025005}{\emph{Phys. Rev.} {\bf
  D74} (2006) 025005}, [\href{https://arxiv.org/abs/hep-th/0501015}{{\tt
  hep-th/0501015}}].

\bibitem{Gaiotto:2008cd}
D.~Gaiotto, G.~W. Moore and A.~Neitzke, \emph{Four-dimensional wall-crossing
  via three-dimensional field theory},
  \href{http://dx.doi.org/10.1007/s00220-010-1071-2}{\emph{Commun.Math.Phys.}
  {\bf 299} (2010) 163--224}, [\href{https://arxiv.org/abs/0807.4723}{{\tt
  0807.4723}}].

\bibitem{Alday:2009aq}
L.~F. Alday, D.~Gaiotto and Y.~Tachikawa, \emph{{Liouville Correlation
  Functions from Four-dimensional Gauge Theories}},
  \href{http://dx.doi.org/10.1007/s11005-010-0369-5}{\emph{Lett.Math.Phys.}
  {\bf 91} (2010) 167--197}, [\href{https://arxiv.org/abs/0906.3219}{{\tt
  0906.3219}}].

\bibitem{Gomis:2010kv}
J.~Gomis and B.~Le~Floch, \emph{'t hooft operators in gauge theory from toda
  cft}, \href{http://dx.doi.org/10.1007/JHEP11(2011)114}{\emph{JHEP} {\bf 11}
  (2011) 114}, [\href{https://arxiv.org/abs/1008.4139}{{\tt 1008.4139}}].

\bibitem{Ito:2011ea}
Y.~Ito, T.~Okuda and M.~Taki, \emph{{Line operators on $S^1xR^3$ and
  quantization of the Hitchin moduli space}},
  \href{http://dx.doi.org/10.1007/JHEP03(2016)085,
  10.1007/JHEP04(2012)010}{\emph{JHEP} {\bf 04} (2012) 010},
  [\href{https://arxiv.org/abs/1111.4221}{{\tt 1111.4221}}].

\bibitem{Gomis:2011pf}
J.~Gomis, T.~Okuda and V.~Pestun, \emph{{Exact Results for 't Hooft Loops in
  Gauge Theories on $S^4$}},
  \href{http://dx.doi.org/10.1007/JHEP05(2012)141}{\emph{JHEP} {\bf 05} (2012)
  141}, [\href{https://arxiv.org/abs/1105.2568}{{\tt 1105.2568}}].

\bibitem{Nekrasov:2010ka}
N.~Nekrasov and E.~Witten, \emph{The omega deformation, branes, integrability,
  and liouville theory},
  \href{http://dx.doi.org/10.1007/JHEP09(2010)092}{\emph{JHEP} {\bf 09} (2010)
  092}, [\href{https://arxiv.org/abs/1002.0888}{{\tt 1002.0888}}].

\bibitem{Gaiotto:2010be}
D.~Gaiotto, G.~W. Moore and A.~Neitzke, \emph{{Framed BPS States}},
  \href{http://dx.doi.org/10.4310/ATMP.2013.v17.n2.a1}{\emph{Adv. Theor. Math.
  Phys.} {\bf 17} (2013) 241--397},
  [\href{https://arxiv.org/abs/1006.0146}{{\tt 1006.0146}}].

\bibitem{Brennan:2018yuj}
T.~D. Brennan, A.~Dey and G.~W. Moore, \emph{{On 't Hooft Defects, Monopole
  Bubbling and Supersymmetric Quantum Mechanics}},
  \href{https://arxiv.org/abs/1801.01986}{{\tt 1801.01986}}.

\bibitem{Kapustin:2006pk}
A.~Kapustin and E.~Witten, \emph{Electric-magnetic duality and the geometric
  langlands program},
  \href{http://dx.doi.org/10.4310/CNTP.2007.v1.n1.a1}{\emph{Commun.Num.Theor.Phys.}
  {\bf 1} (2007) 1--236}, [\href{https://arxiv.org/abs/hep-th/0604151}{{\tt
  hep-th/0604151}}].

\bibitem{Nekrasov:2002qd}
N.~A. Nekrasov, \emph{{Seiberg-Witten prepotential from instanton counting}},
  \href{http://dx.doi.org/10.4310/ATMP.2003.v7.n5.a4}{\emph{Adv.Theor.Math.Phys.}
  {\bf 7} (2004) 831--864}, [\href{https://arxiv.org/abs/hep-th/0206161}{{\tt
  hep-th/0206161}}].

\bibitem{Hori:2014tda}
K.~Hori, H.~Kim and P.~Yi, \emph{{Witten Index and Wall Crossing}},
  \href{http://dx.doi.org/10.1007/JHEP01(2015)124}{\emph{JHEP} {\bf 01} (2015)
  124}, [\href{https://arxiv.org/abs/1407.2567}{{\tt 1407.2567}}].

\bibitem{Hwang:2014uwa}
C.~Hwang, J.~Kim, S.~Kim and J.~Park, \emph{{General instanton counting and 5d
  SCFT}}, \href{http://dx.doi.org/10.1007/JHEP07(2015)063,
  10.1007/JHEP04(2016)094}{\emph{JHEP} {\bf 07} (2015) 063},
  [\href{https://arxiv.org/abs/1406.6793}{{\tt 1406.6793}}].

\bibitem{Brennan:2018rcn}
D.~T. Brennan, A.~Dey and G.~W. Moore, \emph{{'t Hooft Defects and Wall
  Crossing in SQM}},  \href{https://arxiv.org/abs/1810.07191}{{\tt
  1810.07191}}.

\bibitem{Brennan:2018moe}
T.~D. Brennan, \emph{{Monopole Bubbling via String Theory}},
  \href{https://arxiv.org/abs/1806.00024}{{\tt 1806.00024}}.

\bibitem{Aharony:1997bh}
O.~Aharony, A.~Hanany and B.~Kol, \emph{{Webs of (p,q) five-branes,
  five-dimensional field theories and grid diagrams}},
  \href{http://dx.doi.org/10.1088/1126-6708/1998/01/002}{\emph{JHEP} {\bf 01}
  (1998) 002}, [\href{https://arxiv.org/abs/hep-th/9710116}{{\tt
  hep-th/9710116}}].

\bibitem{Assel:2018rcw}
B.~Assel and A.~Sciarappa, \emph{{Wilson loops in 5d $\mathcal{N}=1$ theories
  and S-duality}}, \href{http://dx.doi.org/10.1007/JHEP10(2018)082}{\emph{JHEP}
  {\bf 10} (2018) 082}, [\href{https://arxiv.org/abs/1806.09636}{{\tt
  1806.09636}}].

\bibitem{Kim:2016qqs}
H.-C. Kim, \emph{{Line defects and 5d instanton partition functions}},
  \href{http://dx.doi.org/10.1007/JHEP03(2016)199}{\emph{JHEP} {\bf 03} (2016)
  199}, [\href{https://arxiv.org/abs/1601.06841}{{\tt 1601.06841}}].

\bibitem{Nekrasov:2015wsu}
N.~Nekrasov, \emph{{BPS/CFT correspondence: non-perturbative Dyson-Schwinger
  equations and qq-characters}},
  \href{http://dx.doi.org/10.1007/JHEP03(2016)181}{\emph{JHEP} {\bf 03} (2016)
  181}, [\href{https://arxiv.org/abs/1512.05388}{{\tt 1512.05388}}].

\bibitem{Tong:2014yna}
D.~Tong, \emph{{The holographic dual of $AdS_{3} \times S^{3} \times S^{3}
  \times S^{1}$}}, \href{http://dx.doi.org/10.1007/JHEP04(2014)193}{\emph{JHEP}
  {\bf 04} (2014) 193}, [\href{https://arxiv.org/abs/1402.5135}{{\tt
  1402.5135}}].

\bibitem{Tong:2014cha}
D.~Tong and K.~Wong, \emph{{Instantons, Wilson lines, and D-branes}},
  \href{http://dx.doi.org/10.1103/PhysRevD.91.026007}{\emph{Phys. Rev.} {\bf
  D91} (2015) 026007}, [\href{https://arxiv.org/abs/1410.8523}{{\tt
  1410.8523}}].

\bibitem{Witten:1979ey}
E.~Witten, \emph{{Dyons of Charge e theta/2 pi}},
  \href{http://dx.doi.org/10.1016/0370-2693(79)90838-4}{\emph{Phys. Lett.} {\bf
  B86} (1979) 283--287}.

\bibitem{Hanany:1996ie}
A.~Hanany and E.~Witten, \emph{{Type IIB superstrings, BPS monopoles, and
  three-dimensional gauge dynamics}},
  \href{http://dx.doi.org/10.1016/S0550-3213(97)00157-0}{\emph{Nucl.Phys.} {\bf
  B492} (1997) 152--190}, [\href{https://arxiv.org/abs/hep-th/9611230}{{\tt
  hep-th/9611230}}].

\bibitem{Gaiotto:2008ak}
D.~Gaiotto and E.~Witten, \emph{{S-Duality of Boundary Conditions In N=4 Super
  Yang-Mills Theory}},
  \href{http://dx.doi.org/10.4310/ATMP.2009.v13.n3.a5}{\emph{Adv. Theor. Math.
  Phys.} {\bf 13} (2009) 721--896},
  [\href{https://arxiv.org/abs/0807.3720}{{\tt 0807.3720}}].

\bibitem{Alday:2009fs}
L.~F. Alday, D.~Gaiotto, S.~Gukov, Y.~Tachikawa and H.~Verlinde, \emph{{Loop
  and surface operators in N=2 gauge theory and Liouville modular geometry}},
  \href{http://dx.doi.org/10.1007/JHEP01(2010)113}{\emph{JHEP} {\bf 01} (2010)
  113}, [\href{https://arxiv.org/abs/0909.0945}{{\tt 0909.0945}}].

\bibitem{Drukker:2009id}
N.~Drukker, J.~Gomis, T.~Okuda and J.~Teschner, \emph{{Gauge Theory Loop
  Operators and Liouville Theory}},
  \href{http://dx.doi.org/10.1007/JHEP02(2010)057}{\emph{JHEP} {\bf 02} (2010)
  057}, [\href{https://arxiv.org/abs/0909.1105}{{\tt 0909.1105}}].

\end{thebibliography}\endgroup
\bibliographystyle{JHEP}

\end{document}